\documentclass[12pt]{amsart}

\newtheorem{Theorem}{Theorem}[section]

\newtheorem{Assumption 2}[Theorem]{Assumption 2}

\numberwithin{equation}{section}

\usepackage{float}
\usepackage{amsmath}
\usepackage{amssymb}
\usepackage{epsfig}
\usepackage{caption}
\usepackage{yhmath}

\usepackage{stmaryrd}
\usepackage{ulem}
\usepackage{yfonts}
\usepackage{amsbsy}

\def\F{{\mathcal F}}
\def\DD{{\mathbb D}}

\def\KK{{\mathbb K}}
\def\NN{{\mathbb N}}
\def\MM{{\mathbb M}}
\def\P{{\mathbb{P}}}
\def\RR{{\mathbb{R}}}
\def\HH{{\mathbb{H}}}
\def\ZZ{{\mathbb Z}}

\def\fin{{\begin{flushright}$\square$\end{flushright}}}

\begin{document}
\include{00README.txt}
\title{Waves on accelerating dodecahedral universes}
\author{A. BACHELOT-MOTET and A. BACHELOT}
\address{Universit\'e de Bordeaux, Institut de Math\'ematiques, UMR CNRS 5251, F-33405 Talence Cedex}
\email{marie-agnes.bachelot@u-bordeaux.fr, alain.bachelot@u-bordeaux.fr}

\begin{abstract}
We investigate the wave propagation on a compact $3$-manifold of constant
positive curvature with a non trivial topology, the Poincar\'e
dodecahedral space, if the scale factor is exponentially increasing. We prove
the existence of a limit state as $t\rightarrow+\infty$ and we get its
analytic expression. The deep sky is described by this asymptotic
profile thanks to the Sachs-Wolfe formula. We transform the Cauchy problem into a
mixed problem posed on a fundamental domain determined by the
quaternionic calculus. We perform an accurate scheme of computation: we
employ a variational method using a space
of second order finite elements that is invariant under the action of the binary
icosahedral group.

\end{abstract}

\maketitle
\section{Introduction}
The topology and the geometry of our universe is a fundamental open
problem. There are a lot of articles about these topics (for example
\cite{Lumi-1,Lumi-2,Wee-1}). Recently the data by WMAP (Wilkinson Microwave
Anisotropy Probe) mission have provided strong evidence suggesting
that the observable universe is nearly flat, with the ratio of its total
matter-energy density to the critical value very close to one
\cite{WMAP}, but without fixing the sign of its curvature. So the
study of the wave propagation in universes involving the Poincar\'e dodecahedral space continues to
be relevant in particular for the search of the circles-in-the-sky
signature \cite{Aur}, \cite{Aur2006}, \cite{mota}. This model has received a lot of interest and generated numerous works (see {\it e.g.} \cite{Aur,PDS, C-LR-L-L-R-W,Lach,Lumi-Wee,Rouk,Wee-2}). These spacetimes are
Lorentzian manifolds $(\RR_t\times\mathbf{K},g)$ that are globally
hyperbolic and multi-connected. Here $\mathbf K$ is the
Poincar\'e dodecahedral space that is a three dimensional $C^{\infty}$
 compact manifold, without boundary, defined as the quotient of the 3-sphere $\mathcal{S}^3 $ under
the action of the binary
icosahedral group $\mathcal{I}^*$ (\cite{G-L-Lu-Uz-W, Sco, We-Sei,Wolf}). In \cite{PDS} we have investigated the
propagation of the scalar waves in the case of
the stationary universe for which $ds^2=dt^2-ds^2_{\mathbf{K}}$ where
$ds^2_{\mathbf{K}}$ is
the Euclidean metric induced by the three-sphere $\mathcal{S}^3$ on
$\mathbf{K}$. In this paper we study the wave propagation on
dynamical universes $(\mathcal{U},g)$ of Robertson-Walker type
\begin{equation}
 \label{U}
 \mathcal{U}=\RR_t\times\mathbf{K},\;\;ds_{\mathcal{U}}^2=g_{\mu\nu}dx^{\mu}dx^{\nu}=dt^2-a^2(t)ds^2_{\mathbf{K}},
\end{equation}
where the scale factor $a(t)$ is a smooth positive function. We are
mainly interested in considering the accelerating spacetimes for which
the scale factor is an increasing function with an exponential
behaviour, $a(t)\sim e^{\alpha t}$ as $t\rightarrow\infty$. In the
study of the asymptotics of the fields at the infinity, and for the numerical experiments, we treat two important
cases. In the first one, we choose $a(t)=H^{-1}\cosh(Ht)$ where $H>0$ is the
Hubble constant. Then $g$ is solution of the vacuum Einstein
equations with cosmological constant $\Lambda=3H^2$ and $\mathcal{U}$
is locally isometric to the de Sitter spacetime $d\mathcal{S}_4$ and
differs globally from $d\mathcal{S}_4$ by its topology that is not trivial. We
denote this universe $d\mathcal{S}_4\left(\mathbf K\right)$. In particular,
while $d \mathcal{S}_4$ is spatially
homogeneous and globally isotropic, $d \mathcal{S}_4(
  \mathbf K)$ is homogeneous but not
globally isotropic \cite{McInnes}. In the second model we choose
$a(t)=e^{t}$. Then $g$ is no longer a solution of the vacuum Einstein
equations but it is a toy model for a universe with an
inflation
generated by a scalar field or a
solution in $f(T)$ theory (see {\it e.g.} \cite{hanafy}).\\

In this paper we investigate the scalar waves on $(\mathcal{U},g)$
that are the solutions of the D'Alembert equation
$$
{\Box}_{g}\Psi:=\frac{1}{\sqrt{|g|}} \partial_\mu\left(g^{\mu \nu}
\sqrt{|g|}\partial_\nu\Psi\right)=0.
$$
For the spacetime (\ref{U}), this equation has the form
\begin{equation}
 \label{eq}
 \left[\partial_t^2+3\frac{a'(t)}{a(t)}\partial_t-\frac{1}{a^2(t)}\Delta_{\mathbf K}\right]\Psi=0,
\end{equation}
where $\Delta_{\mathbf K}$ is the Laplace-Beltrami operator on
$\mathbf{K}$. This equation plays a fundamental role in the theory of
cosmological perturbations (see {\it e.g.} \cite{mukhanov}). It arises in the
linearized Einstein equations in the vacuum, written in harmonic coordinates, and in this case $\Psi$ is the fluctuation around the
ground metric, {\it i.e.} $\Psi$ describes the gravitational
waves. On the other hand the Sachs-Wolfe theorem assures that in the
Friedmann-Robertson-Walker universe with an ultrarelativistic gas, the
scalar perturbations obey to this acoustic equation. This result was
proved originally for the spatially flat metric, and has been extended to
arbitrary space curvature \cite{czaja}, \cite{golda}. In this context,
considered in \cite{Aur}, \cite{Aur2006},
we write the ground metric (\ref{U}) as
$$
ds^2_{\mathcal
  U}=A(\eta)\left[d\eta^2-ds_{\mathbf{K}}^2\right],\;\;ds^2_{\mathbf{K}}=d\chi^2+\sin^2\chi\left(d\theta^2+\sin^2\theta
  d\varphi^2\right),
$$
where $\eta=\int a^{-1}(t) dt$ is the conformal time, and $A(\eta)=a(t)$ is
the cosmic scale factor. Then, the metric with the scalar perturbation
$\Psi$ takes the form in conformal Newtonian gauge:
\begin{equation}
 \label{gperturb}
 ds^2=A(\eta)\left[(1+2\Psi)d\eta^2-(1-2\Psi)ds_{\mathbf{K}}^2\right].
\end{equation}
The knowledge of $\Psi$ allows to adress a fundamental issue of the cosmology: the study of
the temperature fluctuations $\delta T(\theta,\varphi)$ of the
temperature $T(\theta,\varphi)$ of the cosmic microwave background
radiation, observed in the direction $(\theta,\varphi)\in S^2$. The
integrated Sachs-Wolfe formula (see {\it i.e.} \cite{Aur}) gives a link
between $\delta T/T$ observed at the conformal time $\eta_0$ of our
present epoch, and $\Psi$ between $\eta_0$ and $\eta_{ls}$, the
conformal time at recombination (\cite{Aur}, formula (31)). Therefore
we must know the time evolution of $\Psi$ to be able to use it. An
approximated formula for low spherical harmonics is also given by the
ordinary Sachs-Wolfe formula,
\begin{equation}
 \label{swapprox}
 \frac{\delta T(\theta,\varphi)}{T}\simeq\frac{1}{3}\Psi(\eta_{ls},\chi=\eta_0-\eta_{ls},\theta,\varphi).
\end{equation}
If $\eta_0-\eta_{ls}>\arccos\left(\frac{\sigma^2}{2\sqrt{2}}\right)$
where $\sigma$ is the golden ratio, the horizon radius of the observer
is larger than the injectivity radius, hence repeating images of the
sphere horizon intersect each other. Since the self-intersections of the sphere
of last scattering are circles, the fluctuations of a realistic
$\Psi(\eta_{ls},\chi=\eta_0-\eta_{ls},.)$, for instance if the
primordial fluctuation amplitude $\Psi(0,.)$ is a Gaussian random variable,
would be correlated around circles: the observer can see
circles-in-the-sky \cite{cornish}.

Besides, we remark that, as regards the propagation of the electromagnetic fields in the curved
space-times, the D'Alembert operator
${\Box}_{g}=\nabla^{\alpha}\partial_{\alpha}$ is also the principal part of
the Maxwell equation for the vector potential
$\nabla^{\alpha}\nabla_{\alpha}A^{\beta}-R^{\beta}_{\lambda}A^{\lambda}=J^{\beta}$. In concluding,
there are numerous motivations to study this equation, represent and
compute the solutions,
investigate their properties.\\

The explicit form of the solutions is known for the classic
F-R-W space-time with the $\RR\times S^3$ topology
\cite{klainerman}. The case of the dodecahedral space is much more
complicated. In this work we solve the Cauchy problem, we
represent the solutions in terms of expansions involving the
eigenmodes of the Laplacian on $\mathbf{K}$, then we investigate the
asymptotic behaviour of $\Psi$ at the time infinity, and finally we develop a numerical scheme of
accurate computation. A crucial result is the exponentially fast
convergence of $\Psi(t,X)$ to a final
asymptotic state $\Psi_{\infty}(X)$ when $t$ tends to the
infinity. The link between $\Psi_{\infty}$ and the initial data of
$\Psi$ is given by a pseudodifferential operator. In particular for
$d\mathcal{S}_4\left(\mathbf K\right)$, we have
\begin{equation}
 \label{zeglop} \Psi_{\infty}=\left(-\Delta_{\mathbf{K}}+1\right)^{-\frac{1}{2}}\sin\left(\frac{\pi}{2}\sqrt{-\Delta_{\mathbf{K}}+1}\right)\Psi(0,.)+\frac{\sqrt{15}}{4}\int_{\mathbf{K}}\partial_t\Psi(0,.)\omega_{\mathbf{K}},
\end{equation}
where $\omega_{\mathbf{K}}$ is the volume form on
$\mathbf{K}$. To compute $\Psi_{\infty}$ with this formula we should
use the eigenfunctions of the Laplacian. Despite significant progress
for the determination of these modes (see in particular
\cite{bellon}, \cite{Lach}),
the implementation of this strategy seems to be challenging. Instead,
we solve numerically the initial value problem, and taking advantage
of the fast convergence, we simply find the
asymptotic profile as
$$
\Psi_{\infty}(X)=\lim_{t\rightarrow+\infty}\Psi(t,X)\sim \Psi(T,X),\;\;T>>1.
$$
From the point of view of the theory of cosmological perturbations, this result has a great interest if $\Psi$ is the scalar perturbation
in the metric (\ref{gperturb}). If the conformal time of recombination
is large enough, we have $\Psi(\eta_{ls},.)\simeq\Psi_{\infty}$. Then
the approximated Sachs-Wolfe formula becomes
\begin{equation}
 \label{}
  \frac{\delta T(\theta,\varphi)}{T}\simeq\frac{1}{3}\Psi_{\infty}(\chi=\eta_0-\eta_{ls},\theta,\varphi),
\end{equation}
and the circles-in-the-sky have simply to be found in $\Psi_{\infty}$.\\

\begin{figure}[!h]
\centerline{\includegraphics[scale=0.2]{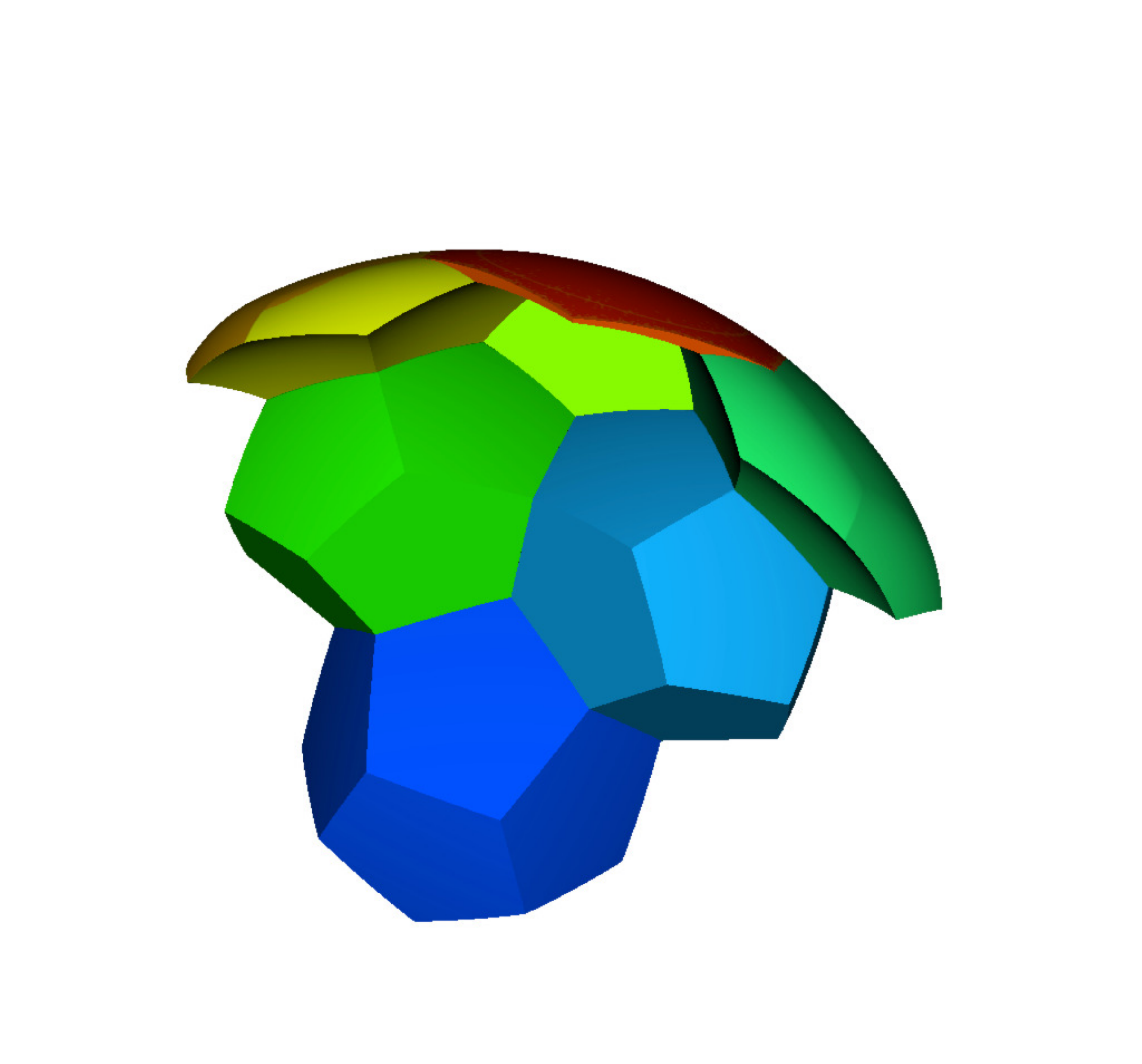}}
\caption{\label{DODIM} The Fundamental Domain (dark blue).
We display $\mathcal{F}_v$ (in dark blue) and the projection $p$ in
$\RR^3$ of the images
of $\mathcal{F}$ by six elements of $\mathcal{I}^*$. It is the
representation of a part, included in ${\mathcal B}'$, of the tiling of $\mathcal{S}^3$.} 
\end{figure}

We briefly describe the following sections. In part 2 we solve the global Cauchy problem for
(\ref{eq}) in the functional framework associated to the energy
\begin{equation}
 \label{energy}
  E(\Psi,t):=\Vert\partial_t\Psi(t)\Vert^2_{L^2(\mathbf
   K)}+\frac{1}{a^2(t)}\Vert\nabla_{\mathbf K}\Psi(t)\Vert^2_{L^2(\mathbf K)}.
\end{equation}
We represent the Poincar\'e dodecahedral space by its fundamental
domain $\mathcal{F}$ that is the geodesic convex hull of twenty vertices
on $\mathcal{S}^3$ that are included in
$\{x_0=\frac{\sigma^2}{2\sqrt{2}}\}\times\RR^3$ where $\sigma$ is the
golden ratio.  We introduce the domain of vizualisation
$\mathcal{F}_v\subset\RR^3$ that is the projection
$(x_0,x_1,x_2,x_3)\in \mathcal{S}^3\mapsto (x_1,x_2,x_3)\in\RR^3$ of
the fundamental domain. $\mathcal{F}_v$ is a dodecahedron with
pentagonal curved faces included in ellipsoids (figure(\ref{DODIM})). Defining $u(t,x_1,x_2,x_3):=\Psi(t,x_0,x_1,x_2,x_3)$
the wave equation (\ref{eq}) is equivalent to
\begin{equation}
 \label{equ}
 \left[\partial_t^2+3\frac{a'(t)}{a(t)}\partial_t-\frac{1}{a^2(t)}\Delta_{\mathcal{F}_v}\right]u=0,\;\;t\in\RR,\;\;x\in\mathcal{F}_v,
\end{equation}
together with the boundaries conditions
\begin{equation}
 \label{clu}
 \forall\tau\in\mathcal{I}^*,\; (x'_0,x'_1,x'_2,x'_3)=\tau(x_0,x_1,x_2,x_3)\Rightarrow u(t,x'_1,x'_2,x'_3)=u(t,x_1,x_2,x_3).
\end{equation}
We solve the initial value problem associated to (\ref{equ}) and
(\ref{clu}) and we state an equivalent  variational problem. We
discuss the existence of a future horizon and we compute the radius of the horizon sphere.

In the third section we investigate the asymptotic behaviour of
$\Psi(t,.)$ as $t\rightarrow+\infty$. We prove the existence of an asymptotic profile
$\Psi_{\infty}:=\lim_{t\rightarrow\infty}\Psi(t)$ if the scale
factor is exponentially increasing, and we compute an analytic
expression of $\Psi_{\infty}$ in the cases  $a(t)=H^{-1}\cosh(Ht)$ and
$a(t)=e^t$. These results extend those of \cite{Rat} regarding the
steady state universe (see also \cite{vasy}). We estimate the norms of
$\Psi_{\infty}$ by the energy of $\Psi$.

In the fourth part we
numerically investigate the mixed problem (\ref{equ},
\ref{clu}).  We are mainly
interested in computing the waves for a long time, hence the exponential behaviour of the coefficients in the
equation leads to a hard numerical challenge. To overcome this
difficulty we have to perform very precise computations. To reach this
goal we adopt the variational method associated to (\ref{equ}) with
suitable finite elements satisfying constraint (\ref{clu}). In the
case of the stationary universes we used in \cite{PDS} finite elements of
$\P_1$ type. To be able to treat the accelerating universes, we need more
precision and we use elements of second order. Our numerical results
are in excellent agreement with the theoretical results on the radius
of the future horizon and on the
asymptotic behaviours of the waves.

\section{Waves on dynamical universes}
In this section we solve the global Cauchy problem for the D'Alembert
equation on the dynamical universes (\ref{U}). We assume the scale
factor $a$ is a $C^{\infty}$ function that is strictly positive. As a
direct consequence of the theorem of Choquet-Bruhat, Cotsakis (Theorem 11.10 of \cite{choquet2009}),
we can see that $(\mathcal{U},g)$ is a globally hyperbolic manifold. Therefore
according to Leray \cite{leray}, the global Cauchy problem is well
posed in $C^{\infty}$. Now it is natural to look for the solutions in
the scale of the Hilbert spaces associated to the energy
(\ref{energy}). Given
$m\in\NN$, we introduce the Sobolev space
\begin{equation*}
  H^m(\mathbf{K}):=\left\{\Psi\in
    L^2(\mathbf{K}),\;\nabla_{\mathbf{K}}^{\alpha}\Psi\in
    L^2(\mathbf{K}),\;\mid\alpha\mid\leq m\right\}
\end{equation*}
endowed with the norm
\begin{equation*}
 \Vert\Psi\Vert_{H^m(\mathbf{K})}^2:=\sum_{\mid\alpha\mid\leq m}\Vert\nabla_{\mathbf{K}}^{\alpha}\Psi\Vert^2_{L^2(\mathbf{K})}.
 \end{equation*}
Here $\nabla_{\mathbf{K}}$  are the covariant derivatives and $
L^2(\mathbf{K})$ is the usual $L^2$-space of Lebesgue associated to
the volume form on $\mathbf{K}$. We can also interpret
this space as the set of the distributions $\Psi$ that belong to the
usual Sobolev space 
$H^m(\mathcal{S}^3)$ such that $\Psi\circ \tau=\Psi$ for any
$\tau\in\mathcal{I}^*$. We use also the space $H^{-1}(\mathbf{K})$
that is the dual space of $H^{1}(\mathbf{K})$. $C^{\infty}$ is dense
in all these spaces and we have for $\Phi,\Psi\in C^{\infty}$
\begin{equation*}
<\Psi,\Phi>_{H^{-1}(\mathbf{K}),H^1(\mathbf{K})}=<\Psi,\Phi^*>_{L^2(\mathbf{K})}.
 \end{equation*}
We also know that $-\Delta_{\mathbf{K}}+1$ is an isometry from
$H^{1}(\mathbf{K})$ onto $H^{-1}(\mathbf{K})$. We shall use an
orthonormal basis in $L^2(\mathbf{K})$, formed of eigenfunctions
$\left(\Psi_q\right)_q\subset H^1(\mathbf{K})$ of the densely defined self-adjoint
operator $-\Delta_{\mathbf K}$ associated to the eigenvalues $q^2$, i.e.
$
-\Delta_{\mathbf K} \Psi_q=q^2\Psi_q.
$
If $\Psi(X)=\sum_qA_q\Psi_q(X)$, we have
$$
\Vert \Psi\Vert_{H^m(\mathbf{K})}^2=\sum_q(q^2+1)^m\mid A_q\mid^2.
$$

A  {\it ``finite energy
  solution''} is a solution $\Psi$ of  (\ref{eq}) that belongs to $C^0\left(\RR_t;
   H^{1}(\mathbf{K})\right)\cap C^1\left(\RR_t;
   L^2(\mathbf{K})\right)$. We remark that such a solution belongs
to $C^2\left(\RR_t;H^{-1}(\mathbf K)\right)$.

\begin{Theorem}
 Given $m\in\NN$,  $\Psi_0\in H^{m+1}(\mathbf{K})$, $\Psi_1\in
 H^{m}(\mathbf{K})$, $t^*\in\RR$,
 there exists a unique solution $\Psi$ of the wave equation (\ref{eq})
 satisfying
\begin{equation}
 \label{regul}
  \Psi\in C^0\left(\RR_t;
   H^{m+1}(\mathbf{K})\right)\cap C^1\left(\RR_t;
   H^{m}(\mathbf{K})\right),
\end{equation}
\begin{equation}
 \label{condinitpsi}
 \Psi(t^*,.)=\Psi_0(.),\;\;\partial_t\Psi(t^*,.)=\Psi_1(.).
\end{equation}
For $m=0$, $\Psi$ is the unique solution satisfying together (\ref{regul}),
(\ref{condinitpsi}), and for all $\Phi\in H^{1}(\mathbf{K})$
\begin{equation}
 \label{pbvariapsi}
 \frac{d^2}{dt^2}<\Psi(t),\Phi>_{L^2(\mathbf
  K)}+3\frac{a'(t)}{a(t)}\frac{d}{dt}<\Psi(t),\Phi>_{L^2(\mathbf K)}
+\frac{1}{a^2(t)}<\nabla_{\mathbf{K}}\Psi(t),\nabla_{\mathbf{K}}\Phi>_{L^2(\mathbf
  K)}=0.
\end{equation}
If the scale factor $a(t)$ is increasing on $(t^*,\infty)$, then the energy defined by
(\ref{energy}) is decreasing on this interval, and if $a(t)$ also tends to the infinity
as $t\rightarrow\infty$, then the energy tends to zero.
 \label{cauchypsi}
\end{Theorem}

{\bf Remarks}
If $m=0$, 
$<\Psi(t),\Phi>_{H^{-1}(\mathbf
  K),H^{1}(\mathbf
  K)}=<\Psi(t),\Phi^*>_{L^2(\mathbf
  K)}\in C^2(\RR_t)$ and the first term in (\ref{pbvariapsi}) is well
defined. If $\Psi_0,\Psi_1\in C^{\infty}(\mathbf K)$, then $\Psi\in
C^{\infty}(\RR_t\times{\mathbf K})$ by the Sobolev embedding theorem
$\bigcap_mH^m(\mathbf{K})=C^{\infty}(\mathbf{K})$, and we say that $\Psi$ is a {\it
  ``smooth solution''}. If the energy tends to zero, we know that
$\Vert\partial_t\Psi(t)\Vert_{L^2(\mathbf K)}$ tends to zero, but we
have no control on $\Vert\Psi(t) \Vert_{L^2(\mathbf K)}$. We shall
prove in the next section that $\Psi(t)$ tends to an asymptotic state
$\Psi_{\infty}$ and the norms $\Vert\Psi_{\infty}\Vert_{L^2(\mathbf
  K)}$ and $\Vert\nabla_{\mathbf K}\Psi_{\infty}\Vert_{L^2(\mathbf
  K)}$ can be estimated by the norms of the initial data.

{\it Proof.} There are a lot of ways to obtain the theorem from well
known results. To construct the solutions we could invoke the
existence of the smooth solutions by the theorem of Leray
\cite{leray}, then we extend by density of $C^{\infty}$ in
$H^{m}(\mathbf{K})$ by using the energy estimate satisfied by the
regular solutions:
\begin{equation}
 \label{equener}
 \frac{d}{dt}E(\Psi,t)=-6\frac{a'(t)}{a(t)}\Vert\partial_t\Psi(t)\Vert^2_{L^2(\mathbf k)}-2\frac{a'(t)}{a^3(t)}\Vert\nabla_{\mathbf{K}}\Psi(t)\Vert^2_{L^2(\mathbf k)}.
\end{equation}
Then by the Gr\"onwall lemma we can control the energy at any time,
hence we get the global solutions for
$m=0$ (see also Theorem 2.29 in Appendix III of \cite{choquet2009}). Finally commuting the covariant derivatives and the D'Alembert
operator, we deduce the existence of solutions for any
$m\in\NN$. Another way
consists in employing the spectral method of Kato (see {\it e.g.}
\cite{tanabe}). In fact the more convenient approach to directly
obtain the existence, the uniqueness and the equivalence with the
variational problem is the variational method by Lions.There are
obtained by a direct application of the results of chapter XVIII of
\cite{dautray}. Finally if $a'$ is positive, equation (\ref{equener}) implies
that for all $t\geq t^*$
$$
 \frac{d}{dt}E(\Psi,t)\leq -\frac{a'(t)}{a(t)}E(\Psi,t),
$$
hence the energy is decreasing and
$\frac{d}{dt}\left(a(t)E(\Psi,t)\right)\leq 0$. We deduce that 
$$
E(\Psi,t)\leq\frac{a(t^*)}{a(t)}E(\Psi,t^*).
$$
We conclude that the energy tends to zero if $a(t)\rightarrow\infty$.
\fin

To perform a computational method to solve the Cauchy problem, it is
convenient to represent the Poincar\'e dodecahedron $\mathbf K$ by a
solid $\mathcal{F}_v$ in
$\RR^3$.
We recall that $\mathbf K$ is the quotient of the 3-sphere $\mathcal{S}^3 $ under
the action of the discrete fixed-point free subgroup $\mathcal{I}^*$
of isometries of $\mathcal{S}^3$, with $\mathcal{I}^*$ acting by left
multiplication (see \cite{Aur, PDS, C-LR-L-L-R-W,Lach,Lumi-Wee,Rouk}). The subgroup $\mathcal{I}^*$ of $SO(4)$ is the binary
icosahedral group (\cite{G-L-Lu-Uz-W, Sco, We-Sei,Wolf}). It is a two sheeted covering of the icosahedral
group $\mathcal{I}\subset SO(3)$ consisting of all
orientation-preserving symmetries of a regular icosahedron. To
visualize $\mathcal{S}^3$ we use the parametrisation: 
\begin{equation}
 \label{coordsphere}
\begin{array}{ll}
x_0=& \cos \chi\\
x_1=& \sin \chi \sin \theta \sin\varphi\\
x_2=& \sin \chi \sin \theta \cos\varphi\\
x_3=& \sin \chi \cos \theta.
\end{array}
\end{equation}
Hence $\mathcal{S}^3$ can be seen as two unit balls, $\mathcal{B}'$, $\mathcal{B}''$ in $\RR^3$ glued
together by their boundary $\mathcal{S}^2$.
\begin{equation}
 \label{strois}
\mathcal{S}^3 = \left(\mathcal{B}' \cup\mathcal{B}''\right)/\mathcal{S}^2.
\end{equation}
The radial coordinate in these balls is
given by $r=\sin\chi$. The $\theta$ and $\varphi$ coordinates are the standard ones for $\mathcal{S}^2$. For the first ball $\chi$ runs from $0$ at the center through $\frac{\pi}{2}$ at the surface. For the second ball $\chi$ runs from $\frac{\pi}{2}$ at the surface through $\pi$ at the center.

To construct $\mathcal{F}_v$, we represent $\mathcal{S}^3 / \mathcal{I}^*$ by a fundamental domain $\mathcal{F}\subset\mathcal{S}^3$ and an equivalence relation $\backsim$
such that 
\begin{equation}
\mathcal{S}^3 / \mathcal{I}^*=\mathcal{F}/ \backsim.
\label{SIM}
\end{equation}
The construction is based on the quaternionic calculus. It is
detailled in \cite{PDS}. Here we just present the main features. $\mathcal{F}$ is such that:
\begin{equation}
\label{SIM2}
\mathcal\displaystyle{{\mathcal{S}}^3=\bigcup_{\tau\in \mathcal{I}^*} \tau(\mathcal{F})},\qquad {\mbox and} \qquad
 \forall \tau\in \mathcal{I}^*,\,\forall \tau' \in \mathcal{I}^*,\;\tau\neq \tau'\Rightarrow\tau\left(\stackrel{\circ}{\mathcal{F}}\right)\cap \tau'\left(\stackrel{\circ}{ \mathcal{F}}\right)=\emptyset.
\end{equation}
$\mathcal{F} $ is a regular spherical dodecahedron (dual of a regular
icosahedron). One hundred twenty such spherical
dodecahedra tile the $3$-sphere in the pattern of a regular
$120$-cell. More specifically we choose  the
fundamental domain such that $(1,0,0,0)\in\mathcal{F}$, then $\mathcal{F}$ is the geodesic convex hull in $\mathcal{S}^3 $ of these $20$ vertices: 
$\frac{1}{2\sqrt{2}} (\sigma ^2,\pm \frac{1}{\sigma ^2},0,\pm 1)$, 
$\frac{1}{2\sqrt{2}} (\sigma ^2,0,\pm 1,\pm \frac{1}{\sigma ^2})$, 
$\frac{1}{2\sqrt{2}} (\sigma ^2,\pm\frac{1}{\sigma},\pm\frac{1}{\sigma},\pm\frac{1}{\sigma})$ and
$\frac{1}{2\sqrt{2}} (\sigma ^2,\pm 1,\pm \frac{1}{\sigma ^2},0)$,
where $\sigma=(1+\sqrt{5})/2$ is the golden number. Each of the $12$
faces of $\mathcal{F}$ is a regular pentagon in $\mathcal{S}^3
$  (see \cite{PDS}, for
details).\\

The equivalence relation $\backsim$ is defined by specifying the
equivalence class $\dot{q}$ of $q\in\mathcal{F}$. It is built from
twelve Cliffort translations $g_i$ that have been used for the construction
of $\mathcal{F}$ (see the Appendix and \cite{PDS}):
\[
\dot{q}=\left\{\frac{}{}g_i(q),\; i\in \{1,...,12\} \right\}\cap {\mathcal{F}}.
\]
It follows that: (1) if $q$ belongs to $\stackrel{\circ}{\mathcal{F}}$, then $\dot{q}$ has only one element,
(2) if $q$ is a vertex of $\mathcal{F}$, then $\dot{q}$ has four elements,
(3) if $q$ belongs to an edge of a face, without being a vertex, then $\dot{q}$ has three elements,
(4) if $q$ belongs to a face and does not belong to an edge, then $\dot{q}$ has two elements.\\

Finally to define $\mathcal{F}_v$  we use the projection
\begin{equation}
 \label{projectionp}
 p:\;(x_0,x_1,x_2,x_3)\in\mathcal{S}^3\longmapsto
(x_1,x_2,x_3)\in\RR^3,
\end{equation}
and we choose
\begin{equation}
 \label{}
 \mathcal{F}_v:=p\left(\mathcal{F}\right).
\end{equation}
With our choice of
fundamental domain $p$ is one-to-one on $\mathcal{F}$ and we have
\begin{equation}
 \label{F-v}
 (x,y,z)\in \mathcal{F}_v \Longleftrightarrow (\sqrt{1-x^2-y^2-z^2},x,y,z) \in \mathcal{F}.
\end{equation}
In figure 1, $\mathcal{F}_v$ is depicted in dark blue, as a part of
${\mathcal B}'$.
The vertices of $\mathcal{F}_v$ are those of a centered regular
dodecahedron in $\RR^3$ and they are given by
$\frac{1}{2\sqrt{2}} (\pm \frac{1}{\sigma ^2},0,\pm 1)$, 
$\frac{1}{2\sqrt{2}} (0,\pm 1,\pm \frac{1}{\sigma ^2})$, 
$\frac{1}{2\sqrt{2}} (\pm\frac{1}{\sigma},\pm\frac{1}{\sigma},\pm\frac{1}{\sigma})$ and
$\frac{1}{2\sqrt{2}} (\pm 1,\pm \frac{1}{\sigma ^2},0)$. Each face of
$\mathcal{F}_v$ is included in an ellipsoid \cite{PDS}.
The equivalence relation $\backsim$ on  $\mathcal{F}$ induces
canonically an
equivalence relation $\sim$ on  $\mathcal{F}_v$:
\begin{equation}
 \label{}
 \forall X,X'\in\mathcal{F},\;X\backsim X'\Leftrightarrow p(X)\sim p(X').
\end{equation}
$\sim$ leaves invariant the interior of $\mathcal{F}_v$ and identifies any pentagonal
face of $\mathcal{F}_v$ with its opposite face, after rotating by
$\frac{\pi}{5}$ clockwise around the outgoing normal
at the center of this last face.

Obviously $\Psi$ is entirely determined on $\RR_t\times\mathbf{K}$ by
its restriction to $\RR_t\times\mathcal{F}$, hence we are going to work
on $\RR_t\times\mathcal{F}_v$. We introduce the map $f:\;(x,y,z)\in\mathcal{F}_v\mapsto f(x,y,z)=\left( \sqrt{1-x^2-y^2-z^2}, x,y,z
\right)\in \mathcal{S}^3.
$
$f$ is one-to-one from  $\mathcal{F}_v\subset \RR^3$ onto  $\mathcal{F}\subset \mathcal{S}^3$ and we put
\begin{equation}
u(t,x,y,z):=\Psi(t,f(x,y,z)).
 \label{psipsi}
\end{equation}
It is clear that $\Phi\in H^1(\mathbf{K})$ iff $v:=\Phi\circ f$ belongs
to the usual Sobolev space $H^1$ on $\mathcal{F}_v$ for the euclidean metric of $\RR^3$,
$
H^1(\mathcal{F}_v)=\{v\in L^2(\mathcal{F}_v),\;\;\nabla_{x,y,z}v\in
L^2(\mathcal{F}_v)\},
$
and satisfies the boundary condition
$$
\forall X,X'\in\partial\mathcal{F}_v,\;\;X\sim X'\Rightarrow v(X)=v(X').
$$
We deduce that
$\Psi\in C^0\left(\RR_t;H^1(\mathbf{K})\right)\cap
C^1\left(\RR_t;L^2(\mathbf{K})\right)$ is solution of (\ref{eq}) iff $u$ belongs to $C^0\left(\RR_t;H^1(\mathcal{F}_v)\right)\cap
C^1\left(\RR_t;L^2(\mathcal{F}_v)\right)$ and satisfies the equation (\ref{equ})
where
\begin{equation}
 \label{delav}
 \Delta_{\mathcal{F}_v}=(1-x^2) \,\partial_{11}^2+(1-y^2)\, \partial_{22}^2+(1-z^2)\, \partial_{33}^2
-2 xy\, \partial_{12}^2 -2 xz \,\partial_{13}^2 -2 yz \,\partial_{23}^2-3 x\,\partial_1-3y\,\partial_2-3z\,\partial_3,
\end{equation}
and the boundary conditions
\begin{equation}
\forall (t,X,X')\in\RR\times\partial\mathcal{F}_v\times\partial\mathcal{F}_v,\;\;X\sim X'\Rightarrow u(t,X)=u(t,X').
  \label{cl}
\end{equation}
To handle the boundary when applying the finite element method, it is
very convenient to take into account the boundary condition (\ref{cl})
by a suitable choice of the functional space. In the sequel we denote
$X:=(x,y,z)$, $\mid X\mid^2:=x^2+y^2+z^2$ and $dX:=dx\,dy\,dz$.
For $m=-1,0,1,$ we introduce
the spaces $W^m(\mathcal{F}_v)$ that are isometric to the spaces
$H^m(\mathbf{K})$ :
\begin{equation}
 \label{}
 W^m(\mathcal{F}_v):=\left\{u=\Phi\circ f,\;\Phi\in  H^{m}(\mathbf
  K)\right\},\;\;\Vert u\Vert_{W^m}:=\Vert \Phi\Vert_{H^m}.
\end{equation}
We have
\begin{equation}
 W^0(\mathcal{F}_v):=L^2(\mathcal{F}_v,(1-\mid
  X\mid^{2})^{-\frac{1}{2}}dX) \label{wzero},
\end{equation}
\begin{equation}
 \label{}
 W^1(\mathcal{F}_v):=\left\{u\in H^1(\mathcal{F}_v),\;\;X\sim X'\Rightarrow
  u(X)=u(X')\right\},\;
\|u\|_{W^1}^2=\sum_{\mid\alpha\mid\leq
  1}\|\partial^{\alpha}u\|^2_{W^0(\mathcal{F}_v)} \label{}.
\end{equation}
Moreover
$
u\in C^k\left(\RR_t, W^m(\mathcal{F}_v)\right)$ iff $\Psi\in C^k\left(\RR_t; H^m(\mathbf{K})\right)
$
and Theorem \ref{cauchypsi} can be expressed in terms of $u$ as the
following

\begin{Theorem}
Given $u_0\in W^1(\mathcal{F}_v)$, $u_1\in W^0(\mathcal{F}_v)$,
$t^*\in\RR$, there
exists a unique $u$ solution of the equation (\ref{equ}) and
satisfying
\begin{equation}
 \label{regulu}
 u\in C^0\left(\RR_t^+;W^1(\mathcal{F}_v)\right)\cap
C^1\left(\RR_t^+;W^0(\mathcal{F}_v)\right)\cap
C^2\left(\RR_t^+;W^{-1}(\mathcal{F}_v)\right),
\end{equation}
\begin{equation}
u(t^*,.)=u_0(.),\;\;\partial_tu(t^*,.)=u_1(.).
\label{condinitu}
\end{equation}
$u$ is the unique function satisfying (\ref{regulu}), (\ref{condinitu}) and such that for any $\phi\in W^1(\mathcal{F}_v)$, we have:
\begin{eqnarray}
 0=\frac{d^2}{dt^2}\int_{\mathcal{F}_v}&(1-\mid X\mid^2)^{-\frac12}u(t,X)\phi(X)\,dX +3\frac{a'(t)}{a(t)}\frac{d}{dt}\int_{\mathcal{F}_v}(1-\mid X\mid^2)^{-\frac12}u(t,X)\phi(X)\,dX \nonumber \\
+\frac{1}{a^2(t)}\int_{\F_v} &(1-\mid X\mid^2)^{-\frac12}\nonumber\\
&\left[\nabla_X u(t,X) \cdot \nabla_X \phi(X)-\left(X\cdot \nabla_{X} u(t,X)\right)\left(X\cdot \nabla_{X} \phi(X)\right)\right]dX \label{pbvaria}
\end{eqnarray}
where $\nabla_X
u=\left(\partial_{x}u,\partial_{y}u,\partial_{z}u\right)$, $dX=dxdydz$.
\end{Theorem}

We end this part by discussing the possible horizons in the dynamical
universe (\ref{U}). We recall that the comoving spatial distance $d$ between any two points $x$ and $y$ on $\mathcal{S}^3$ is given by:
$d(x,y)=\arccos [ \delta_{ij}x^i y^j] $. $\mathcal{F}_v$ is endowed with the
distance induced by $d$. Moreover the largest distance $d_{max}$
between the origin and the boundary of $\mathcal{F}_v$ is reached at
any of its vertices. So we have $d_{max}=\arccos \left(\frac{\sigma
    ^2}{2\sqrt 2}\right)\simeq 0.388$. The metric on $\RR_t\times\mathcal{F}_v$ can be
expressed in spherical coordinates $\rho=\mid X\mid$, $\omega=\rho^{-1}X$, by $ds^2=dt^2-a^2(t)\left(\frac{1}{1-\rho^2}
d\rho^2+\rho^2 d\omega^2\right)$ where $\omega \in \mathcal{S}^2$,
$0\leq \rho\leq \rho_{max}(\omega)$. We have
\begin{equation}
 \label{rmax}
 R_{max}:=\sup_{\omega\in\mathcal{S}^2}\rho_{max}(\omega)=\sin(d_{max})=\sqrt{1-\frac{\sigma^4}{8}}\simeq
0.378.
\end{equation}
Given $t^*\in\RR$, $0\leq R<R_{max}$ we investigate the future
causal domain $\mathcal{O}^+(t^*,R)$ of $\{t^*\}\times(\mathcal{F}_v\cap(
[0\leq\rho\leq R]\times{\mathcal S}^2))$.
Solving
\begin{equation}
 \label{requdiff}
\frac{\rho'(t)}{\sqrt{1-\rho^2(t)}}=\frac{1}{a(t)},
\end{equation}
we get that
$$
\mathcal{O}^+(t^*,R)=\bigcup_{t\geq t^*}\{t\}\times(\mathcal{F}_v\cap(
[0\leq\rho\leq R(t)]\times{\mathcal S}^2))
$$
where
\begin{equation}
 \label{R(t)}
 R(t)=\sin\left(\arcsin(R)+\int_{t^*}^t\frac{1}{a(s)}ds\right)
\end{equation}
provided that
$\arcsin(R)+\int_{t^*}^t\frac{1}{a(s)}ds\leq\pi/2$. Therefore an
interesting phenomenon appears if
\begin{equation}
 \label{condihor}
 \arcsin(R)+\int_{t^*}^{\infty}\frac{1}{a(s)}ds<\arcsin(R_{max})=d_{max}.
\end{equation}
In this case, a future horizon exists:
\begin{equation}
 \label{}
 \mathcal{O}^+(t^*,R)\subset [t^*,\infty)\times [0,R_h]\times\mathcal{S}^2,
\end{equation}
where the radius of this future horizon is
\begin{equation}
 \label{rayonRh}
 R_h:=\sin\left(\arcsin(R)+\int_{t^*}^{\infty}\frac{1}{a(s)}ds\right).
\end{equation}
Returning to the scalar waves, we conclude that if $u_0$ and $u_1$ are
supported in $\rho\leq R$, then for any $t\geq t^*$, $u(t)$ is
supported in $\rho\leq R_h$.\\

Similarly, if $t_{ls}$ is the time of recombination, there exists a
past horizon, called horizon sphere or last scattering surface,  for
an observer located at $t_{obs}>t_{ls}$. In the universal covering
${\mathcal S}^3$ of $\mathbf{K}$ described in spherical coodinates $(\xi,\theta,\varphi)$, it is the submanifold
$\{\chi=\eta_{obs}-\eta_{ls}\}\times S^2_{\theta,\varphi}$ where
$\eta$ is the conformal time,
\begin{equation}
 \label{}
 \eta_{obs}-\eta_{ls}=\int_{t_{ls}}^{t_{obs}}\frac{1}{a(s)}ds<d_{max}.
\end{equation}
Then, the horizon sphere on which this observer can see the
primordial plasma in the sky $S^2_{\theta,\varphi}$ has the radius
\begin{equation}
 \label{}
 R_{t_{ls},t_{obs}}:=\sin\left(\int_{t_{ls}}^{t_{obs}}\frac{1}{a(s)}ds\right).
\end{equation}
If $\eta_{obs}-\eta_{ls}<d_{max}$ we have
\begin{equation}
 \label{}
R_{t_{ls},t_{obs}}<R_{max}=\sqrt{1-\frac{\sigma^4}{8}}
\end{equation}
hence the observable universe of radius $R_{t_{ls},t_{obs}}$ is strictly
included in the whole universe ${\mathbf K}$. In this case, there is
no multiple copies of images.
In contrast, if
\begin{equation}
 \label{cici}
 R_{t_{ls},t_{obs}}>R_{max}=\sqrt{1-\frac{\sigma^4}{8}},
\end{equation}
the horizon radius is larger that the injectivity radius. Then the
horizon sphere wraps all the way around the universe and intersects
itself. The case (\ref{cici}) has a great interest in cosmology: the observer could detect multiple images of radiating
sources,  some circles-in-the-sky appear, leading evidence of a
non-trivial topology of the Universe \cite{Aur2006}, \cite{mota},
\cite{Rouk}, \cite{Wee-2}. We present a numerical experiment of this
situation in part 4.2.\\

Finally we give the expression of these
quantities for the two accelerating universes that we consider, $d
\mathcal{S}_4(\mathbf{K})$ and the inflationary spacetime. For $a(t)=H^{-1}\cosh
(Ht)$, we have
$$R(t)=\sin\left(\arcsin
  R-2\arctan(e^{Ht^{\star}})+2 \arctan(e^{H t})\right),
$$
\begin{equation}
 \label{horch}
 R_{t_{ls},t_{obs}} = \sin\left(2\arctan\left(e^{Ht_{obs}}\right)-2\arctan\left(e^{Ht^{\star}}\right)\right),
\end{equation}
and if $a(t)=e^t$ we have
\begin{equation}
 \label{horet}
 R(t)=\sin\left(\arcsin R+e^{-t^{\star}}-e^{-t}\right),
\;\;
R_{t_{ls},t_{obs}} = \sin\left(e^{-t_{ls}}-e^{-t_{obs}}\right).
\end{equation}

\section{Asymptotic behaviours of smooth solutions}
In this part, we investigate the asymptotic behaviour of the fields as
$t\rightarrow+\infty$. We prove that the asymptotic profile
$\Psi_{\infty}(.):=\lim_{t\rightarrow\infty}\Psi(t,.)$ exists if the scale
factor is exponentially increasing. In the case of the de Sitter type
universe $d\mathcal{S}_4(\mathbf K)$, and in the case of the exponentially inflating
model for which $a(t)=e^t$ we are able to compute its expression, and the first terms of the asymptotic expansion,
in terms of the Cauchy data. Similar results have been
obtained in \cite{Rat} for the steady state universe
$\left(\RR_t\times \RR_{\mathrm x}^3,
  \;ds^2=dt^2-e^{2t}d\mathrm{x}^2\right)$ (see also \cite{vasy}).

\subsection{Asymptotic profile.} We prove the existence of $\Psi_{\infty}$
if the scale factor is at least exponentially increasing, {\it i.e.}
satisfies an assumption of the type:
\begin{equation}
 \label{hypex}
 \exists\gamma>0,\;\; \exists t^*\in\RR,\;\;\forall t\geq t^*\;\;\gamma\leq\frac{a'(t)}{a(t)}.
\end{equation}
\begin{Theorem}
 We assume (\ref{hypex}) is satisfied. Then for any finite energy
 solution $\Psi$ of (\ref{eq}), the energy (\ref{energy}) is exponentially decreasing
\begin{equation}
 \label{decener}
 \forall t\geq t^*,\;\;E(\Psi,t)\leq e^{-2\gamma(t-t^*)}E(\Psi,t^*),
\end{equation}
and there exists
 $\Psi_{\infty}\in L^2(\mathbf{K})$ such that
\begin{equation}
 \label{psiinfini}
 \forall t\geq t^*,\;\;\Vert \Psi(t)-\Psi_{\infty}\Vert_{L^2(\mathbf{K})}\lesssim e^{-\gamma
   t}E^{\frac{1}{2}}(\Psi,t^*).
\end{equation}
 \label{profile}
\end{Theorem}

{\it Proof.}
(\ref{equener}) and (\ref{hypex}) imply that $dE(\Psi,t)/dt\leq
-2\gamma E(\Psi,t)$, hence (\ref{decener}) is proved. To establish the existence of the profile $\Psi_{\infty}$, our method relies on the spectral expansion of $\Psi(t,.)$ and the
energy estimate (or an explicit resolution) for the ODE. More precisely, since the Laplace-Beltrami operator $\Delta_{\mathbf{K}}$ on $\mathbf{K}$ is a non
positive, self-adjoint elliptic operator on a compact manifold, its
spectrum is a discrete set of eigenvalues $-q^2\leq 0$. Moreover, by the Hilbert-Schmidt theorem there exists an orthonormal basis in $L^2(\mathbf{K})$, formed of eigenfunctions $\left(\Psi_q\right)_q\subset H^1(\mathbf{K})$  associated to $q^2$, i.e.
\begin{equation}
-\Delta_{\mathbf K} \Psi_q=q^2\Psi_q.
  \label{eqlaplac}
\end{equation}
Moreover the eigenvalues are explicitly known (see
\cite{Aur,Ike,Lach,Leh-W-U-G-Lu}). One has:
\begin{equation} 
0\leq q,\;\;q^2=\beta ^2 -1,\label{valpra}
\end{equation}
\begin{equation}
\mbox{with }\beta \in
\{1,13,21,25,31,33,37,41,43,45,49,51,53,55,57\} \cup\{2n+1,\, n\geq
30\}.
\label{valprb}
 \end{equation}

In particular, since the volume of $\mathbf{K}$ is $\pi^2/60$ (see
{\it e.g.} \cite{G-L-Lu-Uz-W} p. 5160), we
have:
\begin{equation}
 \label{psizero}
\Psi_0(X)=\frac{2\sqrt{15}}{\pi} .
\end{equation}
Any finite energy solution $\Psi(t,X)$ of  equation (\ref{eq}) has an
expansion of the  form
\begin{equation}
 \label{expansion}
 \Psi(t,X)=\sum_{q}u_q(t) \Psi_q(X),\;\;t\in\RR,\;\;X\in\mathbf{K},
\end{equation}
where $u_q$ is
solution of the ordinary differential equation
\begin{equation}
 \label{lequdiff}
 u_q''+3\frac{a'(t)}{a(t)}u'+\frac{q^2}{a^2(t)}u_q=0.
\end{equation}
For the finite energy solution we have
\begin{equation}
 \label{estfini}
 E(\Psi,t)=\sum_{q}\left\vert u'_q(t)\right\vert^2+ \frac{q^2}{a^2(t)}\left\vert
   u_q(t)\right\vert^2,\;\;
\sum_{q}\left\vert u'_q(t)\right\vert^2+ q^2\left\vert u_q(t)\right\vert^2<\infty,
\end{equation}
and if $\Psi$ is a smooth solution we have
$$
\forall k,p\in\NN,\;\;\sum_{q}q^p\left\vert \frac{d^k}{dt^k}u_q(t)\right\vert <\infty.
$$
These series converge locally uniformly with respect to $t$. Now,
multiplying  (\ref{lequdiff}) by $u'_q$, we
get an energy type equality:
$$
\mid u'_q(t)\mid^2+\frac{q^2}{a^2(t)}\mid u_q(t)\mid^2=-2\frac{a'(t)}{a(t)}\left(3\mid u'_q(t)\mid^2+\frac{q^2}{a^2(t)}\mid u_q(t)\mid^2\right).
$$
We get from hypothese (\ref{hypex}) that
$$
\frac{d}{dt}\left(\mid u'_q(t)\mid^2+\frac{q^2}{a^2(t)}\mid
  u_q(t)\mid^2\right)\leq -2\gamma \left(\mid u'_q(t)\mid^2+\frac{q^2}{a^2(t)}\mid
  u_q(t)\mid^2\right),
$$
hence
$$
\mid u'_q(t)\mid^2+\frac{q^2}{a^2(t)}\mid
  u_q(t)\mid^2\leq e^{-2\gamma(t-t^*)}\left(\mid u'_q(t^*)\mid^2+\frac{q^2}{a^2(t^*)}\mid
  u_q(t^*)\mid^2\right),
$$
and we deduce that for any $t\geq t^*$ we have:
\begin{equation}
 \label{expdec}
 \mid u'_q(t)\mid\leq e^{-\gamma t}\left(\mid u'_q(t^*)\mid^2+\frac{q^2}{a^2(t^*)}\mid
  u_q(t^*)\mid^2\right)^{\frac{1}{2}}.
\end{equation}
Therefore we can introduce
$$
u_q(\infty):=u_q(t^*)+\int_{t^*}^{\infty}u'_q(t)dt,\;\;\Psi_{\infty}:=\sum_{q}u_q(\infty) \Psi_q.
$$
Now (\ref{psiinfini}) follows from (\ref{estfini}) and (\ref{expdec}).
\fin

We compute $\Psi_{\infty}$ in the next sections. We remark that the
profile $\Psi_{\infty}(.)$ is more regular than the field
$\Psi(t,.)$. In short, the Theorems \ref{reguchinfini} and \ref{reguexpinfini} show that
$\Psi_{\infty}\in H^{m+1}$ if $\Psi(0,.)\in H^m$,
$\partial_t\Psi(0,.)\in H^{m-1}$. This phenomenon has been noticed in
the case of the steady-state universe \cite{Rat}.

\subsection{$d\mathcal{S}_4(\mathbf K)$ universe.}
We consider the $d\mathcal{S}_4(\mathbf K)$ universe for which
$a(t)=H^{-1}\cosh(Ht)$. We calculate an explicit expression of the asymptotic profile
and we prove that $\Delta_{\mathbf K} \Psi(t)
-\frac{1}{2H} \left(1-\tanh Ht\right)^{-1} \partial_t \Psi(t)$
tends to zero.

\begin{Theorem}
Given a smooth solution $\Psi$ of (\ref{eq}), there exist constants $A^+_q$ and $A^-_q$
depending only on the initial data such that
\begin{equation}
\label{expresspsi}
\begin{split}
\Psi(t,X)=&\left[-\sqrt{\frac{2}{\pi}} A^+_0+\left(\cosh H t
  \right)^{-\frac32} A^-_0 \mathrm{P}^{-\frac32}_{\frac12}(\tanh
  H t)\right]\Psi_0(X)\\
&+\left(\cosh H t
  \right)^{-\frac32} \sum_{q\neq 0} \left[A^+_q \mathrm{P}^{\frac32}_{\beta-\frac12}(\tanh H t)+A^-_q
  \mathrm{Q}^{\frac32}_{\beta-\frac12}(\tanh H t)\right]\Psi_q(X), 
\end{split}
\end{equation}
\begin{equation}
 \Psi_{\infty}(X)=-\sqrt{\frac{2}{\pi}} \sum_q A^+_q \Psi_q(X),
\end{equation}
\begin{equation}
\Psi(t,X)-\Psi_{\infty}(X)=-\sqrt{\frac{2}{\pi}} \sum_{q\neq 0} q^2A^+_q
  \Psi_q(X)\left(1-\tanh Ht\right) +0\left(\left(1-\tanh
  Ht\right)^{\frac32}\right) \label{vcoshinfty1},
\end{equation}
\begin{equation}
\begin{split}
\label{deltach}
 \Delta_{\mathbf K} \Psi(t,X)-&\frac{1}{2H} \left(1-\tanh
  Ht\right)^{-1} \partial_t\Psi(t,X)=
\frac{2}{\sqrt{\pi}}A^-_0 \left(1-\tanh
  Ht\right)^{\frac12}\\
&\qquad +\sqrt{\frac{2}{\pi}}\left[\sum_{q\neq 0}q^2(q^2+\frac34) A^+_q \Psi_q(X)\right] \left(1-\tanh Ht\right)+ 0\left(1-\tanh Ht\right)^{\frac32} 
\end{split}
\end{equation}
where $q$ and $\beta$ are given by equation (\ref{valpra}), equation (\ref{valprb}). The
coefficients $A_q^{\pm}$ are given by
\begin{equation}
 \label{aplusch}
  A^+_0=-\sqrt{\frac{\pi}{2}}\left(u_0(0)+\frac{\pi}{4}u'(0)\right),\;\;A^+_q=-\sqrt{\frac{\pi}{2}}\,\frac{\sin\left(\frac{\pi}{2}\sqrt{q^2+1}\right)}{\sqrt{q^2+1}}u_q(0),\;q\neq 0,
\end{equation}
\begin{equation}
 \label{amoinsch}
  A_0^-=-\sqrt{\frac{\pi}{2}}u_0'(0),\;\;A^-_q=-\sqrt{\frac{2}{\pi}}\,\frac{\sin\left(\frac{\pi}{2}\sqrt{q^2+1}\right)}{q^2}u'_q(0),\;q\neq 0,
\end{equation}
where $u_q(t):=<\Psi(t,.),\Psi_q>_{L^2(\mathbf{K})}$.
\label{propch}
\end{Theorem}

{\it Proof.}
The differential equation (\ref{lequdiff}) has the form:
\begin{equation}
u''_q(t)+3 H\tanh (H t)
\, u'_q(t)+\frac{q^2H^2}{\cosh ^2(H
  t)}u_q(t)=0.
 \label{chq}
\end{equation}
For each $q$ we introduce a new function $w_q$ defined by
\[ u_q(t):=(\cosh H t)^{-\frac32 }\, w_q(t).\]
Then $u_q$ satisfies equation (\ref{chq}) iff $w_q$ is solution of 
\begin{equation}
H^{-2} w_q''(t)+\left( (q^2+\frac34)\cosh ^{-2}(H
  t)-\frac94 \right) w_q(t)=0.\;
 \label{chw}
\end{equation}
Now we put
$w_q(t)=\tilde{w}_q(\tanh H
  t)$. Thanks to the relation $q^2+\frac34=(\beta -\frac12)(\beta +\frac12)$ we deduce that $w_q$ satisfies equation (\ref{chw}) iff
$\tilde{w}_q$ is solution of the associated Legendre equation
\begin{equation}
(1-x^2) \tilde{w}''_q(x)-2x \tilde{w}'_q(x)+\left(\nu(\nu+1)-\frac{\mu^2}{1-x^2}\right) \tilde{w}_q(x)=0,
\end{equation}
where $\nu=\beta -\frac12$, $\mu=\pm \frac32$ and $x\in ]-1,+1[$.  In
the sequel, we use several formulae, all from \cite{nist}. The
first eigenvalue $q=0$ is a particular case where
$\nu=\frac12$. Thanks to formula 14.2.7 the Wronskian of $\mathrm{P}^{-\frac32}_{\frac12}
(x)$ and $\mathrm{P}^{\frac32}_{\frac12}
(x)$ is not zero. For the other eigenvalues, $\beta$ is greater than $13$ and is an odd
integer. So $\mathrm{Q}^{\frac32}_{\beta-\frac12}$ is defined and, thanks to
formula 14.2.4, the Wronskian of $\mathrm{P}^{\frac32}_{\beta-\frac12}
(x)$ and $\mathrm{Q}^{\frac32}_{\beta-\frac12}
(x)$ is not zero. So there exist constants $A^+_q$ and $A^-_q$
depending on initial data such that
\[
\tilde{w}_0(x)= A^+_0 \mathrm{P}^{\frac32}_{\frac12}(x)+A^-_0
  \mathrm{P}^{-\frac32}_{\frac12}(x) ,\;\mbox{and }\;
\tilde{w}_q(x)= A^+_q \mathrm{P}^{\frac32}_{\beta-\frac12}(x)+A^-_q
  \mathrm{Q}^{\frac32}_{\beta-\frac12}(x) \;\mbox{for } q\neq 0.
\]
Then
\begin{eqnarray} 
&u_0(t)=\left(\cosh H t
  \right)^{-\frac32} \left[A^+_0 \mathrm{P}^{\frac32}_{\frac12}(\tanh H t) +A^-_0 \mathrm{P}^{-\frac32}_{\frac12}(\tanh H t) 
 \right] ,\\
& u_q(t)=\left(\cosh H t
  \right)^{-\frac32} \left[A^+_q \mathrm{P}^{\frac32}_{\beta-\frac12}(\tanh H t)+A^-_q
  \mathrm{Q}^{\frac32}_{\beta-\frac12}(\tanh H t)\right] \;\mbox{for } q\neq 0.
\label{eqvv}
\end{eqnarray}

To obtain asymptotic profiles we use hypergeometric
representation of the Ferrers
functions (see formulae 14.3.i). We recall that for $x\in ]-1,+1[$
\begin{eqnarray*} 
\mathrm{P}^{\mu}_{\nu}(x)&=\left(\frac{1+x}{1-x}\right)^{\frac{\mu}{2}}\mathbf{F}(\nu+1,-\nu;1-\mu;\frac12-\frac12
                             x)\\
&=\left(\frac{1+x}{1-x}\right)^{\frac{\mu}{2}}\sum_{s=0}^{+\infty} \frac{(\nu+1)_s (-\nu)_s}{\Gamma(1-\mu+s)} \frac{1}{s!}\left(\frac12\right)^s (1-x)^s,
\end{eqnarray*}
where $(\alpha)_n$ is the Pochhammer's symbol:
$\alpha(\alpha+1)(\alpha+2)···(\alpha+n-1)$ if $n=1,2,3,...$, and $1$
if $n=0$. First we consider the case $q\neq 0$, that is
$\nu=\beta-\frac12 \neq \frac12$. For $t$ in the neighbourhood of $+\infty$, the
regularity of the previous series at $0$ leads to 
\begin{eqnarray*}
 \mathrm{P}^{\frac32}_{\nu}(\tanh H
  t)=&\left(\frac{1+\tanh H t}{1-\tanh H
  t}\right)^{\frac34}\,  \left[\frac{1}{\Gamma(-\frac12)}+\frac{(\nu+1)(-\nu)}{2\Gamma(\frac12)} \left(1-\tanh Ht\right)\right.\\
&\left.+\frac{(\nu+1)(\nu+2)(-\nu)(-\nu+1)}{8\Gamma(\frac32)} \left(1-\tanh Ht\right)^2+0\left(\left(1-\tanh Ht\right)^3\right)\right].
\end{eqnarray*}
So
\begin{eqnarray}
\left(\cosh H t
  \right)^{-\frac32} \mathrm{P}^{\frac32}_{\beta-\frac12}(\tanh
  H t)&=\left(1+\tanh Ht\right)^{\frac32}
\,
\left[\frac{1}{\Gamma(-\frac12)}-(\beta^2-\frac14)\frac{1}{2\Gamma(\frac12)}
  \left(1-\tanh Ht\right)\right.\nonumber\\
&\left.+\frac{(\beta^2-\frac14)(\beta^2-\frac94)}{8\Gamma(\frac32)}\left(1-\tanh Ht\right)^2+0\left(\left(1-\tanh Ht\right)^3\right)\right].
\label{asympP}
\end{eqnarray}
Likewise for $x\in ]-1,+1[$ 
\begin{eqnarray*}
\mathrm{Q}^{\frac32}_{\beta-\frac12}(x)&=\frac{\pi}{2\sin(\frac32 \pi)}\left(\cos(\frac32 \pi)\left(\frac{1+x}{1-x}\right)^{\frac34}\mathbf{F}(\nu+1,-\nu;-\frac12;\frac12-\frac12 x)\right.\\
&\qquad \qquad \qquad\left.-\frac{\Gamma(\nu+\frac52)}{\Gamma(\nu-\frac12)} \left(\frac{1-x}{1+x}\right)^{\frac34}\mathbf{F}(\nu+1,-\nu;\frac52;\frac12-\frac12 x)\right)\\
&=\frac{\pi}{2}\frac{\Gamma(\nu+\frac52)}{\Gamma(\nu-\frac12)}\left(\frac{1-x}{1+x}\right)^{\frac34}\sum_{s=0}^{+\infty} \frac{(\nu+1)_s (-\nu)_s}{\Gamma(\frac52+s)} \frac{1}{s!}\left(\frac12\right)^s (1-x)^s.
\end{eqnarray*}
Then, for $t$ in the neighbourhood of $+\infty$ we have
\begin{equation*}
\begin{split}
\mathrm{Q}^{\frac32}_{\beta-\frac12}(\tanh H
t)&=\frac{\pi}{2}\frac{\Gamma(\beta+2)}{\Gamma(\beta-1)}\\
&\left(\frac{1-\tanh H t}{1+\tanh H t}\right)^{\frac34}
\left[\frac{1}{\Gamma(\frac52)}-(\beta^2-\frac14)
  \frac{1}{2\Gamma(\frac72)}\left(1-\tanh
  Ht\right)+0\left(\left(1-\tanh
    Ht\right)^2\right)\right],
\end{split}
\end{equation*}
and so
\begin{equation}
\left(\cosh H t
  \right)^{-\frac32} \mathrm{Q}^{\frac32}_{\beta-\frac12}(\tanh H t)=\frac{\pi}{2}\frac{1}{\Gamma(\frac52)}\frac{\Gamma(\beta+2)}{\Gamma(\beta-1)}\left(1-\tanh Ht\right)^{\frac32}+0\left(\left(1-\tanh Ht\right)^{\frac52}\right).
\label{asympQ}
\end{equation}
From equation (\ref{eqvv}), equation (\ref{asympP}) and equation (\ref{asympQ}) we deduce for
$t\rightarrow+\infty$ and $q\neq 0$
\begin{eqnarray*}
 u_q(t)=-A^+_q \frac{1}{2\sqrt{\pi}}\left(1+\tanh Ht\right)^{\frac32}-A^+_q \frac{\sqrt{\pi}}{2 } (\beta^2-\frac14)\left(1+\tanh Ht\right)^{\frac32} \left(1-\tanh Ht\right)\\
\; +0\left(\left(1-\tanh Ht\right)^{\frac52}\right).
\end{eqnarray*}
Hence  we have got for $q\neq 0$
\begin{equation} 
u_q(t)\sim -A^+_q \sqrt{\frac{2}{\pi}},{\mbox { and } } \;u_q(t)+A^+_q
\sqrt{\frac{2}{\pi}}=-q^2A^+_q \sqrt{\frac{2}{\pi}} \left(1-\tanh
  Ht\right)+0\left(\left(1-\tanh
    Ht\right)^{\frac32}\right).
\label{asympvq}
\end{equation}
Using formula 14.10.4 we deduce derivatives of $u_q$ from
equation (\ref{eqvv}). For $q\neq 0$ we have
\begin{eqnarray} 
 u'_q(t)&=H\left(\cosh H t \right)^{-\frac32}\left(\tanh
  H t \right)(\beta -1)\left[A^+_q\mathrm{P}^{\frac32}_{\beta-\frac12}(\tanh H t) +A^-_q \mathrm{Q}^{\frac32}_{\beta-\frac12}(\tanh H t) \right]\nonumber\\
 &\qquad  -H\left(\cosh H t \right)^{-\frac32}(\beta -1)\left[ A^+_q \mathrm{P}^{\frac32}_{\beta+\frac12}(\tanh H t) +A^-_q \mathrm{Q}^{\frac32}_{\beta+\frac12}(\tanh
  H t) \right]\nonumber\\
&=H\left(\cosh H t \right)^{-\frac32}(\beta -1)A^+_q\left[ \left(\tanh H t \right)\mathrm{P}^{\frac32}_{\beta-\frac12}(\tanh H t)- \mathrm{P}^{\frac32}_{\beta+\frac12}(\tanh H t)\right]\nonumber\\
&\qquad +H\left(\cosh H t \right)^{-\frac32}(\beta -1)A^-_q\left[\left(\tanh
  H t \right)\mathrm{Q}^{\frac32}_{\beta-\frac12}(\tanh H t)- \mathrm{Q}^{\frac32}_{\beta+\frac12}(\tanh H t)\right].
\label{vprimeq}
\end{eqnarray}
And from equation (\ref{vprimeq}), equation (\ref{asympP}) and equation (\ref{asympQ}) we deduce
for $\nu=\beta-\frac12\neq \frac12$ and $t\rightarrow+\infty$
\begin{equation*}
\begin{split}
 &u'_q(t)=H(\beta -1)A^+_q \left(1+\tanh Ht\right)^{\frac32}\\
&\quad  \left[  \frac{1}{\Gamma(-\frac12)} \left(-1+\tanh Ht\right) +\left(-(\beta^2-\frac14) \tanh Ht+(\beta+\frac32)(\beta +\frac12)\right)\frac{1}{2\Gamma(\frac12)}  \left(1-\tanh Ht\right) \right.\\
&\qquad \qquad\left. +\left((\beta^2-\frac14)(\beta^2-\frac94) \tanh Ht-(\beta+\frac32)(\beta+\frac52)(\beta^2-\frac14)\right) \frac{1}{8\Gamma(\frac32)}\left(1-\tanh Ht\right)^2\right]\\
&\qquad \qquad+0\left(\left(1-\tanh Ht\right)^{\frac52}\right).
\end{split}
\end{equation*}
That is for $q\neq 0$
\begin{equation*}
\begin{split}
 u'_q(t)&=A^+_q \frac{H}{ \sqrt{\pi}}(\beta^2 -1)\left(1+\tanh Ht\right)^{\frac32}\\
&\qquad \left[\left(1-\tanh Ht\right)-(\beta^2 -\frac14) \left(1-\tanh Ht\right)^2+0\left(1-\tanh Ht\right)^{\frac52}\right],
\end{split}
\end{equation*}
and then
\begin{equation}
\begin{split}
 -2Hq^2 u_q(t)-&\left(1-\tanh Ht\right)^{-1} u'_q(t)=\nonumber\\
&q^2(q^2+\frac34) A^+_q \frac{H}{\sqrt{\pi}}\left(1+\tanh Ht\right)^{\frac32}\left(1-\tanh Ht\right)+ 0\left(1-\tanh Ht\right)^{\frac32}.
\label{asympvprimeq}
\end{split}
\end{equation}
It only remains to consider the case $q=0$. Actually $u_0$ can be
written more simply by explaining $\mathrm{P}^{\frac32}_{\frac12}$. We
use formula 14.10.4 :
$(1-x^2)\left(\mathrm{P}^{\frac32}_{\frac12}\right)'(x)=(\frac32
-\frac12-1)\mathrm{P}^{\frac32}_{\frac32}(x)+\frac32 x\mathrm{P}^{\frac32}_{\frac12}(x)=\frac32 x\mathrm{P}^{\frac32}_{\frac12}(x)$. So $\mathrm{P}^{\frac32}_{\frac12}$ has the form $C \left(1-x^2\right)^{-\frac34}$ where $C$ is a constant. Since $\mathrm{P}^{\frac32}_{\frac12}(0)=\mathbf{F}(\frac32,-\frac12;-\frac12;\frac12)=-\sqrt{\frac{2}{\pi}}$ we can write 
$\mathrm{P}^{\frac32}_{\frac12}(x)=-\sqrt{\frac{2}{\pi}}\left(1-x^2\right)^{-\frac34}$. So
with equation (\ref{eqvv}) we obtain a new expression of $u_0$:
\[ u_0(t)=-\sqrt{\frac{2}{\pi}} A^+_0+\left(\cosh H t
  \right)^{-\frac32} A^-_0 \mathrm{P}^{-\frac32}_{\frac12}(\tanh H t).\] 
Moreover the hypergeometric
representation of $\mathrm{P}^{-\frac32}_{\frac12}$ gives us for $t$ at $+\infty$
\begin{equation*}
\begin{split}
 \left(\cosh H t \right.
&\left.  \right)^{-\frac32} \mathrm{P}^{-\frac32}_{\frac12}(\tanh H t)=\\
 \left(1-\tanh Ht\right)^{\frac32} \left[\frac{1}{\Gamma(\frac52)}-\frac38\frac{1}{\Gamma(\frac72)}\left(1-\tanh
  Ht\right)-\frac{15}{128} \frac{1}{\Gamma(\frac92)} \left(1-\tanh Ht\right)^2+0\left(\left(1-\tanh Ht\right)^3\right)\right],
\end{split}
\end{equation*}
so
\begin{equation}
u_0(t)=-\sqrt{\frac{2}{\pi}}
A^+_0+A^-_0\frac{1}{\Gamma(\frac52)}\left(1-\tanh
  Ht\right)^{\frac32}+0\left(\left(1-\tanh
    Ht\right)^{\frac52}\right).
\label{asympv0}
\end{equation}
Then equation (\ref{asympvq}) and equation (\ref{asympv0}) prove equation (\ref{vcoshinfty1}). 
On the other hand we can simplify $u'_0$ since $\mathrm{P}^{-\frac32}_{\frac32}$ has a simple writting (see formula
14.5.18 with $\theta \in ]0,\pi[$ such that $\cos \theta =\tanh
H t$). We have:
\begin{equation*}
\begin{split}
u'_0(t)&=-3H\left(\cosh H t
  \right)^{-\frac32}A^-_0\mathrm{P}^{-\frac32}_{\frac32}(\tanh
  H t)\\
 &=-3H\left(\cosh H t
  \right)^{-\frac32}\frac{1}{2^{\frac32}}\frac{1}{\Gamma (\frac32
    +1)}A^-_0\left(\sin \theta
  \right)^{\frac32}=H\sqrt{\frac{2}{\pi}}\left(\cosh H t
  \right)^{-3}A^-_0\\
&=-H\sqrt{\frac{2}{\pi}}A^-_0 \left(1+\tanh Ht\right)^{\frac32}\left(1-\tanh Ht\right)^{\frac32},
\end{split}
\end{equation*}
so
\[ -\frac{1}{2H} \left(1-\tanh Ht\right)^{-1} u'_0(t)=\frac{1}{\sqrt{2\pi}}A^-_0 \left(1+\tanh Ht\right)^{\frac32}\left(1-\tanh Ht\right)^{\frac12}.
\]

Finally to compute $A_q^{\pm}$ we remark that since
$\beta=\sqrt{q^2+1}$ is an odd integer number, we have
\begin{equation}
 \label{ppp}
  \mathrm{P}^{-\frac32}_{\frac12}(0)=\frac{1}{2}\sqrt{\frac{\pi}{2}},\;\left(\mathrm{P}^{-\frac32}_{\frac12}\right)'(0)=-\sqrt{\frac{2}{\pi}},
\end{equation}
\begin{equation}
 \label{pppp}
  \mathrm{P}^{\frac32}_{\beta-\frac12}(0)=-\sqrt{\frac{2}{\pi}}\beta\sin\left(\beta\frac{\pi}{2}\right),\;\left(\mathrm{P}^{\frac32}_{\beta-\frac12}\right)'(0)=0,
\end{equation}
\begin{equation}
 \label{qqq}
  \mathrm{Q}^{\frac32}_{\beta-\frac12}(0)=0,\;\left(\mathrm{Q}^{\frac32}_{\beta-\frac12}\right)'(0)=-
\sqrt{\frac{\pi}{2}}q^2\sin\left(\beta\frac{\pi}{2}\right).
\end{equation}
The expressions of the coefficients follows from (\ref{expresspsi}) by
elementary
computations. The proof is complete.
\fin

We deduce from this theorem that any smooth solution $\Psi$ can
be expressed as
$$
\Psi(t,X)=\Psi^+(t,X)+\Psi^-(t,X),
$$
where $\Psi^+$ and $\Psi^-$ are smooth solutions satisfying
\begin{equation}
 \label{}
\Psi^+(t,X)=\Psi_{\infty}(X)+O\left(e^{-Ht}\right),
\end{equation}
\begin{equation}
 \label{}
 \Psi^-(t,X)=O\left(e^{-\frac{3}{2}Ht}\right).
\end{equation}
It is sufficient to define
\begin{equation}
\Psi^+(t,X):=-\sqrt{\frac{2}{\pi}} A^+_0\Psi_0(X)
+\left(\cosh H t
  \right)^{-\frac32} \sum_{q\neq 0} A^+_q \mathrm{P}^{\frac32}_{\beta-\frac12}(\tanh H t)\Psi_q(X), 
\end{equation}
\begin{equation}
\Psi^-(t,X):=\left(\cosh H t
  \right)^{-\frac32}\left[A^-_0 \mathrm{P}^{-\frac32}_{\frac12}(\tanh
  H t)\Psi_0(X)
+\sum_{q\neq 0} A^-_q
  \mathrm{Q}^{\frac32}_{\beta-\frac12}(\tanh H t)\Psi_q(X)\right].
\end{equation}

We express the initial data for $\Psi^{\pm}$ in terms of those of
$\Psi$, and we estimate the norms of $\Psi_{\infty}$.

\begin{Theorem}
The smooth solutions $\Psi$, $\Psi^{\pm}$ and the asymptotic profile
$\Psi_{\infty}$ satisfy :
\begin{equation}
 \label{psiplus}
 \Psi^+(0,X)=\Psi(0,X)+\frac{\pi}{4}<\partial_t\Psi(0,.),\Psi_0>_{L^2(\mathbf{K})}\Psi_0(X),\;\partial_t\Psi^+(0,X)=0,
\end{equation} 
\begin{equation}
 \label{psimoins}
 \Psi^-(0,X)=-\frac{\pi}{4}<\partial_t\Psi(0,.),\Psi_0>_{L^2(\mathbf{K})}\Psi_0(X),\;\;\partial_t\Psi^-(0,X)=\partial_t\Psi(0,X),
\end{equation}
\begin{equation}
 \label{profpse}
 \Psi_{\infty}=\left(-\Delta_{\mathbf{K}}+1\right)^{-\frac{1}{2}}\sin\left(\frac{\pi}{2}\sqrt{-\Delta_{\mathbf{K}}+1}\right)\Psi^+(0,.),
\end{equation}
\begin{equation}
 \label{profldeux}
\Vert\Psi_{\infty}\Vert_{L^2(\mathbf{K})}=\Vert\Psi^+(0,.)\Vert_{H^{-1}(\mathbf{K})},
\end{equation}
\begin{equation}
 \label{profhm}
\Vert\nabla_{\mathbf{K}}\Psi_{\infty}\Vert_{H^m(\mathbf{K})}=\Vert\nabla_{\mathbf{K}}\Psi^+(0,.)\Vert_{H^{m-1}(\mathbf{K})}=\Vert\nabla_{\mathbf{K}}\Psi(0,.)\Vert_{H^{m-1}(\mathbf{K})},\;m=0,1.
\end{equation}
 \label{reguchinfini}
\end{Theorem}

In (\ref{profpse}), the pseudodifferential operator
$f\left(\Delta_{\mathbf{K}}\right)$ is defined by the usual functional
calculus of the spectral theory:
$f\left(\Delta_{\mathbf{K}}\right)\Phi$ is defined by
$<f\left(\Delta_{\mathbf{K}}\right)\Phi,\Psi_q>_{L^2(\mathbf{K})}=f(-q^2)<\Phi,\Psi_q>_{L^2(\mathbf{K})}$. We
note that (\ref{valpra}) assures that $\sin\left(\frac{\pi}{2}\sqrt{q^2+1}\right)\in\{+1,\,-1\}$.\\

{\it Proof.} We obtain (\ref{psiplus}) and (\ref{psimoins}) from
(\ref{ppp}), (\ref{pppp}) and (\ref{qqq}).
We get (\ref{profpse}) by (\ref{psiplus}) and the equalities
$$
\Psi^+(0,X)=-\sqrt{\frac{2}{\pi}}\sum_q\beta\sin\left(\beta\frac{\pi}{2}\right)A^+_q\Psi_q=\sqrt{-\Delta_{\mathbf{K}}+1}\sin\left(\frac{\pi}{2}\sqrt{-\Delta_{\mathbf{K}}+1}\right)\Psi_{\infty}.
$$

Finally since $\sin\left(\beta\frac{\pi}{2}\right) =\pm 1$, (\ref{profldeux}) and (\ref{profhm}) follow from
(\ref{psiplus}) and
(\ref{profpse}).

\fin

We remark that $\Psi^-=0$ iff $\partial_t\Psi(0,.)=0$, and the
asymptotic profile $\Psi_{\infty}$ depends only on $\Psi^+(0,.)$, therefore
it is determined by $\Psi(0,.)$ and
$\int_{\mathbf{K}}\partial_t\Psi(0,.)$. Moreover the map
$$
S:\;\Psi^+(0,.)\longmapsto \Psi_{\infty}
$$
is an isomorphism from $H^{-1}(\mathbf{K})$ onto $L^2(\mathbf{K})$
and from $L^2(\mathbf{K})$ onto $H^1(\mathbf{K})$. Since $S$ is a
pseudodifferential operator given by
$\left(-\Delta_{\mathbf{K}}+1\right)^{-\frac{1}{2}}\sin\left(\frac{\pi}{2}\sqrt{-\Delta_{\mathbf{K}}+1}\right)$,
the computation of the asymptotic profile using this formula is a very hard task. In
contrast, the numerical method for the computation of the time
dependent field $\Psi(t,.)$ presented in the
next part, allows to find $\Psi_{\infty}(X)=\lim_{t\rightarrow+\infty}\Psi(t,X)$ with a good accuracy.

\subsection{Inflating universe.} We consider the toy model of an
exponentially inflating universe with $a(t)=e^t$. We calculate the
asymptotic profile of a wave $\Psi$ and we show that $\Delta_{\mathbf K} \Psi(t)
-e^{2t} \partial_t \Psi(t)$ tends to zero.

\begin{Theorem}
Given a  smooth solution $\Psi$ of equation (\ref{eq}), there exist constants $A^+_q$ and $A^-_q$
depending only on initial data such that
\begin{equation}
\Psi(t,X)=\left[A^-_0 e^{-3t}+A^+_0\right]\Psi_0(X)+e^{-\frac32 t}
\sum_{q\neq 0} \left[ A^-_q J_{\frac32}(qe^{-t})+ A^+_q
  Y_{\frac32}(qe^{-t}) \right]\Psi_q(X), 
\end{equation}
\begin{equation}
\begin{split}
 &\Psi_{\infty}(X)=A^+_0 \Psi_0(X) -\sqrt{\frac{2}{\pi}} \sum_{q\neq 0}
q^{-\frac32 } A^+_q \Psi_q(X),\\
 &\Psi(t,X)-\Psi_{\infty}(X)= -e^{-2t} \left(\frac{1}{2}
  \sqrt{\frac{2}{\pi}} \sum_{q\neq 0}  q^{\frac12}A^+_q 
  \Psi_q(X)\right)+0\left(e^{-3t}\right),
\label{vexpinfty1}
\end{split}
\end{equation}
\begin{equation}
\begin{split}
 \Delta_{\mathbf K} \Psi(t,X)-e^{2t}  \partial_t \Psi(t,X)=e^{-t} \left( 3A^-_0 \Psi_0(X) +\sqrt{\frac{2}{\pi}}\sum_{q\neq0}q^{\frac32} A^-_q \Psi_q(X)\right) +0\left(e^{-2t}\right)
\label{vexpinfty2}
\end{split}
\end{equation}
where the set of $q$ is given by equation (\ref{valpra}), equation (\ref{valprb}). The
coefficients $A_q^{\pm}$ are given by:
\begin{equation}
 \label{ainfp}
  A_0^+=\frac{1}{4}\left(3u_0(0)+u'_0(0)\right),\;\;A^+_q=-\sqrt{\frac{\pi}{2}}\left(\left(q^{\frac{1}{2}}\sin
  q\right) u_q(0)+\frac{\sin q-q\cos q}{q^{\frac{3}{2}}} u'_q(0)\right),
\end{equation}
\begin{equation}
 \label{ainfm}
 A_0^-=\frac{1}{4}\left(u_0(0)-u'_0(0)\right),\;\;A^-_q=-\sqrt{\frac{\pi}{2}}\left(\left(q^{\frac{1}{2}}\cos
  q\right) u_q(0)+\frac{\cos q+q\sin q}{q^{\frac{3}{2}}} u'_q(0)\right),
\end{equation}
where $u_q(t):=<\Psi(t,.),\Psi_q>_{L^2(\mathbf{K})}$.
\label{propexp}
\end{Theorem}

{\it Proof.}
If $a(t)=e^t$, equation (\ref{lequdiff}) is just
\begin{equation}
u_q''(t)+3 \, u_q'(t)+q^2e^{-2t}u_q=0.
 \label{expq}
\end{equation}
For each $q$ we introduce a new function $w_q$ defined by
\[ u_q(t):=e^{-\frac32 t} w_q(t).\]
Then $u_q$ satisfies equation (\ref{expq}) iff $w_q$ is solution of 
\begin{equation}
w_q''(t)+\left( q^2e^{-2t}-\frac94\right) w_q=0.
 \label{expw}
\end{equation}
$q=0$ is a particular case. $w_0$ can be written $w_0(t)=A^-_0
e^{-\frac32 t}+A^+_0 e^{+\frac32 t}$ where $A^+_0$ and $A^-_0$ are
constants depending on initial data. So 
\begin{equation} u_0(t)=A^-_0 e^{-3t}+A^+_0.
 \label{vexp0}
\end{equation}
Then we consider $q\neq 0$ and we put
$w_q(t)=\tilde{w}_q(qe^{-t})$. $w_q$ satisfies equation (\ref{expw}) iff
$\tilde{w}_q$ is solution of the Bessel equation
\begin{equation}
x^2\tilde{w}_q''(x)+x\tilde{w}_q'(x)+\left( x^2-\left(\frac32\right)^2\right) \tilde{w}_q(x)=0.
 \label{expw2}
\end{equation}
For $x$ belonging to $]0,+\infty[$, $\tilde{w}_q$ can be written $\tilde{w}_q(x)=A^+_q
Y_{\frac32}(x)+A^-_q J_{\frac32}(x)$ where $A^-_q$ and $A^+_q$ are
constants depending on initial data (see 10.2 of
\cite{nist}). So we have
\[u_q(t)=e^{-\frac32 t} A^-_q J_{\frac32}(qe^{-t})+e^{-\frac32 t} A^+_q Y_{\frac32}(qe^{-t}).\]
 According to 10.6.2 (\cite{nist}) the derivatives of $J_{\frac32}$ and $Y_{\frac32}$
satisfy ${\mathcal C}'_{\frac32}(x)={\mathcal
  C}_{\frac12}(x)-\frac{3}{2x}{\mathcal C}_{\frac32}(x)$. We obtain

\begin{eqnarray*}
 u'_q(t)&=-e^{-\frac32 t} \left[\frac32 A^-_q J_{\frac32}(qe^{-t})+qe^{-t} A^-_q J'_{\frac32}(qe^{-t})+\frac32 A^+_q
        Y_{\frac32}(qe^{-t})+qe^{- t} A^+_q
        Y'_{\frac32}(qe^{-t})\right]\\
 &= -e^{-\frac52 t} q\,A^-_q J_{\frac12}(qe^{-t}) -e^{-\frac52 t} q\,A^+_q Y_{\frac12}(qe^{-t}).
\end{eqnarray*}
In fact, by the
use of some formulae in 10.47ii and 10.49i, these Bessel functions have
simple expressions:
\begin{eqnarray*}
 J_{\frac12}(x)=&\sqrt{\frac{2x}{\pi}} j_0(x)=\sqrt{\frac{2}{\pi x}} \sin x,\quad Y_{\frac12}(x)=\sqrt{\frac{2x}{\pi}} \;y_0(x)=-\sqrt{\frac{2}{\pi x}} \cos x,\\
 J_{\frac32}(x)=&\sqrt{\frac{2x}{\pi}} 
j_1(x)=\sqrt{\frac{2x}{\pi}} \left[ \frac{\sin
    x}{x^2}-\frac{\cos x}{x}\right], \quad Y_{\frac32}(x)=\sqrt{\frac{2x}{\pi}} 
y_1(x)=\sqrt{\frac{2x}{\pi}} \left[ -\frac{\cos
    x}{x^2}-\frac{\sin x}{x}\right]
\end{eqnarray*}
and we deduce (\ref{ainfp}), (\ref{ainfm}) by elementary
calculations. To investigate the behaviour of $u_q$ at $+\infty$ we use the following
asymptotics at $0$ of $Y_{\frac32}$, $J_{\frac32}$, $Y_{\frac12}$
and $J_{\frac12}$.
\begin{eqnarray*}
 J_{\frac12}(x)=\sqrt{\frac{2}{\pi}} \;\left[ x^{\frac12}-\frac{1}{6}x^{ \frac{5}{2}}+\frac{1}{120}x^{ \frac{9}{2}}+O(x^{\frac{13}{2}})
\right],\; \\Y_{\frac12}(x)=\sqrt{\frac{2}{\pi}} \;\left[ -x^{-\frac12}+\frac12  x^{\frac32}-\frac{1}{24} x^{\frac72}+\frac{1}{720} x^{ \frac{11}{2}}+O(x^{\frac{15}{2}})
\right],\\
 J_{\frac32}(x)=\sqrt{\frac{2}{\pi}} \;\left[ \frac13
  x^{\frac32}-\frac{1}{30} x^{\frac72}+\frac{1}{840} x^{ \frac{11}{2}}+O(x^{\frac{15}{2}})
\right], \;\\Y_{\frac32}(x)=\sqrt{\frac{2}{\pi}} \;\left[ -
  x^{-\frac32}-\frac{1}{2} x^{\frac12}+\frac{1}{8} x^{ \frac{5}{2}}-\frac{1}{144}x^{ \frac{9}{2}}+O(x^{\frac{13}{2}})
\right].\\
\end{eqnarray*}
All these relations show that for any $q\neq 0$ we have at $+\infty$
\begin{equation}
\begin{split}
u_q(t)&=\sqrt{\frac{2}{\pi}} e^{-\frac32 t} A^-_q \;\left[ \frac13
  \left(qe^{-t}\right)^{\frac32}\right]+\nonumber\\
&\qquad\sqrt{\frac{2}{\pi}} e^{-\frac32 t} A^+_q \;\left[  -
   \left(qe^{-t}\right)^{-\frac32}-\frac{1}{2}  \left(qe^{-t}\right)^{\frac12}+\frac{1}{8}  \left(qe^{-t}\right)^{ \frac{5}{2}}
\right]+0\left(e^{-4t}\right) \nonumber\\
&=\sqrt{\frac{2}{\pi}} \left[-q^{-\frac32 } A^+_q -\frac{1}{2}
   q^{\frac12}A^+_q  e^{-2t}+\frac13 q^{\frac32 } A^-_q
   e^{-3t}\right]+0\left(e^{-4t}\right),\label{vexpq}
\end{split}
\end{equation}
and
\begin{eqnarray}
 u'_q(t)&= -e^{-\frac52 t} q \sqrt{\frac{2}{\pi}} \,A^-_q 
         \left(qe^{-t}\right)^{\frac12} +e^{-\frac52 t} q \sqrt{\frac{2}{\pi}} \,A^+_q 
  \left(qe^{-t}\right)^{-\frac12} +0\left(e^{-4t}\right)\\
&=\sqrt{\frac{2}{\pi}} \left[  q^{\frac12} A^+_q  e^{-2t}  -
  q^{\frac32} A^-_q  e^{-3t}   \right]+0\left(e^{-4t}\right).
\label{vexpprimeq}
\end{eqnarray}
equation (\ref{vexpinfty1}) is proved thanks to equation (\ref{vexp0}) and
equation (\ref{vexpq}). Moreover for all $q\neq 0$ equation (\ref{vexpprimeq}) gives
\begin{eqnarray*}
-q^2u_q(t)-e^{2t}   u'_q(t)&=\sqrt{\frac{2}{\pi}} q^{\frac12 } A^+_q - \sqrt{\frac{2}{\pi}} \left[  q^{\frac12} A^+_q   -
  q^{\frac32} A^-_q  e^{-t} 
  \right]+0\left(e^{-2t}\right)\\
&=\sqrt{\frac{2}{\pi}} q^{\frac32} A^-_q  e^{-t} +0\left(e^{-2t}\right).
\end{eqnarray*}
Then, thanks to equation (\ref{vexp0}), we deduce 
\begin{eqnarray*}
\Delta_{\mathbf K} \Psi(t,X)-e^{2t}  \partial_t \Psi(t,X)&=-e^{2t}   u'_0(t)
                                         \Psi_0(X)+\sum_{q\neq
                                         0}\left(-q^2u_q(t)- e^{2t}
                                         u'_q(t)\right)\Psi_q(X)\\
&=\left( 3A^-_0 \Psi_0(X) \right) e^{-t} +\left(\sqrt{\frac{2}{\pi}}\sum_{q\neq0}q^{\frac32} A^-_q \Psi_q(X)\right) e^{-t} +0\left(e^{-2t}\right).
\end{eqnarray*}
This achieves the proof.
\fin

This theorem shows that the situation for the inflating universe is
similar to those of the de Sitter space. We can write again
$$
\Psi(t,X)=\Psi^+(t,X)+\Psi^-(t,X),
$$
where $\Psi^+$ and $\Psi^-$ are smooth solutions satisfying
\begin{equation}
 \label{}
\Psi^+(t,X)=\Psi_{\infty}(X)+O\left(e^{-2t}\right),
\end{equation}
\begin{equation}
 \label{}
 \Psi^-(t,X)=O\left(e^{-3t}\right).
\end{equation}
It is sufficient to introduce
\begin{equation}
\Psi^+(t,X):=A^+_0\Psi_0(X)
+e^{-\frac{3}{2}t}\sum_{q\neq 0} A^+_q Y_{\frac{3}{2}}\left(qe^{-t}\right)\Psi_q(X), 
\end{equation}
\begin{equation}
\Psi^-(t,X):=A_0^-e^{-3t}\Psi_0(X)+e^{-\frac{3}{2}t}\sum_{q\neq 0} A^-_q J_{\frac{3}{2}}\left(qe^{-t}\right)\Psi_q(X).
\end{equation}
The coefficients $A^{\pm}_q$ are calculated from the initial data of
$\Psi$ at
$t=0$, but
we can also deduce them from $(\Psi^+(0,.),\partial_t\Psi^+(0,.))$. $\Psi^+(0,.)$ characterizes $A_0^+$ and the family
$A^+_qY_{\frac{3}{2}}(q)$. $\partial_t\Psi^+(0,.)$ characterizes the
sequence $A^+_q\left(\frac{3}{2}Y_{\frac{3}{2}}(q)+q
  Y'_{\frac{3}{2}}(q)\right)$. Since $\mid Y_{\frac{3}{2}}(q)\mid+\mid
Y'_{\frac{3}{2}}(q)\mid\neq 0$, the whole family $A^+_q$ is
determined. To estimate the norms of the asymptotic profile
$\Psi_{\infty}$, we introduce generalized energies at time zero of the solutions
$\Psi$ of (\ref{eq}) by putting
\begin{equation}
 \label{}
 E_m(\Psi,0):=\Vert\nabla_{\mathbf{K}}\Psi(0,.)\Vert_{H^m(\mathbf{K})}^2+\Vert\partial_t\Psi(0,.)\Vert^2_{H^m(\mathbf{K})},\;\;m\in\ZZ.
\end{equation}

\begin{Theorem}
 There exists $C,C'>0$ such that for any smooth solutions we have
\begin{equation}
 \label{} \Vert\Psi^+(0,.)\Vert_{L^2(\mathbf{K})}^2+C^{-1}E_{-2}(\Psi^+,0)\leq\Vert\Psi_{\infty}\Vert_{L^2(\mathbf{K})}^2\leq
\Vert\Psi^+(0,.)\Vert_{L^2(\mathbf{K})}^2+C E_{-2}(\Psi^+,0),
\end{equation}
\begin{equation}
 \label{}
  C^{-1}E_{-1}(\Psi^+,0)\leq\Vert\nabla_{\mathbf{K}}\Psi_{\infty}\Vert_{L^2(\mathbf{K})}^2\leq C E_{-1}(\Psi^+,0),
\end{equation}
\begin{equation}
 \label{}
  C^{-1}E_{0}(\Psi^+,0)\leq\Vert\Delta_{\mathbf{K}}\Psi_{\infty}\Vert_{L^2(\mathbf{K})}^2\leq
 C E_{0}(\Psi^+,0)\leq C'E_0(\Psi,0).
\end{equation}
 \label{reguexpinfini}
\end{Theorem}

{\it Proof.} We expand the norms in terms of series:
$$
\Vert\Psi_{\infty}\Vert_{L^2(\mathbf{K})}^2=\mid
A^+_0\mid^2+\frac{2}{\pi}\sum_{q\neq 0}q^{-3}\mid A^+_q\mid^2,
\;
\Vert\nabla_{\mathbf{K}}\Psi_{\infty}\Vert_{L^2(\mathbf{K})}^2=\frac{2}{\pi}\sum_{q\neq
  0}q^{-1}\mid A^+_q\mid^2,
$$
$$
\Vert\Delta_{\mathbf{K}}\Psi_{\infty}\Vert_{L^2(\mathbf{K})}^2=\frac{2}{\pi}\sum_{q\neq 0}q\mid A^+_q\mid^2,
$$
$$
\Vert\Psi^+(0,.)\Vert^2_{H^m(\mathbf{K})}=\mid
A_0^+\mid^2+\frac{2}{\pi}\sum_{q\neq 0}(q^2+1)^mq^{-3}\mid \cos q-q\sin
q\mid^2\mid A_q^+\mid^2,
$$
$$
E_m(\Psi^+,0)=\frac{2}{\pi}\sum_{q\neq
  0}(q^2+1)^m\left(q-\sin(2q)+\frac{\cos q}{q}\right)\mid A_q^+\mid^2,
$$
and we use the inequalities $c^{-1}q\leq q-\sin(2q)+\frac{\cos
  q}{q}\leq cq$ for some $c>0$, and $\mid A^+_q\mid^2\lesssim q\mid
u_q(0)\mid^2+q^{-1}\mid u'_q(0)\mid^2$.
\fin
\section{Numerical resolution}

We develop in this part an accurate scheme of computation of the wave
propagating on $\mathbf K$ in both cases
$a(t)=H^{-1}\cosh(Ht)$ and $a(t)=e^t$.  The computations are
performed on $\mathcal{F}_v$. The precise computation during a long time is a hard task because of the exponential behaviour
of the scale factor. To overcome this difficulty we use finite
elements of second order. We carefully check that the numerical
diffusion is very weak by computing waves with a future horizon, and
we compare the numerical localization of this horizon with its
theoretical value given by (\ref{rayonRh}). The validity of our
approach is established by comparison with explicit solutions, and by numerically testing the asymptotic results
of the previous section. In particular the asymptotics (\ref{deltach}) and
(\ref{vexpinfty2}) show that
\begin{equation}
 \label{asymptodeltadt}
 \Delta_{\mathbf{K}}\Psi(t,.)-A(t)\partial_t\Psi(t,.)=O\left(e^{-at}\right),\;\;t\rightarrow+\infty,
\end{equation}
for some $a>0$ where $A$ is an {\it exponentially increasing}
function. Since $\partial_t\Psi(t,.)$ tends exponentially to zero, the
numerical checking of this property is challenging: it is an excellent
test of the robusteness and the accuracy of the scheme.
\\

 We solve the variational problem equation (\ref{pbvaria}) using the usual way (see
 {\it e.g.} \cite{christiansen}). We take a family $V_h$, $0<h\leq h_0$, of finite dimensional vector subspaces of
$W^1(\mathcal{F}_v)$. We assume that
\begin{equation*}
\overline{\cup_{0<h\leq h_0}V_h}=W^1(\mathcal{F}_v).
  \label{}
\end{equation*}
We choose sequences $u_{0,h},\;u_{1,h}\in V_h$ such that
$$
u_{0,h}\rightarrow
u_0\;\;in\;\;W^1(\mathcal{F}_v),\;u_{1,h}\rightarrow u_1\;\;in\;\;L^2(\mathcal{F}_v).
$$
We consider the solution $u_h\in C^{\infty}(\RR_t;V_h)$ of
\begin{equation*}
\begin{split}
  \forall\phi_h\in V_h,\qquad 0=&
\frac{d^2}{dt^2} \int_{\F_v}(1-|X|^2)^{-\frac12}u_h(t,X)\phi_h(X)dX \nonumber\\
&+3\frac{a'(t)}{a(t)}\frac{d}{dt} \int_{\F_v}(1-|X|^2)^{-\frac12}u_h(t,X)\phi_h(X)dX\nonumber\\
&+\frac{1}{a^2(t)}\int_{\F_v} (1-|X|^2)^{-\frac12}\nabla u_h(t,X) \cdot \nabla \phi_h(t,X)\, dX\nonumber\\
&-\frac{1}{a^2(t)}\int_{\mathcal{F}_v}(1-|X|^2)^{-\frac12}\left(X\cdot \nabla u_h(t,X)\right)\left(X \cdot \nabla \phi_h(t,X)\right)\, dX,\nonumber\\
\end{split}
\end{equation*}
satisfying $u_h(t^{\star},.)=u_{0,h}(.)$,
$\partial_tu_h(t^{\star},.)=u_{1,h}(.)$. We shall put an initial
data at $t=t^{\star}$ with $t^*$ large enough, rather than at $t=0$, in order to achieve faster limit states.
We consider a basis $\left(e_j^h\right)_{1\leq j\leq N_h}$ of $V_h$
and we
expand $u_h$ on this basis:
$$
u_h(t)=\sum_{j=1}^{N_h}u_j^h(t)e_j^h.
$$
We introduce 
$$
U(t):=
\;^t\!\left(u_1^h,u_2^h,\cdots,u_{N_h}^h\right),\;\;\MM=\left(M_{ij}\right)_{1\leq
  i,j\leq N_h},\;\;\DD=\left(D_{ij}\right)_{1\leq i,j\leq
  N_h},\;\;\KK=\left(K_{ij}\right)_{1\leq i,j\leq N_h}
$$
where
\begin{equation*}
\begin{split}
& M_{ij}:=\int_{\F_v}\frac{1}{\sqrt{1-|X|^2}}e_i^h(X)e_j^h(X) \, dX,\\
& K_{ij}:=\int_{\F_v}\frac{1}{\sqrt{1-|X|^2}}\left(\partial_xe_i^h(X)\partial_xe_j^h(X)+ \partial_ye_i^h(X)\partial_ye_j^h(X) +\partial_ze_i^h(X)\partial_ye_j^h(X)\right) \, dX,\\
&D_{ij}:=-\int_{\F_v}\frac{1}{\sqrt{1-|X|^2}}\left(x\partial_xe_i^h(X)+y\partial_ye_i^h(X)+z\partial_ze_i^h(X)\right)\left(x\partial_xe_j^h(X)+y\partial_ye_j^h(X)+z\partial_ze_j^h(X)\right)\, dX. 
\end{split}
\end{equation*}
Then the variational formulation is equivalent to 
\begin{equation}
\MM \left(U'' +3\frac{a'(t)}{a(t)} U'\right)+\frac{1}{a^2(t)}\left(\KK+\DD\right) U=0.
  \label{SumLaplacien}
\end{equation}
Like in \cite{PDS} we can prove that the solution $u_h$ obtained by
solving (\ref{SumLaplacien}) tends to the solution $u$ of equation (\ref{pbvaria})
as $h\rightarrow 0$.
This differential system is solved very simply by iteration by solving
\begin{eqnarray*}
 0=\;\MM\left(U^{n+1}-2U^n+U^{n-1}\right)&+3\frac{a'(t^{\star}+ n\Delta T)}{a(t^{\star}+ n\Delta T)} \Delta T\; \MM\left(\frac{U^{n+1}-U^{n-1}}{2}\right)\\&+(\Delta T)^2 \frac{1}{a^2( t^{\star}+ n\Delta T)}\left(\KK+\DD\right) U^n
  \label{}
\end{eqnarray*}
with an initial data at $t^{\star}$. The Lax-Richtmyer theorem
assures that the approximate solution given by this scheme tends to
$u_h$ when the time step $\Delta T\rightarrow 0$.
Moreover we compute $E_d(t^{\star}+n\Delta T)$, an approximation at time $t^{\star}+n\Delta T$ of the energy $E(t^{\star}+n\Delta T)$:
\begin{equation}
 E_d(t^{\star}+n\Delta T ):=\left< \MM \;\frac{U^n-U^{n-1}}{\Delta t}\; , \;
  \frac{U^n-U^{n-1}}{\Delta t}\right>\;+\; \frac{1}{a^2(t^{\star}+n\Delta T)}\left< (\KK +\DD)\; U^{n-1}\; , \; U^n\right>.
\label{Enerd}
\end{equation}
We use the same mesh as in our previous article \cite{PDS}. But
we construct the finite element spaces $V_h$ of $\P_2$ type instead of
$\P_1$ type. We perform this choice to gain a better accuracy that is
necessary to overcome the difficulty linked to the exponential
behaviour of $a(t)$. We take into account the boundary condition (\ref{cl}) in the
definition of the finite elements, so that $V_h\subset W^1(\mathcal{F}_v)$.
We note $\mathcal{T}_h$ the set of all tetrahedra of the mesh,
$\mathcal{ F}_{v,h}$ the set $\displaystyle{\cup_{T\in \mathcal{T}_h}
  T}$, and $\P_2(T)$ the set of second degree polynomial functions on
$T$. We choose $P_2$ Lagrange finite elements, so our nodes are among
vertices and middle of edges of $T \in \mathcal{T}_h$. In our construction of the mesh we
took care that vertices belonging to $\partial(\mathcal{ F}_{v,h})$
belong to $\partial(\mathcal{F}_v) $ and that for any vertex $S$ of $T\in
\mathcal{T}_h$,  $S'$ is also a vertex of a tetrahedron of
$\mathcal{T}_h$ if $S'\sim S$ (see
\cite{PDS}). It is therefore easy to take into account the boundary
condition for vertices of $T\in\mathcal{T}_h$.  Moreover middle of edges of peripheral tetrahedra are not on the
border of $\mathcal{F}_v $ thanks to the convexity of edges and faces
of $\mathcal{F}_v $;
so they are only equivalent to themselves.  Then we
introduce:
$$V_h:=\left\{ v: \mathcal{ F}_{v,h} \rightarrow \RR,\, v \in
\mathcal{C}^0(\mathcal{ F}_{v,h}), \forall T \in \mathcal{T}_h,\,
v_{|T}\in \P_2(T),\, M\sim M'\Rightarrow v(M)=v(M')\right\}.$$
$dim\, V_h $ is equal to the number of equivalence classes of vertices plus middle
of edges of tetrahedra of $\mathcal{T}_h$. 
If $j$ is the number of a node associated to a vertex $M_j$ or to the
middle $A_j$ of an edge, we construct a basis $\left(e_j^h\right)_{1\leq j\leq N_h}$ of $V_h$ by:\\
\begin{enumerate}
\item If $j$ is associated to a node $M_j$ that does not belong to $\partial \mathcal{F}_v$ :
$e_j^h(M_i)=\delta _{ij}$, $e_j^h(A_i)=0$.\\
There are $n_{\mathrm{vi}}$ functions of this kind.
\item If $j$ is associated to a node $M_j$ that is a vertex of  $\mathcal{F}_v$: \\
$e_j^h(M_i)=\left\{\begin{array}{ll}
1 & if\ M_i\sim M_j,\\
0 & otherwise
\end{array}
\right.$, $e_j^h(A_i)=0$.\\
There are five functions of this kind.
\item If $j$ is associated to a node $M_j$ that belongs to a face of $\partial \mathcal{F}_v$ and not to an edge:\\
$e_j^h(M_i)=\left\{\begin{array}{ll}
1 & if\ M_i=P_i,\\
0 & otherwise
\end{array}
\right.$, $e_j^h(A_i)=0$.\\
There are $6\times n_{\mathrm{vf}}$ functions of this kind.
\item If $j$ is associated to a node $M_j$ that belongs to an edge of a face of $\partial \mathcal{F}_v$ and is not a vertex of $\mathcal{F}_v$:
$e_j^h(M_i)=\left\{\begin{array}{ll}
1 & if\ M_i=P_i,\\
0 & otherwise
\end{array}
\right.$, $e_j^h(A_i)=0$.\\
There are $10\times n_{\mathrm{ve}}$ functions of this last kind.
\item If $j$ is associated to a node $A_j$ that is a middle of an edge:
$e_j^h(A_i)=\delta _{ij}$, $e_j^h(M_i)=0$.\\
There are as many functions of this kind as different edges in the mesh.
\end{enumerate}
with $n_{\mathrm{ve}}$ the number of mesh vertices on an edge of a
face that are not a vertex of $\mathcal{F}_v $, $n_{\mathrm{vf}}$ the
number of mesh vertices on a face that are not on an edge and $n_{\mathrm{vi}}$ the number of mesh vertices in $
\stackrel{\circ}{\mathcal{F}_v}$. For a given mesh, the number of nodes is important. For example if we
have a mesh with $89$ vertices on each edge of $\partial\mathcal{F}_v$, we
have $730\,309$ nodes for $\P_1$ type finite elements method, and
$6\,377\,052$ nodes for $\P_2$ type method. Matrices are created with a 31
points tetrahedral quadrature formula which is of degree 7
(\cite{quadrature}).\\

With our choice of finite elements, the Laplacian is constant in a
tetrahedron $T$; in the following it will be noted $\Delta_{\mathcal{F}_v}
{u_h}_{|T}(t,G_T)$ where $G_T$ is the gravity center of tetrahedron
$T$. To test equation (\ref{asymptodeltadt}), given by equation (\ref{deltach}) and equation (\ref{vexpinfty2}), we have chosen to
calculate $Norm$ defined respectively by:
$$
Norm :=\left[\sum_{T\in \mathcal{T}_h} \int_T \frac{1}{\sqrt{1-|X|^2}}\left(\Delta_{\mathcal{F}_v}
{u_h}_{|T}(t,G_T)-\frac{1}{2H} \left(1-\tanh H t\right)^{-1} \partial_t
u_h(t,G_T)\right)^2\; dX \right]^{\frac12}
$$
 and
$$
 Norm :=\left[\sum_{T\in \mathcal{T}_h} \int_T \frac{1}{\sqrt{1-|X|^2}}\left(\Delta_{\mathcal{F}_v}
{u_h}_{|T}(t,G_T)-e^{2t}  \partial_t
u_h(t,G_T)\right)^2\; dX \right]^{\frac12}.
$$

\subsection{Numerical results for the exponentially inflating universe.}
\begin{figure}[!h]
\begin{center}
\includegraphics[scale=0.6]{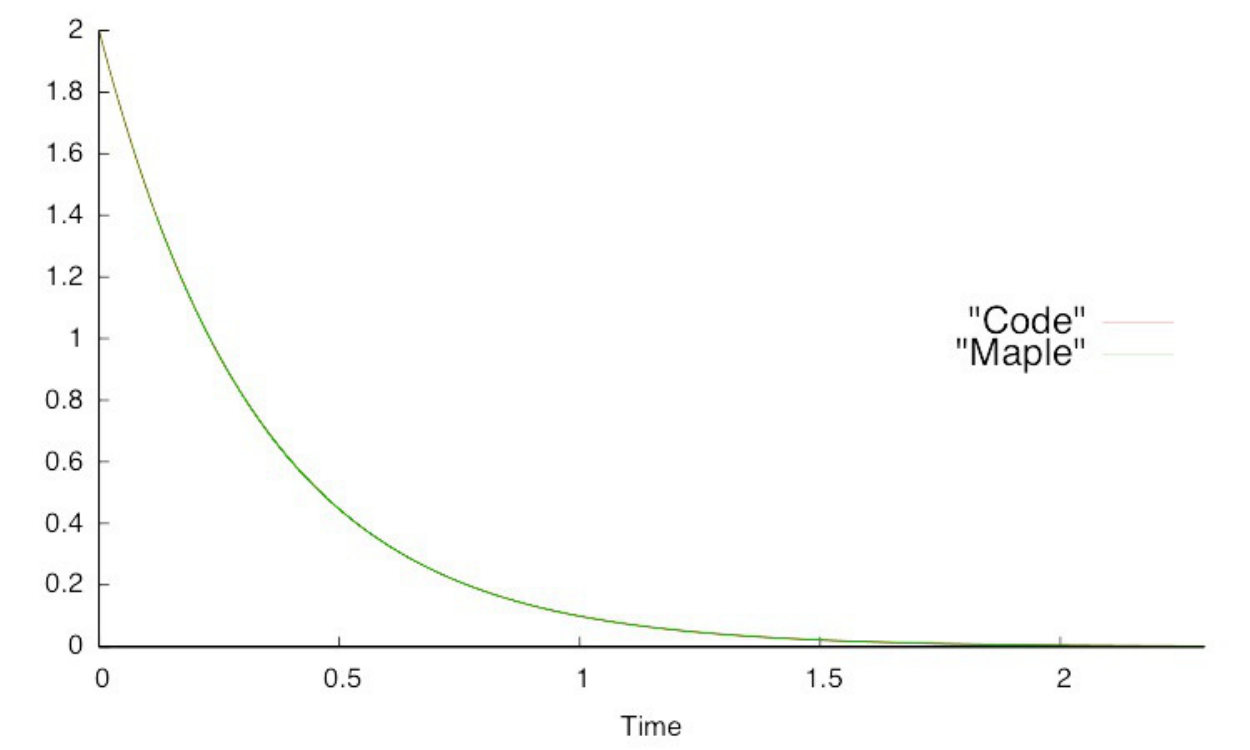}
\caption{\label{courbes-testExp} Computation of $\Psi(t,X)=2e^{-3t}$.}
\end{center}
\end{figure}

For this model the scale factor is $a(t)=e^t$. First we check the
efficiency of our code by the computation of the trivial solution
$\Psi(t,X)=2e^{-3t}$. The figure (\ref{courbes-testExp}) shows that the result
  obtained by using our code matches with the exact solution $e^{-3t}$
  depicted with MAPLE. The results agree at $\pm 10^{-5}$.

\begin{figure}[!h]
\begin{minipage}{10cm}
\includegraphics[scale=0.15]{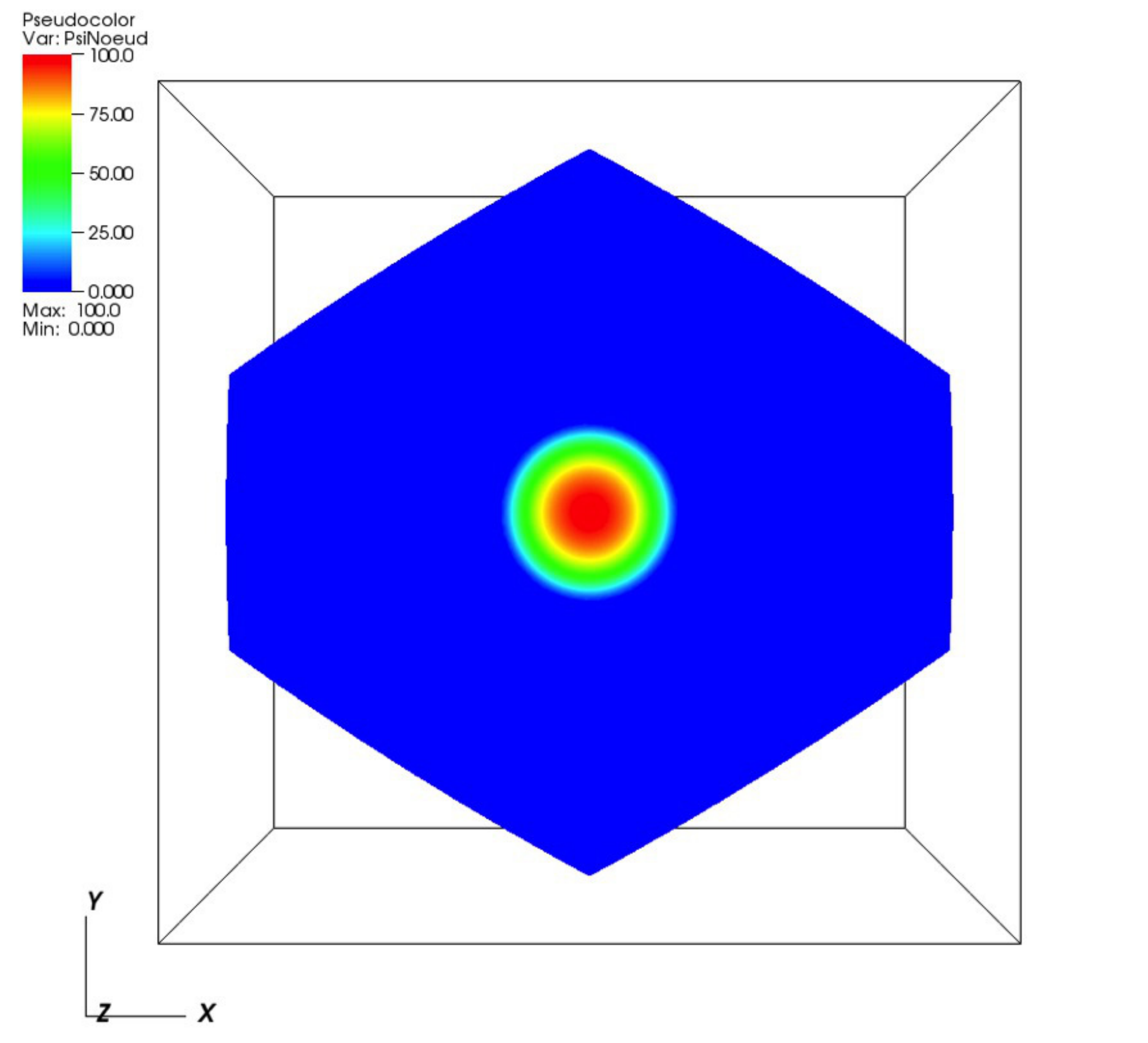}
\includegraphics[scale=0.15]{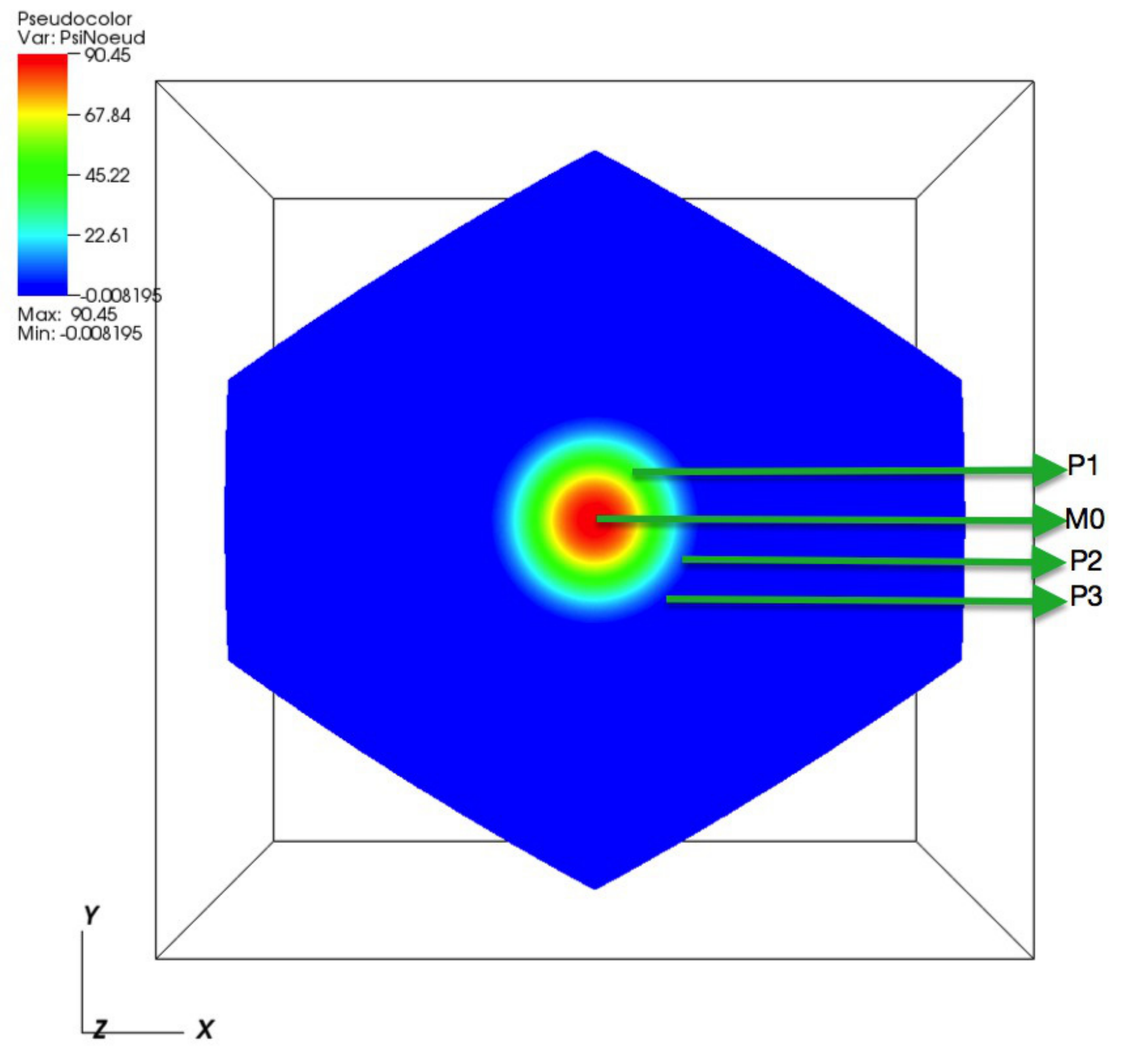}
\caption{\label{Init1}Solution for $Init_1$, on the left at
  $t^{\star}=3.5$, on the right at $t_{obs}=28.7$.}
\end{minipage}\hfill
\begin{minipage}{6cm}
There is a future horizon and $R_h=\sin\left(\arcsin
  0.1+e^{-3.5}\right)\simeq
0.13$. 
\end{minipage}
\end{figure}

\begin{figure}[h!]
\begin{minipage}{10cm}
\includegraphics[scale=0.15]{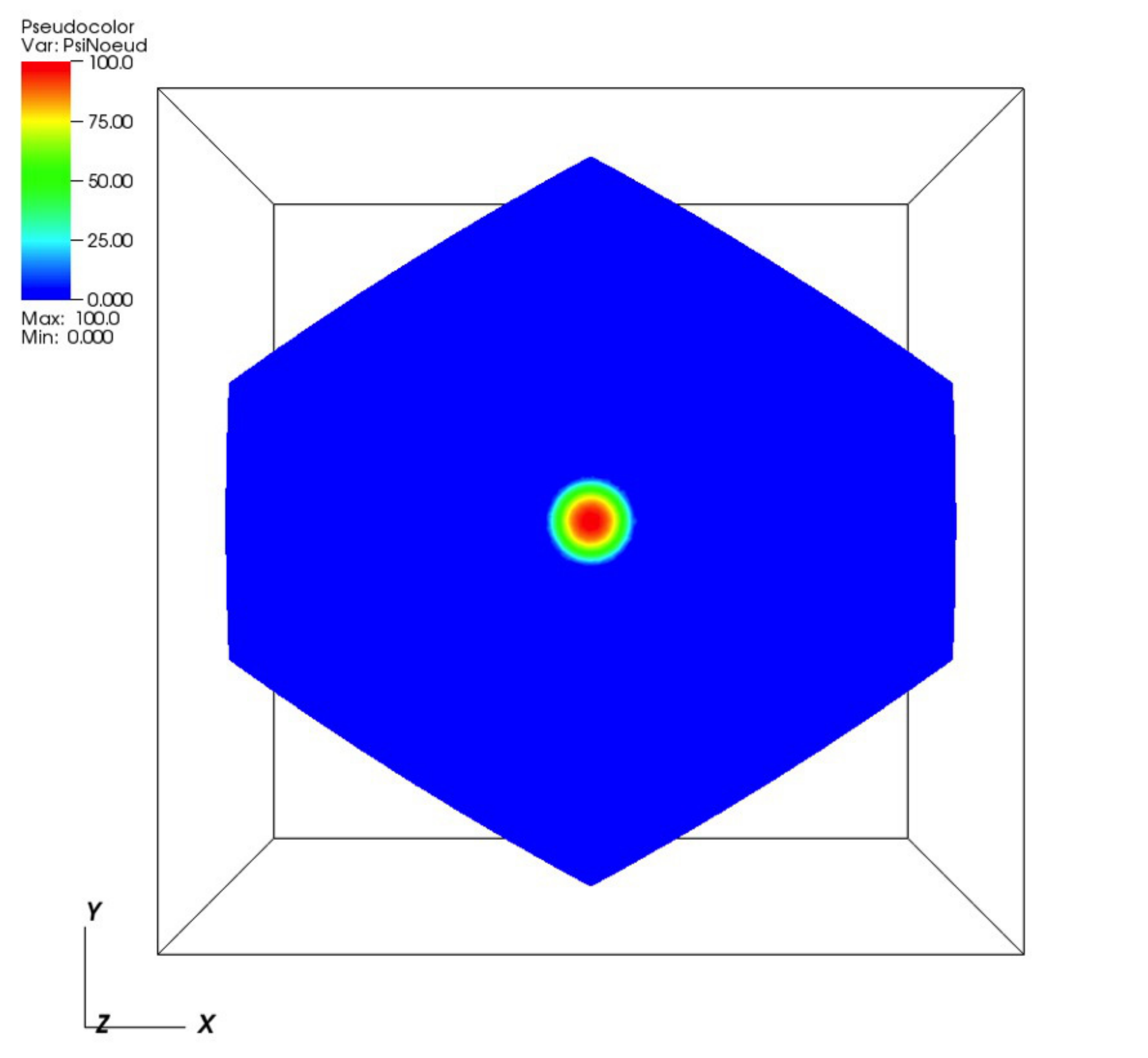}
\includegraphics[scale=0.15]{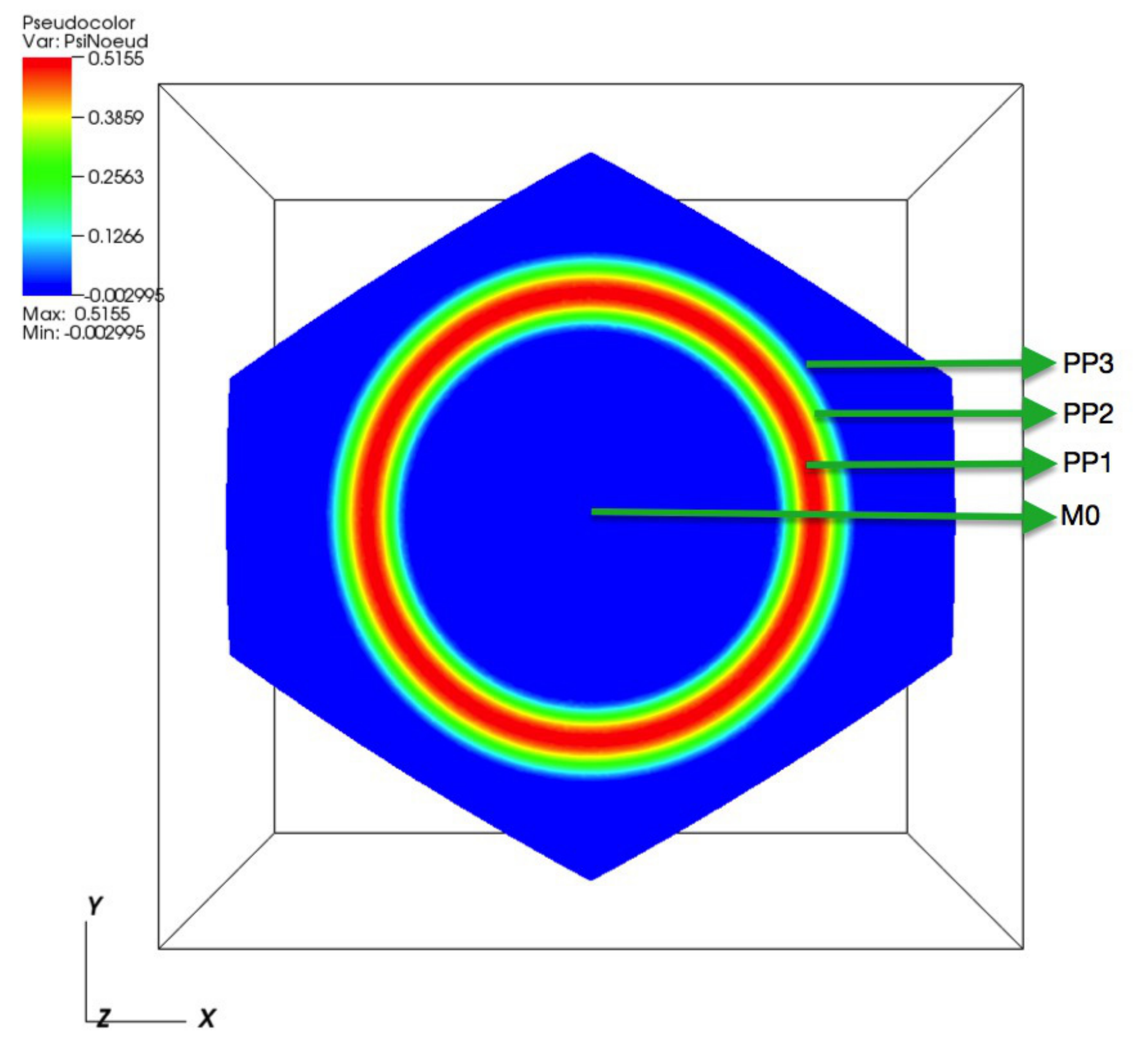}
\caption{\label{Init2}Solution for $Init_2$, on the left at
  $t^{\star}=1.5$, on the right at $t_{obs}=19.7$.}
\end{minipage}\hfill
\begin{minipage}{6cm}
In this case we do have
$R_h=\sin\left(\arcsin
  0.05+e^{-1.5}\right)\simeq 0.27$. 
\end{minipage}
\end{figure}

\begin{figure}[!h]
\begin{center}
\includegraphics[scale=0.3]{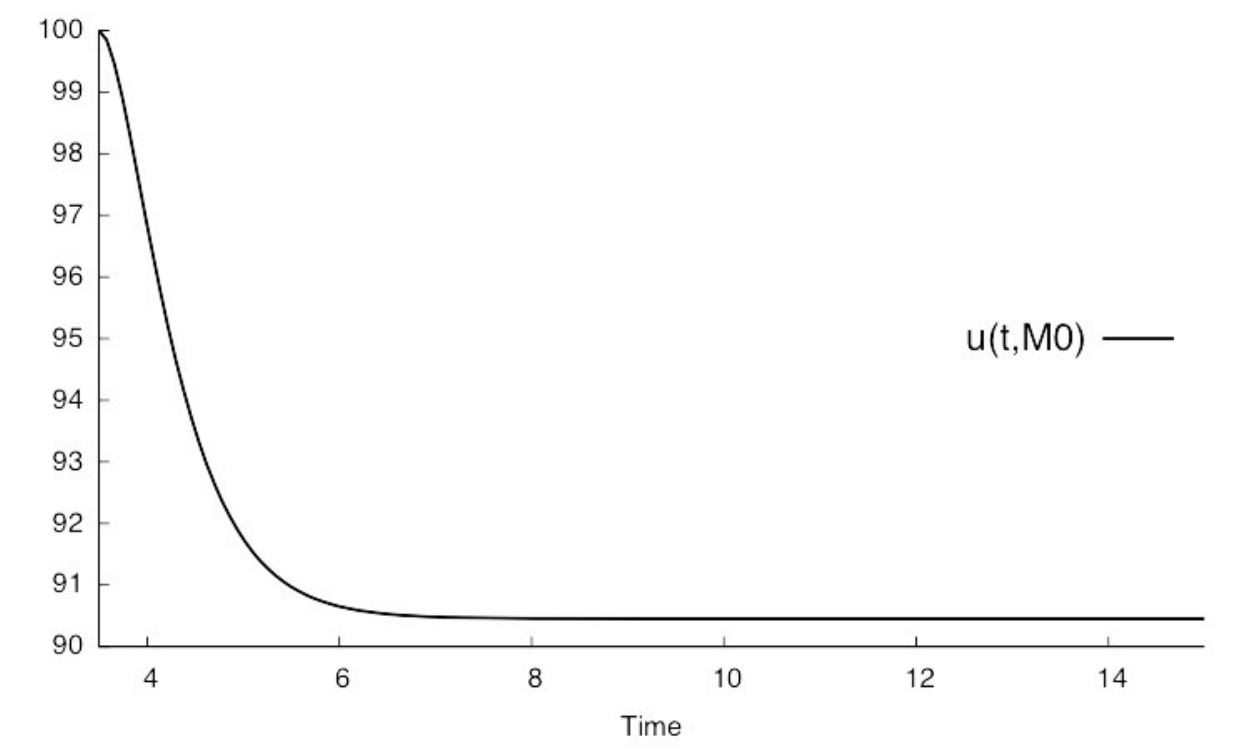}
\includegraphics[scale=0.3]{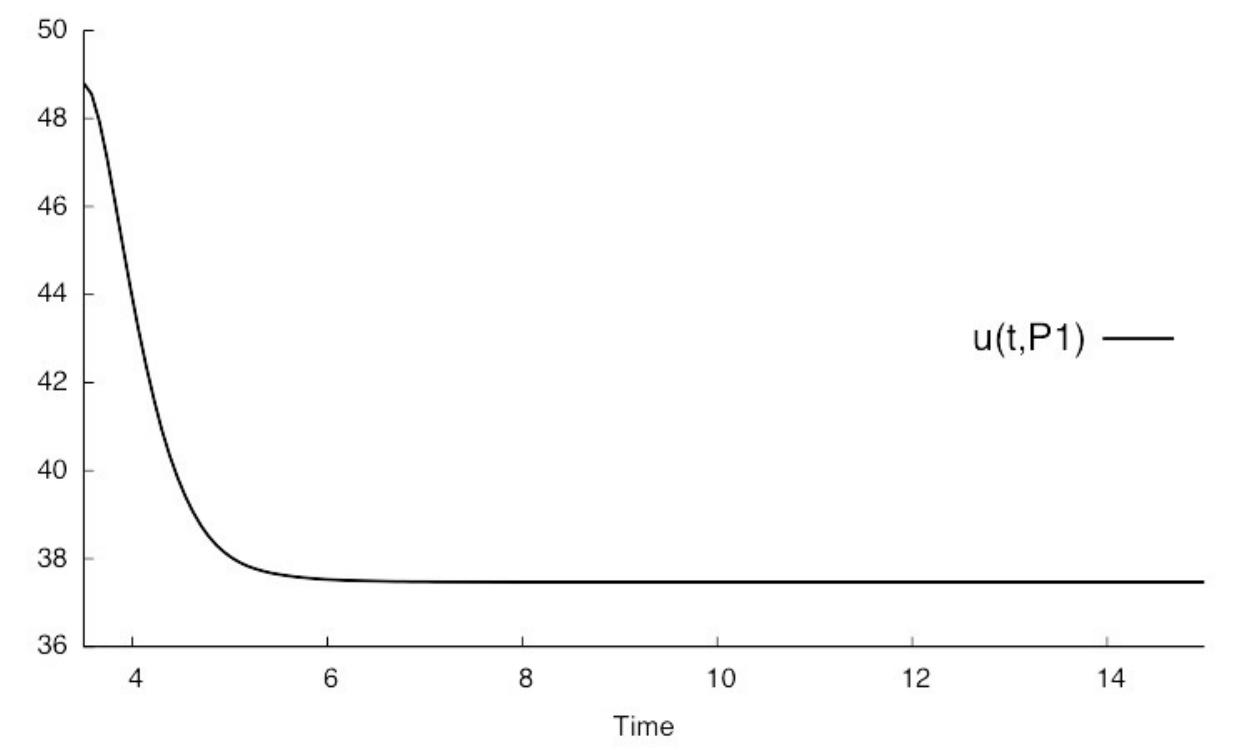}
\includegraphics[scale=0.3]{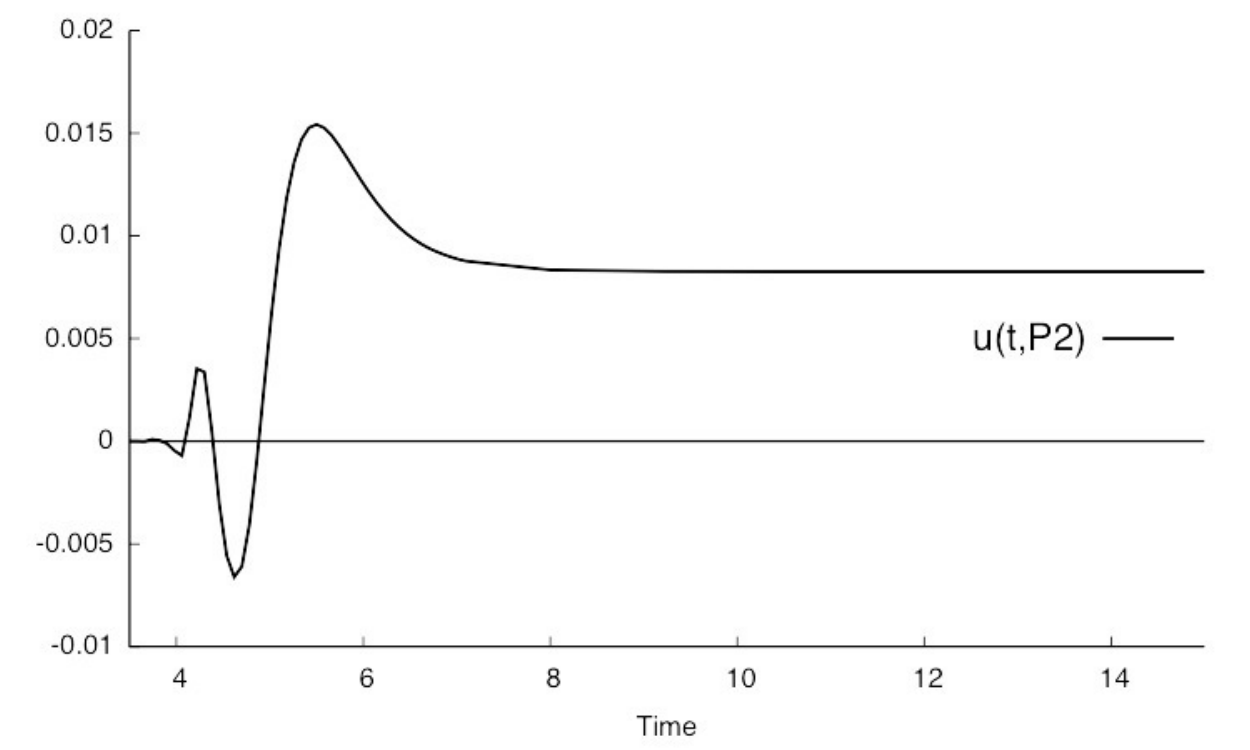}
\includegraphics[scale=0.3]{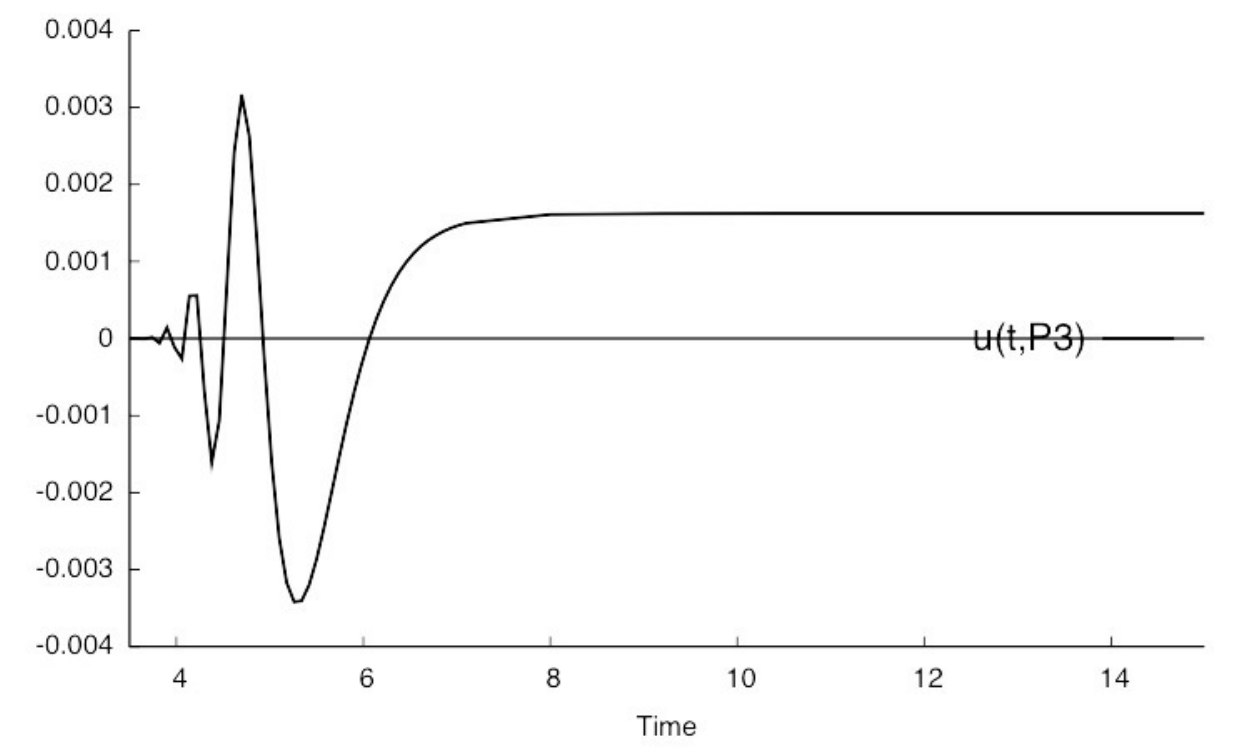}
\caption{\label{Init1-M}The solution $u(t,X)$ at M0, P1, P2, P3 with
  $Init_1$ and $t^{\star}=3.5$.}
\end{center}
\end{figure}

\begin{figure}[!h]
\begin{center}
\includegraphics[scale=0.3]{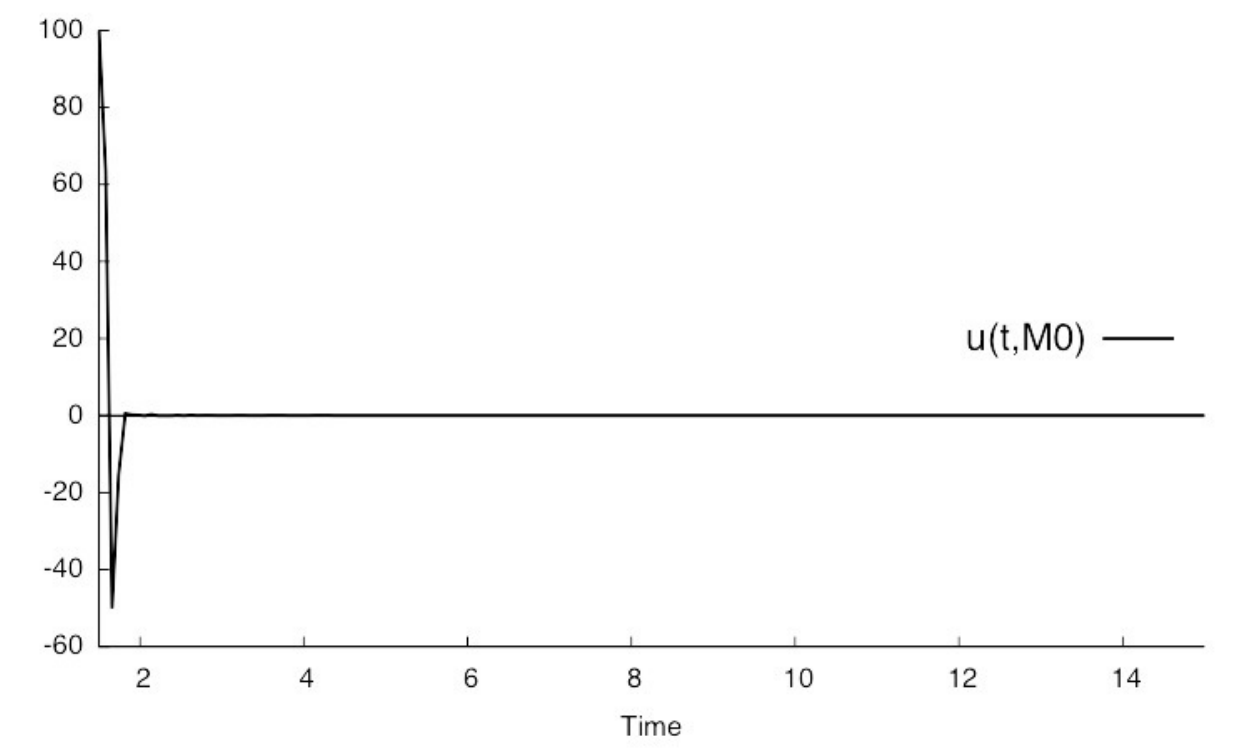}
\includegraphics[scale=0.3]{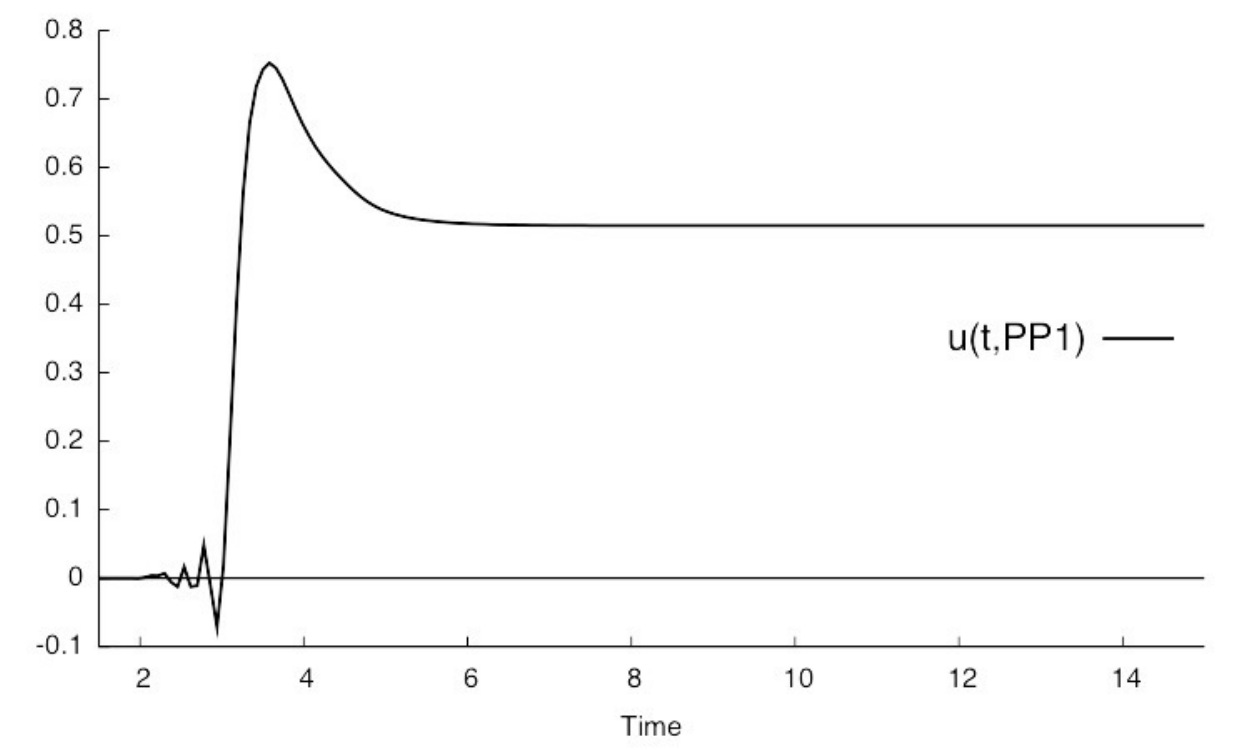}
\includegraphics[scale=0.3]{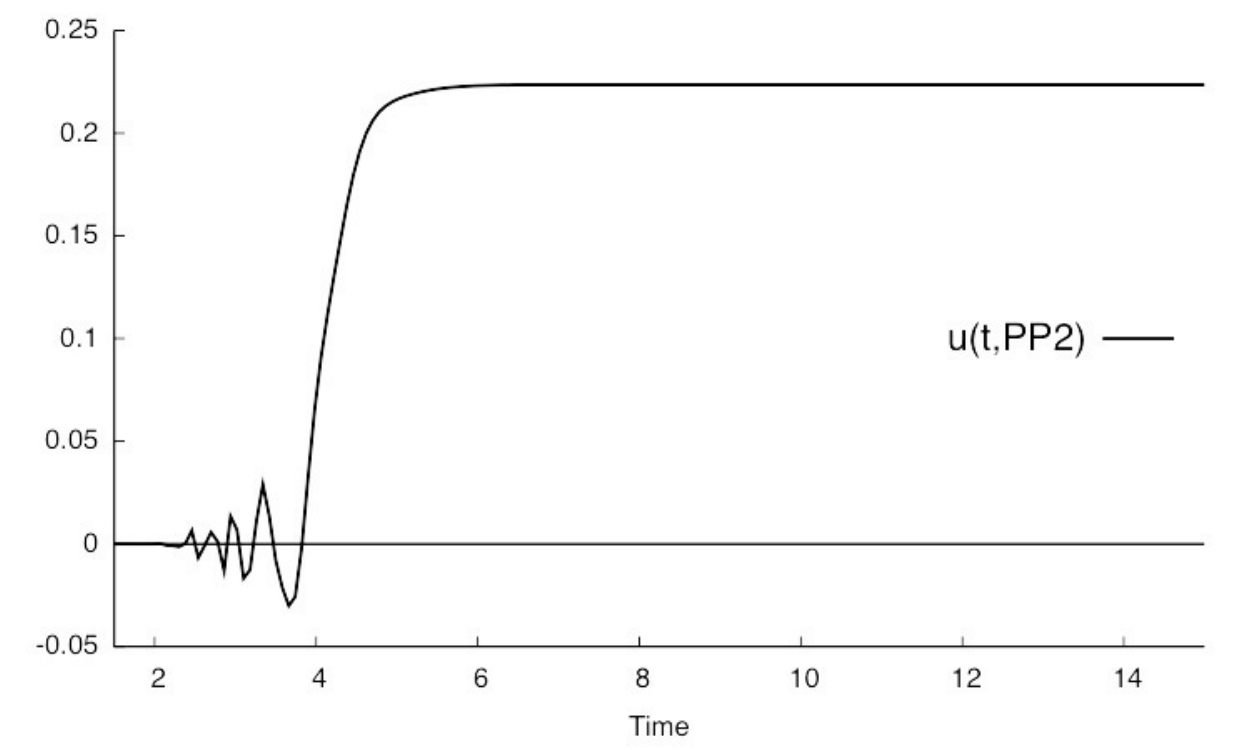}
\includegraphics[scale=0.3]{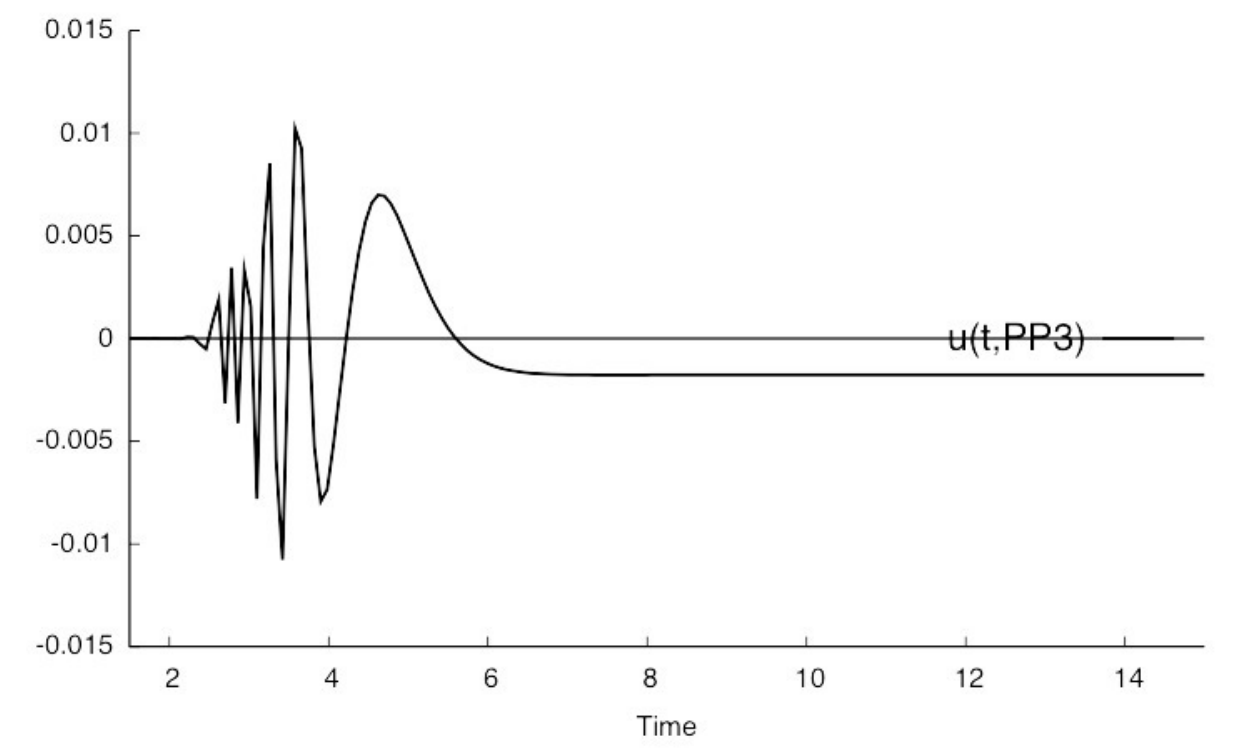}
\caption{\label{Init2-M}The solution $u(t,X)$ at M0, PP1, PP2, PP3 with $Init_2$ and $t^{\star}=1.5$.}
\end{center}
\end{figure}

We now test the accuracy of
our code by the numerical localization of a future horizon and its
robustness by checking the asymptotic behaviour
(\ref{asymptodeltadt}). We choose different initial data and a small time step:
$0.00015$. We present our results for two initial data at time $t^{\star}$
denoted $Init_1$
and $Init_2$, of the form:
\begin{equation}
 u_0(X)=100e^{\frac{d^2(X,0)}{d^2(X,0)
    -R_0^2}},\;\;\mbox{for}\;d(X,0)<
R_0,\;\mbox{and}\;u_0(X)=0,\;\;\mbox{for}\;d(X,0)\geq R_0.
\label{Init}
\end{equation}
For $Init_1$ we have $R_0=0.1$ and for $Init_2$, $R_0=0.05$. In order to
simplify we choose $\partial_tu(t^{\star},.)=u_1(.)=0$. Figure
(\ref{Init1}) and figure
(\ref{Init2}) present the initial data and the
limit state on the slice $z=0$ of $\mathcal{F}_v$ observed at time
$t_{obs}>t^*$ that we choose large enough in order to that the asymptotic
state is reached. Figure (\ref{Init1-M})
shows the time evolution of the solution associated to $Init_1$
for some points $P$ of $\mathcal{F}_v$ pointed out in
figure (\ref{Init1}), and figure (\ref{Init2-M}) shows the time evolution of the
solution associated to $Init_2$
for some point $PP$ of $\mathcal{F}_v$ pointed out in
figure (\ref{Init2}). In the following, $M0$ denotes a node of the mesh that is
the closest to the center $0$ of $\mathcal{F}_v$. 

\begin{table}[!h]
\begin{center}
\begin{minipage}{7cm}
\centering
    \begin{tabular}{|l|c|c|}
\hline
Point&\quad $d(0_{\RR^3},M)$ & $u(15,M)$\\
\hline
M0 & $\quad  0.00008 \quad $ & 90.45\\
P1 & $\quad  0.064 <R \quad $ &    37.47\\
P2 & $\quad  0.126\leq R_h \quad $& 0.008\\
P3 &  $\quad  0.131  \simeq R_h\quad $  & 0.0016  \\
\hline
\end{tabular}
\caption{\label{tab:Dist1}$Init_1$ and $t^{\star}=3.5$, $R_h\simeq 0.1299$}
\end{minipage}
\hfill
\begin{minipage}{7cm}
\centering
 \begin{tabular}{|l|c|c|}
\hline     
Point& $d(0_{\RR^3},M)$ & $u(15,M)$\\
\hline
$M0$ & $0.00008 <R$ &  0.0032\\
$PP1$ & $R<0.218 <R_h$&0.51\\
$PP2$ & $R<0.245 <R_h$&0.22\\
$PP3$ & $R < 0.270 \simeq R_h$&-0.0017\\
\hline
\end{tabular}
\caption{\label{tab:Dist2}$Init_2$ and $t^{\star}=1.5$, $R_h\simeq
  0.2697$}
\end{minipage}
\end{center}
\end{table}

It clearly appears that a limit state is
reached very quickly, as soon as $t=8$ in both cases. We also note the
existence of a future horizon. In table (\ref{tab:Dist1}) and table (\ref{tab:Dist2}) we
can read the distance between previous
points $M(x_M,y_M,z_M)$ of $\mathcal{F}_v$ and its center. The agreement between the value of $R_h$ given by equation (\ref{horet}),
equation (\ref{horch}) is quite good, but it is difficult to have a precise
numerical estimation of this horizon; for example P2 and P3 belong to a same tetrahedron, so the theoretical horizon
passes throught this tetrahedron. For $Init_1$ and $t^{\star}=3.5$ it is found that  outside the
ball of radius $0.129$ the solution is always less than
$0.01$, and outside the ball of radius $0.142$ the solution is always
less than $0.001$. Knowing that the limit state reaches $90$ we
conclude that the numerical estimation of $R_h$ is satisfactory.

\begin{figure}[!h]
\begin{center}
\includegraphics[scale=0.6]{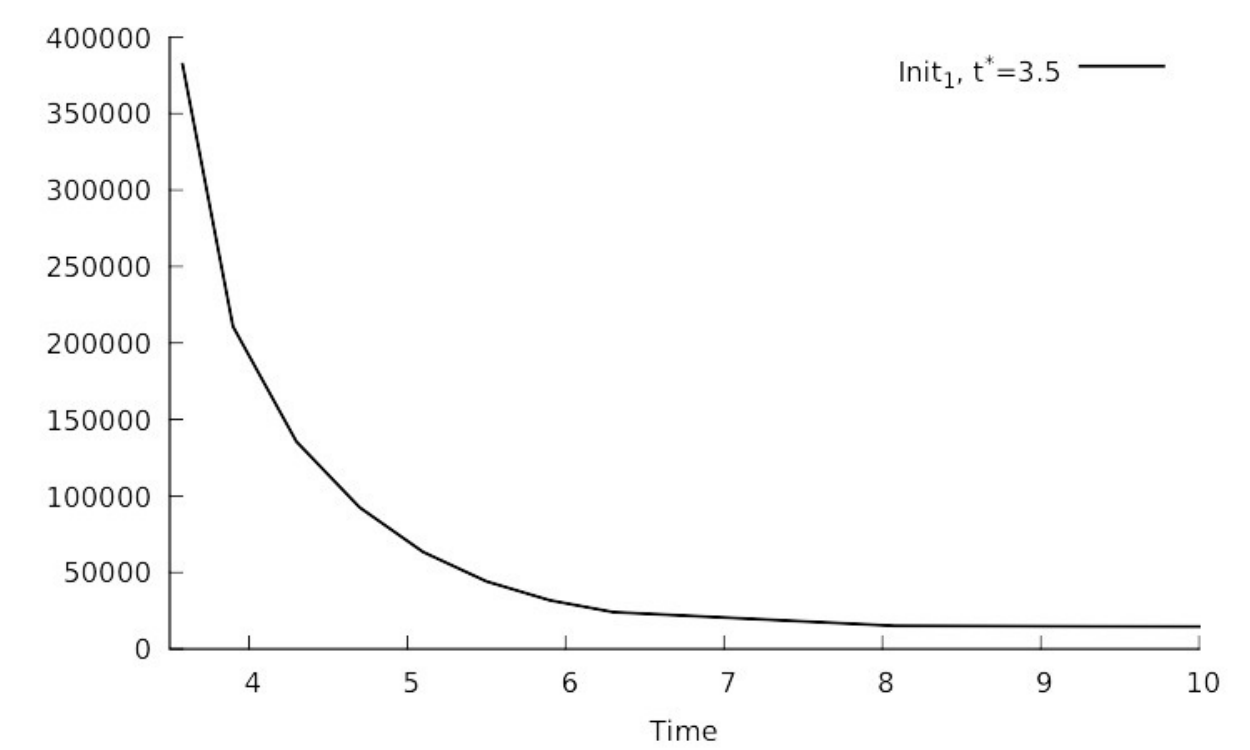}
\includegraphics[scale=0.6]{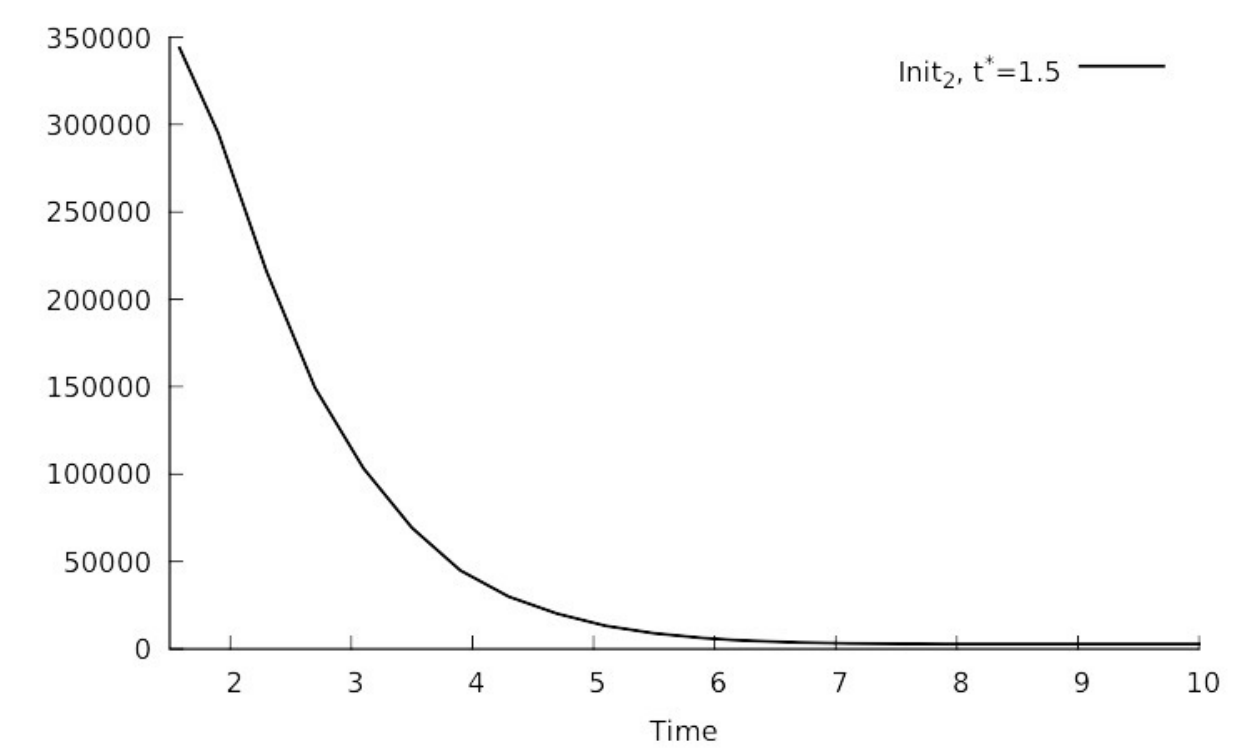}
\caption{\label{Norm2exp} Validation of asymptotic behaviour
  equation (\ref{asymptodeltadt}): $Norm$ with respect to time}
\end{center}
\end{figure}

For large time, the time derivative becomes smaller
and smaller, so the energy continues decreasing and there is no longer evolution even to $t=30$. $E_d $ defined by equation (\ref{Enerd}) is
decreasing. For example for $Init_2$ and $t^{\star}=1.5$ we have obtained:
\[
\begin{array}{|c|c||c|c||c|c||c|c||c|c|}
 \hline
\mbox{Time }t&E_d(t) &\mbox{Time }t&E_d(t)&\mbox{Time }t&E_d(t)&\mbox{Time }t&E_d(t)&\mbox{Time }t&E_d(t)\\
\hline
t^{\star}=1.5 &217.16 &3 &0.56 &4.8 & 1.\; 10^{-3}&11&  2.5\; 10^{-9} &19.5&10^{-16}\\
\hline
\end{array}
\]
The test of the asymptotic behaviour equation (\ref{vexpinfty2}) is presented
in figure (\ref{Norm2exp}) where we note that $Norm$ is rapidly decreasing until $t=10$.
We emphasize that the exponential growth of the coefficients makes
the computations very difficult, and this test shows the
robustness of our numerical method.

\subsection{Numerical results for $d\mathcal{S}_4(\mathbf K)$.}
We assume the Hubble constant is equal to $1$ hence the scale factor
is $a(t)=\cosh(t)$. First we test the accuracy of our scheme in the trivial case of
the constant solution $\Psi(t,X)=2$. Taking $\Psi(0,X)=2$,
$\Psi(\Delta t,X)=2$ with $\Delta t= 1.5\;10^{-4}$, our scheme gives a
constant solution
$\Psi(t,X)=2\pm 10^{-12}$ for $t\in[0,3.5]$ and the energy is around $10^{-15}$.

\begin{figure}[!h]
\begin{center}
\includegraphics[scale=0.6]{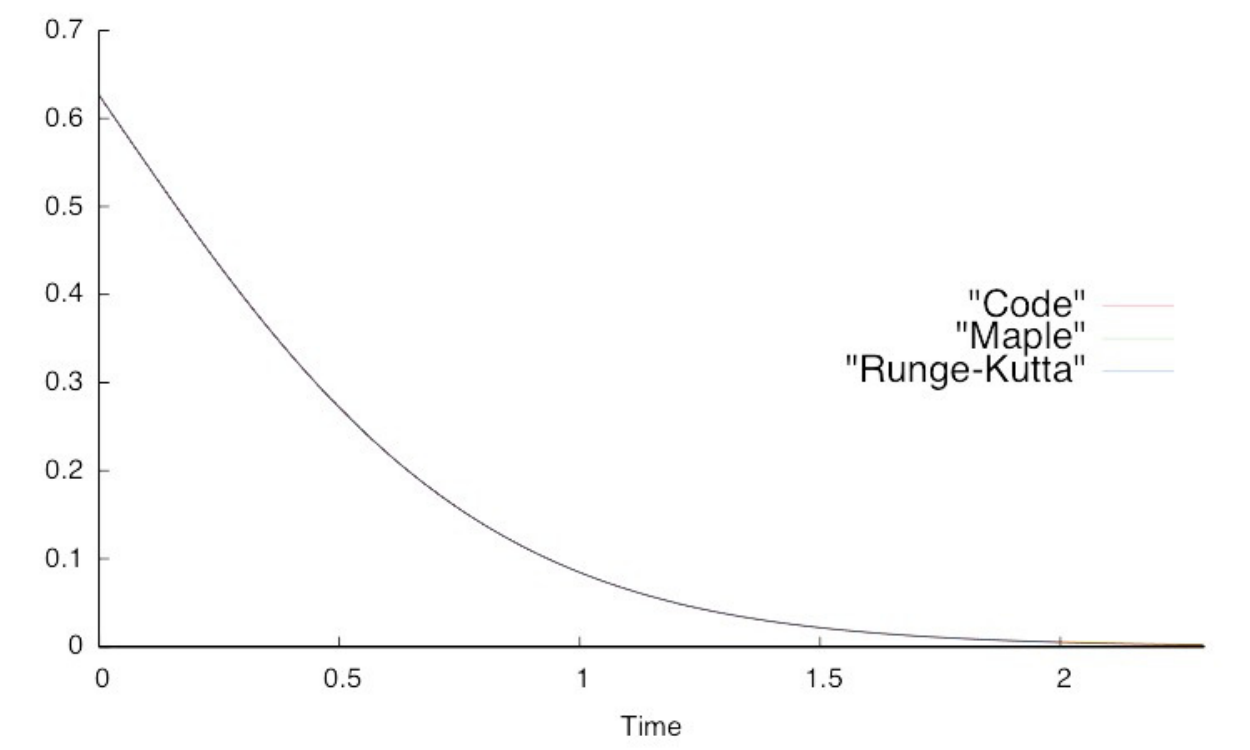}
\caption{\label{courbes-test} The black curve is the superposition of
  the three colored curves.}
\end{center}
\end{figure}

On the other hand we test our scheme by comparing it with two
other numerical methods for computing the solution
\begin{equation}
 \label{}
 \Psi(t,X)=A_0^-\left(\cosh
   t\right)^{-\frac{3}{2}}\mathrm{P}_{\frac{1}{2}}^{-\frac{3}{2}}\left(\tanh
   t\right).
\end{equation}
In figure (\ref{courbes-test}) we present the results computed by:

(i) our finite elements
scheme with initial data
\begin{equation}
 \label{datak}
 \Psi(0,X)=\frac{1}{2}\sqrt{\frac{\pi}{2}},\;\Psi(\Delta
t,X)=\Psi(0,X)-\sqrt{\frac{2}{\pi}}\Delta t,\;\;\Delta t=1.5\;10^{-4}.
\end{equation}

(ii) a Runge-Kunta scheme for the ordinary differential
equation (\ref{chq}) with the same data (\ref{datak}).

(iii) a symbolic calculation of $\left(\cosh
  t\right)^{-\frac{3}{2}}\mathrm{P}_{\frac{1}{2}}^{-\frac{3}{2}}\left(\tanh
  t\right)$ by using MAPLE.

The three curves perfectly matche and we
can see only a unique graph. In fact the numerical values are the same
at $\pm 5\;10^{-6}$.

\begin{figure}[!h]
\begin{minipage}{10cm}
\includegraphics[scale=0.15]{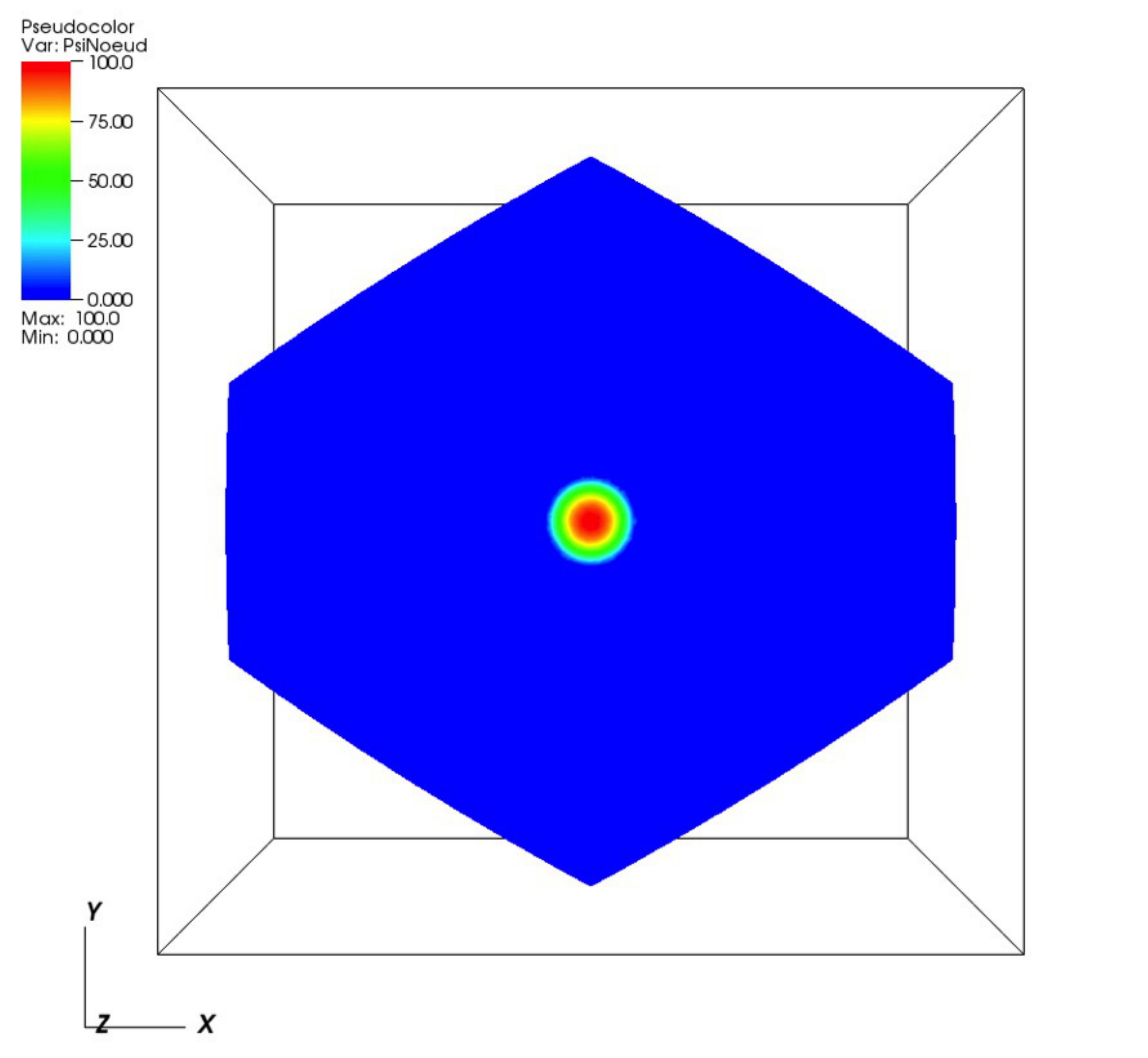}
\includegraphics[scale=0.15]{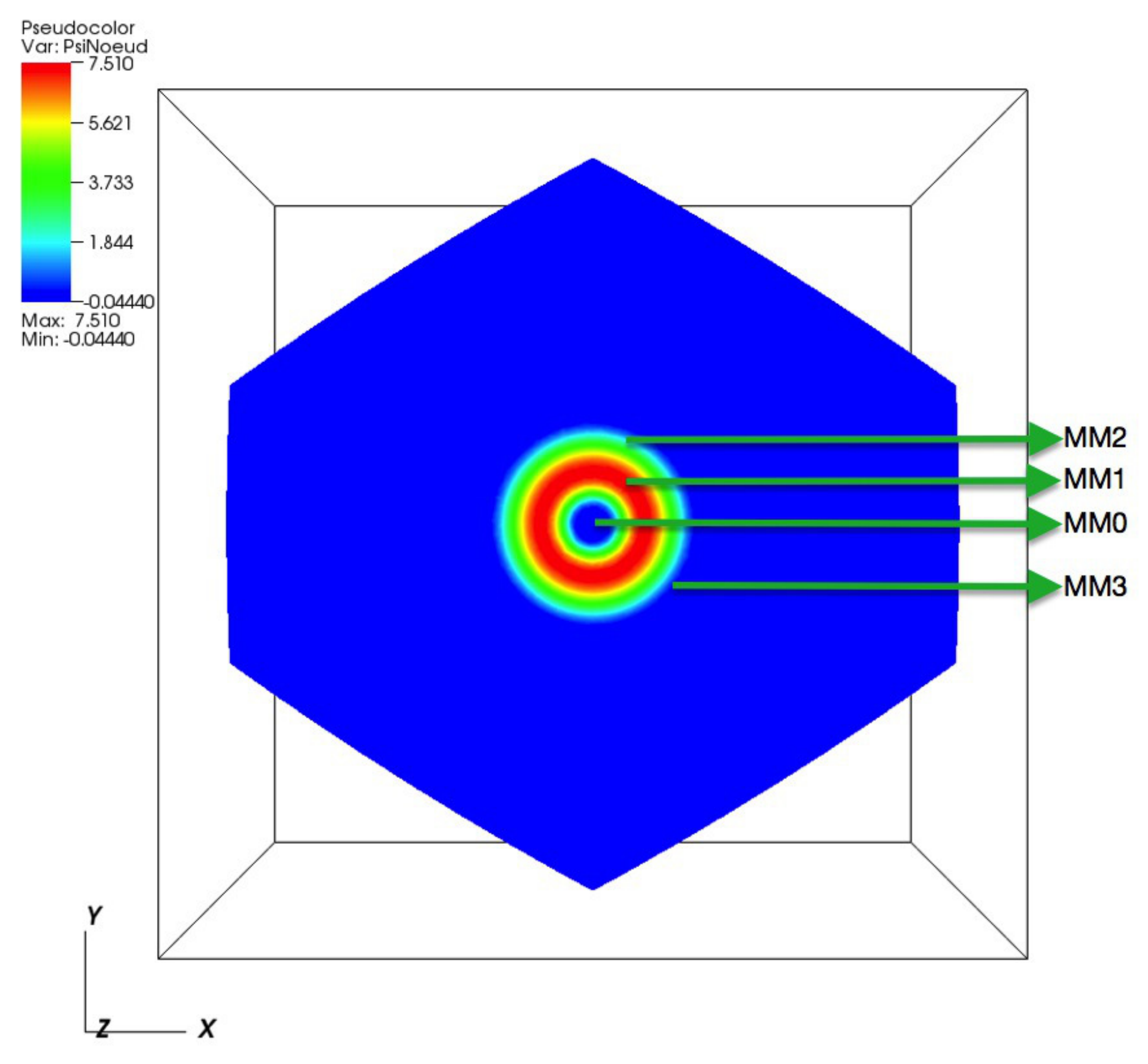}
\caption{\label{Init2ch2}Solution for $Init_2(R_0=0.05)$, on the left at $t^{\star}=3.5$, on the right at $t_{obs}=20.08$.}
\end{minipage}\hfill
\begin{minipage}{6cm}

There is a horizon and
$R_h=\sin\left(\arcsin
  0.05-2\arctan(e^{3.5})+\pi\right)\simeq 0.11$.  

\end{minipage}
\end{figure}

\begin{figure}[!h]
\includegraphics[scale=0.3]{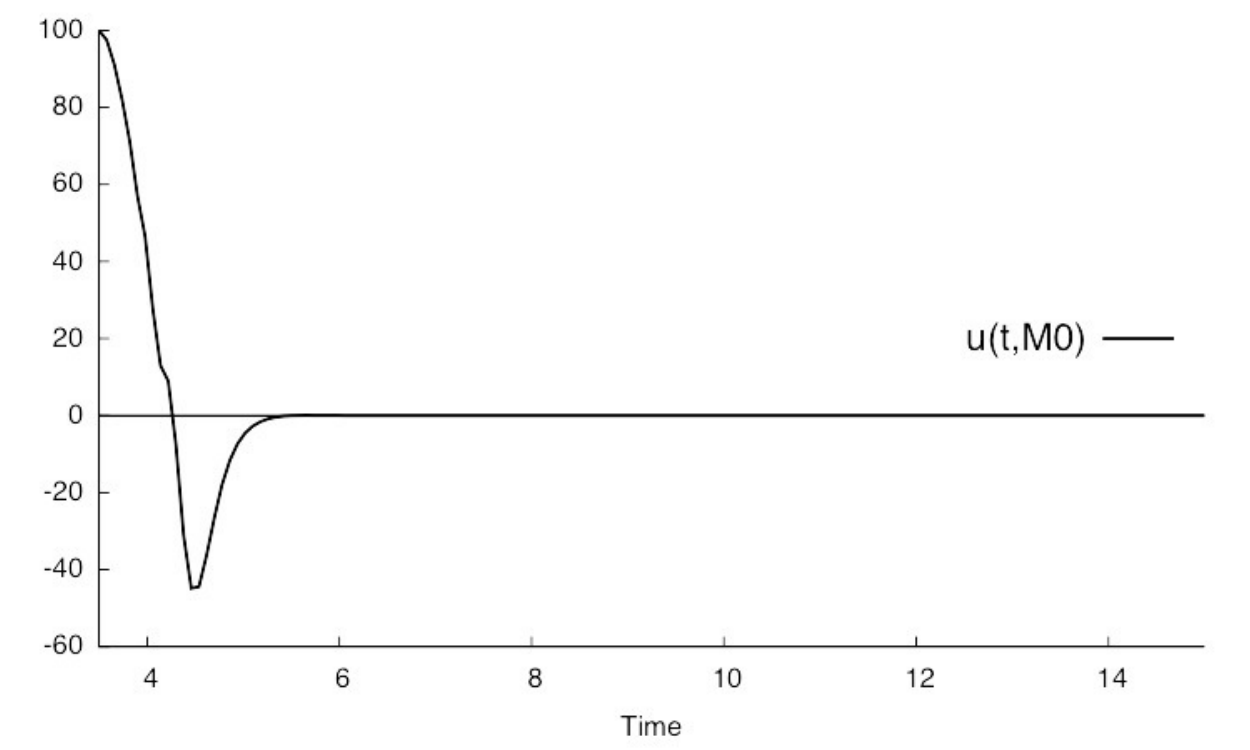}
\includegraphics[scale=0.3]{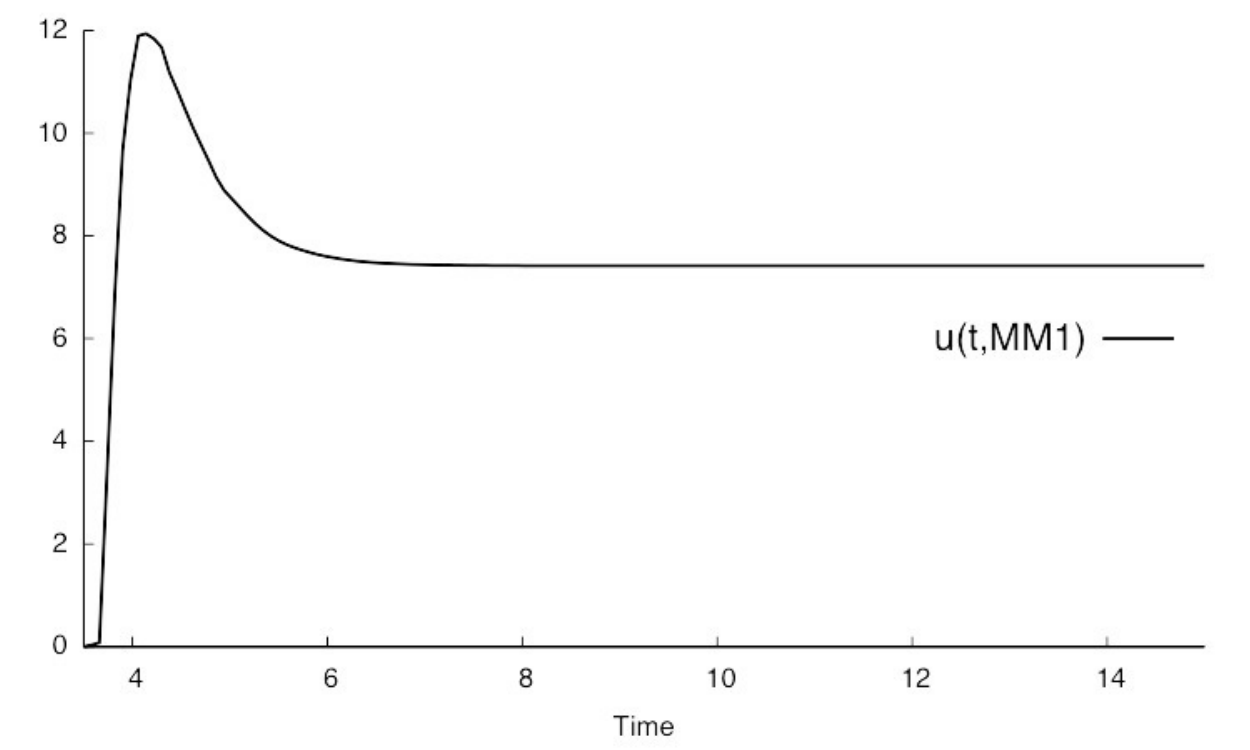}
\includegraphics[scale=0.3]{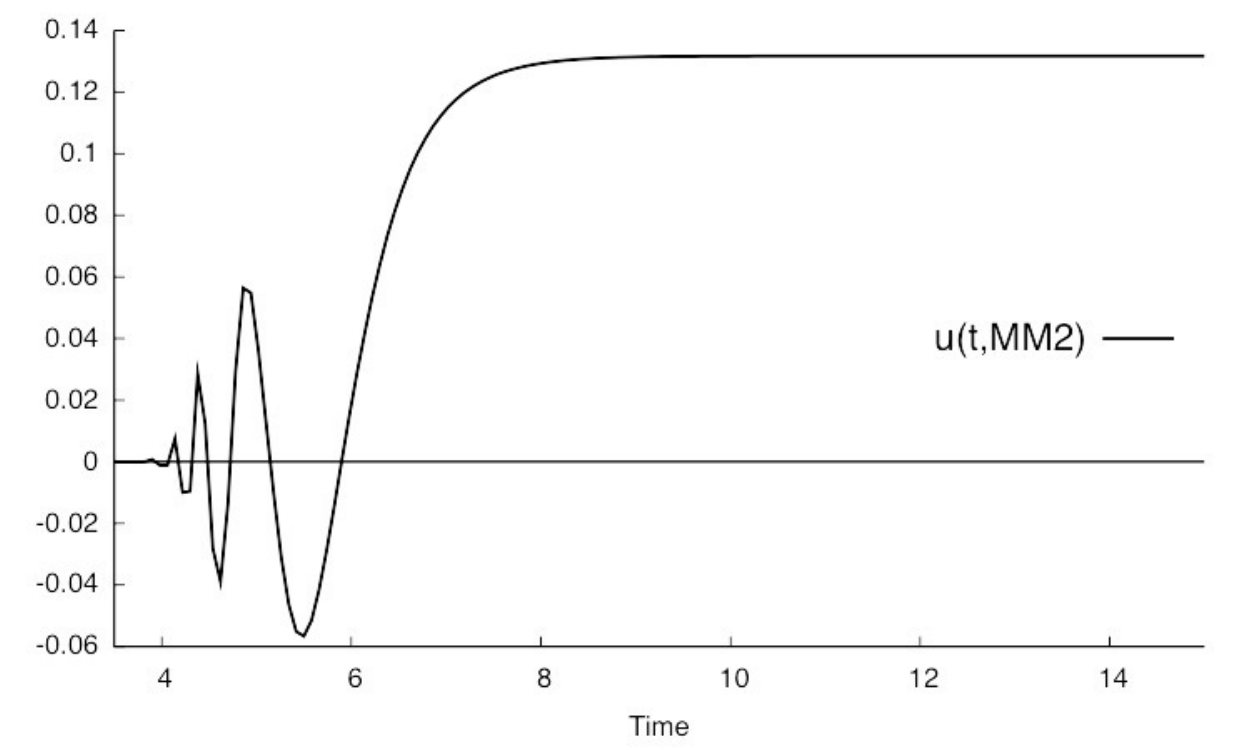}
\includegraphics[scale=0.3]{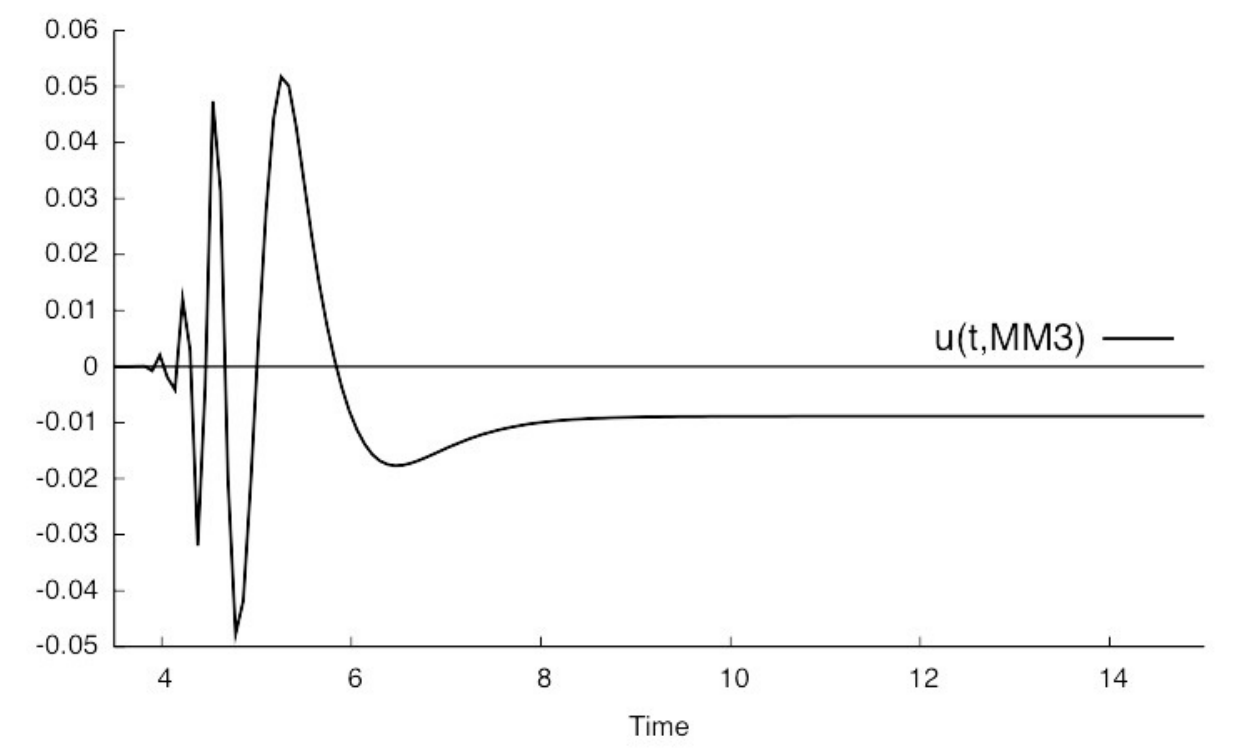}
\caption{\label{Init2ch2-M}}{Solution $u(t,X)$ with initial data
  $Init_2$ at $t^{\star}=3.5$.}
\end{figure}

In the sequel  we take initial data
$Init_2$ and $\partial_tu(t^{\star},.)=0$, for various $t^*$. This
is the initial data presented in the first picture of figure (\ref{Init2}).
First we check the accuracy of the numerical localization of a future
horizon. We choose $t^{\star}=3.5$.
Then $R_h\simeq 0.11$.  Results are presented in figure (\ref{Init2ch2}) and
figure (\ref{Init2ch2-M}). Figure (\ref{Init2ch2}) shows the initial data at time
$t^*=3.5$ and at time $t_{obs}=20.08$ for which the asympotic state is
reached. The pictures are on the slice $z=0$ of the
dodecahedron. There is a future horizon and the support of the field
is strictly included in ${\mathbf K}$. Figure (\ref{Init2ch2-M})
presents the time evolution of the solution
at some points $MM$ of $\mathcal{F}_v$ depicted in  figure
(\ref{Init2ch2}). In table (\ref{tab:Distch}) and table (\ref{tab:Distch2}) we can read the distance between previous
points $M(x_M,y_M,z_M)$ of $\mathcal{F}_v$ and its center, and all these
numerical results are in good agreement with the theoretical results
on the future horizon (\ref{rayonRh}) and the asymptotic profile.

As for $a(t)=e^t$ we see that the limit state is quickly reached: as soon
as $t=9$ in both cases. $E_d $ defined by equation (\ref{Enerd}) is decreasing as we can see for example for $Init_2$ and $t^{\star}=1.5$:
\[
\begin{array}{|c|c||c|c||c|c||c|c||c|c|}
 \hline
\mbox{Time }t&E_d(t) &\mbox{Time }t&E_d(t)&\mbox{Time }t&E_d(t)&\mbox{Time }t&E_d(t)&\mbox{Time }t&E_d(t)\\
\hline
t^{\star}=1.5 & 4361& 3.4& 0.49&5 & 1.2\; 10^{-3}&11.5&  10^{-9} &19.6&9. \; 10^{-17}\\
\hline
\end{array}
\]

\begin{table}[!h]
\begin{center}
\begin{minipage}{7cm}
\centering
    \begin{tabular}{|l|c|c|}
\hline
Point&\quad $d(0_{\RR^3},M)$ & $u(15,M)$\\
\hline
M0 & $\quad  0.00008 \quad $ & -0.0005\\
M1 & $\quad  0.270 \quad $ &    0.415\\
M2 & $\quad  0.182  \quad $& 0.004\\
M3 &  $\quad  0.354 \quad $  & 0.013  \\
\hline
\end{tabular}
\caption{\label{tab:Distch}$Init_2$ and $t^{\star}=1.5$, no horizon.}
\end{minipage}
\hfill
\begin{minipage}{7cm}
\centering
 \begin{tabular}{|l|c|c|}
\hline     
Point& $d(0_{\RR^3},M)$ &$u(15,M)$
\\
\hline
$M0$ & 0.00008 & -0.016\\ 
$MM1$ & 0.0544 &    7.41\\
$MM2$ & 0.1004&0.131\\
$MM3$ & 0.1055& -0.008\\
\hline
\end{tabular}
\caption{\label{tab:Distch2}$Init_2$ and $t^{\star}=3.5$, $R_h\simeq 0.11$}
\end{minipage}
\end{center}
\end{table}

\begin{figure}[!h]
\begin{center}
\includegraphics[scale=0.6]{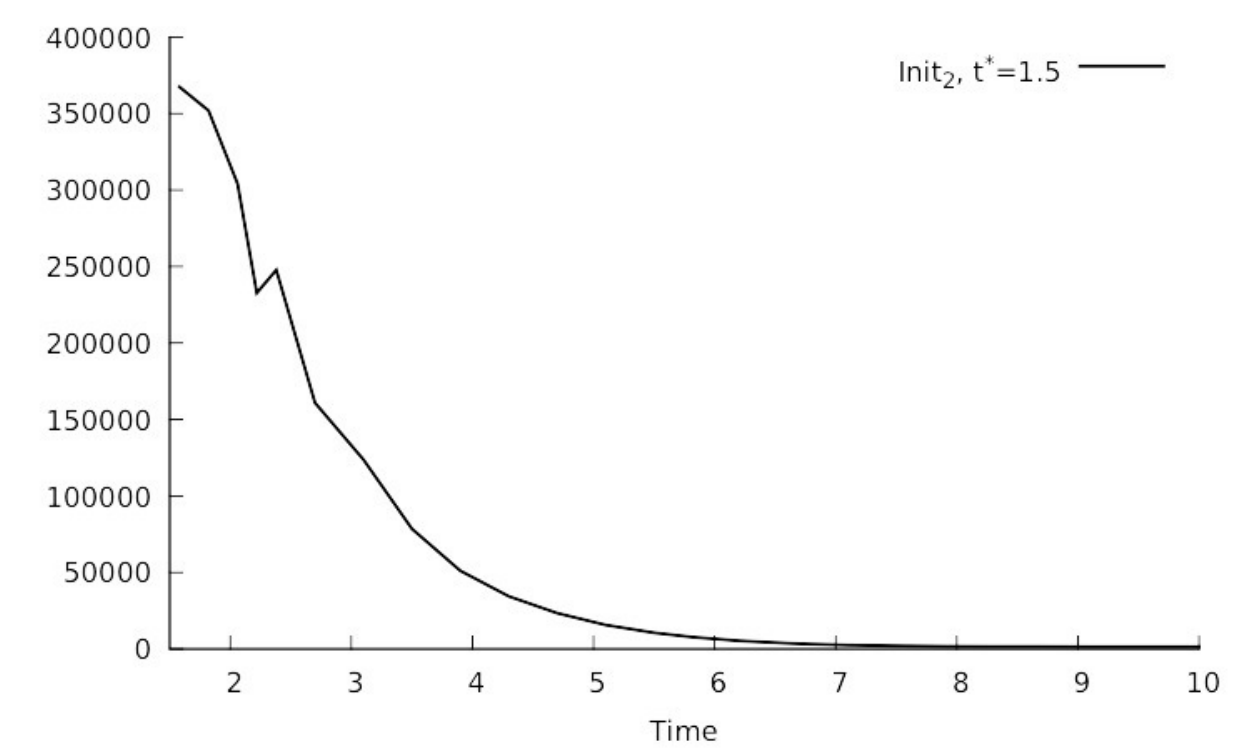}
\includegraphics[scale=0.6]{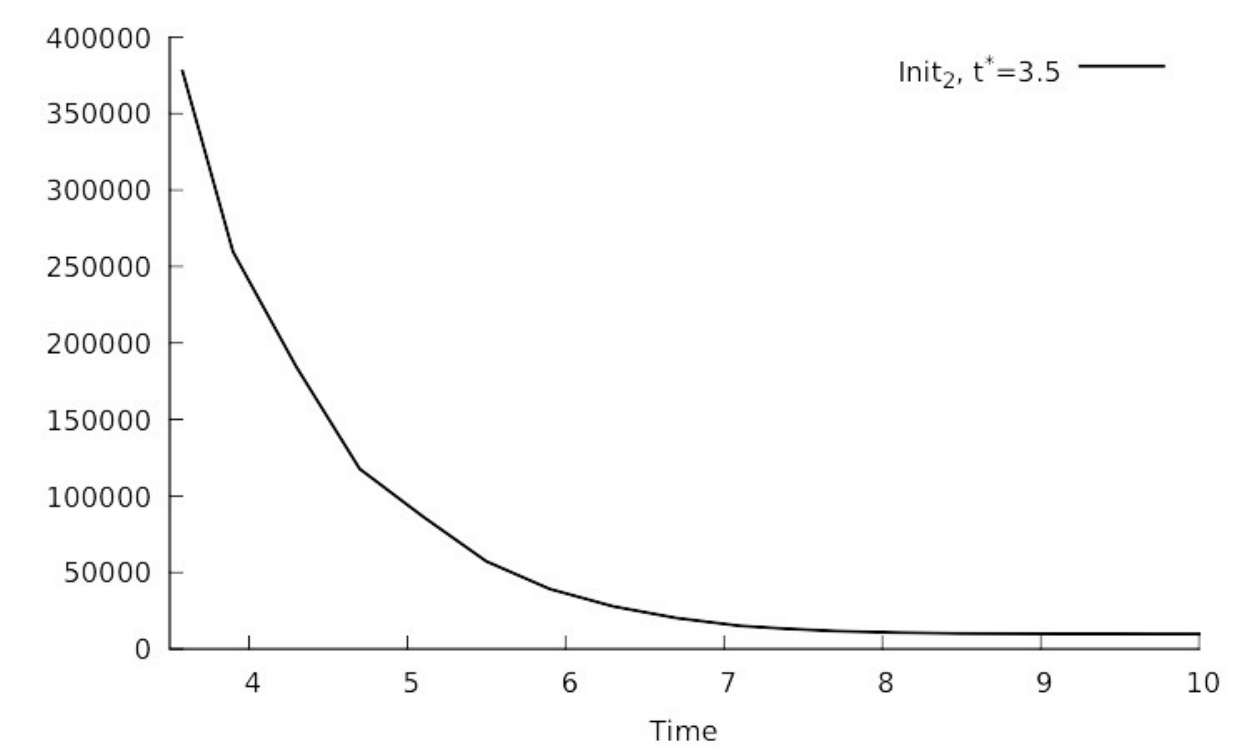}
\caption{\label{Norm2ch}Validation of equation (\ref{asymptodeltadt}) : $Norm$ with respect to time}
\end{center}
\end{figure}

Finally we test the robustness of our code by the hard computation of
equation (\ref{asymptodeltadt}). We consider the solution with initial data
given at $t^*=1.5$ and $t^*=3.5$. In all cases the quantity $Norm$ that tests equation (\ref{deltach}) 
quickly tends to zero. On figure (\ref{Norm2ch}) we remark that for $t^{\star}=1.5$ $Norm$
is decreasing except when the support of the solution reaches the
border of $\mathcal{F}_v$ at $t\simeq 2.3$ (we detail the peculiar
properties of this solution below). We conclude that as for
$a(t)=e^t$, the tests of equation (\ref{deltach}) by the computation of $Norm$ prove that our numerical scheme is accurate and robust.

 We end this part by the most interesting case: we present two solutions with ``the
 circles-in-the-sky''. Firstly we consider the previous solution with data
 $\Psi(t^*,.)$ described by 
$Init_2$ and $\partial_t\Psi(t^{\star},.)=0$ given at $t^{\star}=1.5$.
Then $R_h$
is equal to $\sin\left(\arcsin
  0.05-2\arctan(e^{1.5})+\pi\right)\sim 0.469$ which is greater than
$\sin(d_{max})\simeq 0.378$. Therefore the support of the solution is the whole
$\mathcal{F}_v$ for $t$ large enough and the constraint (\ref{cici}) is
fulfilled: we shall be able to see multiple copies of the asymptotic profile.

\begin{figure}[!h]
\includegraphics[scale=0.25]{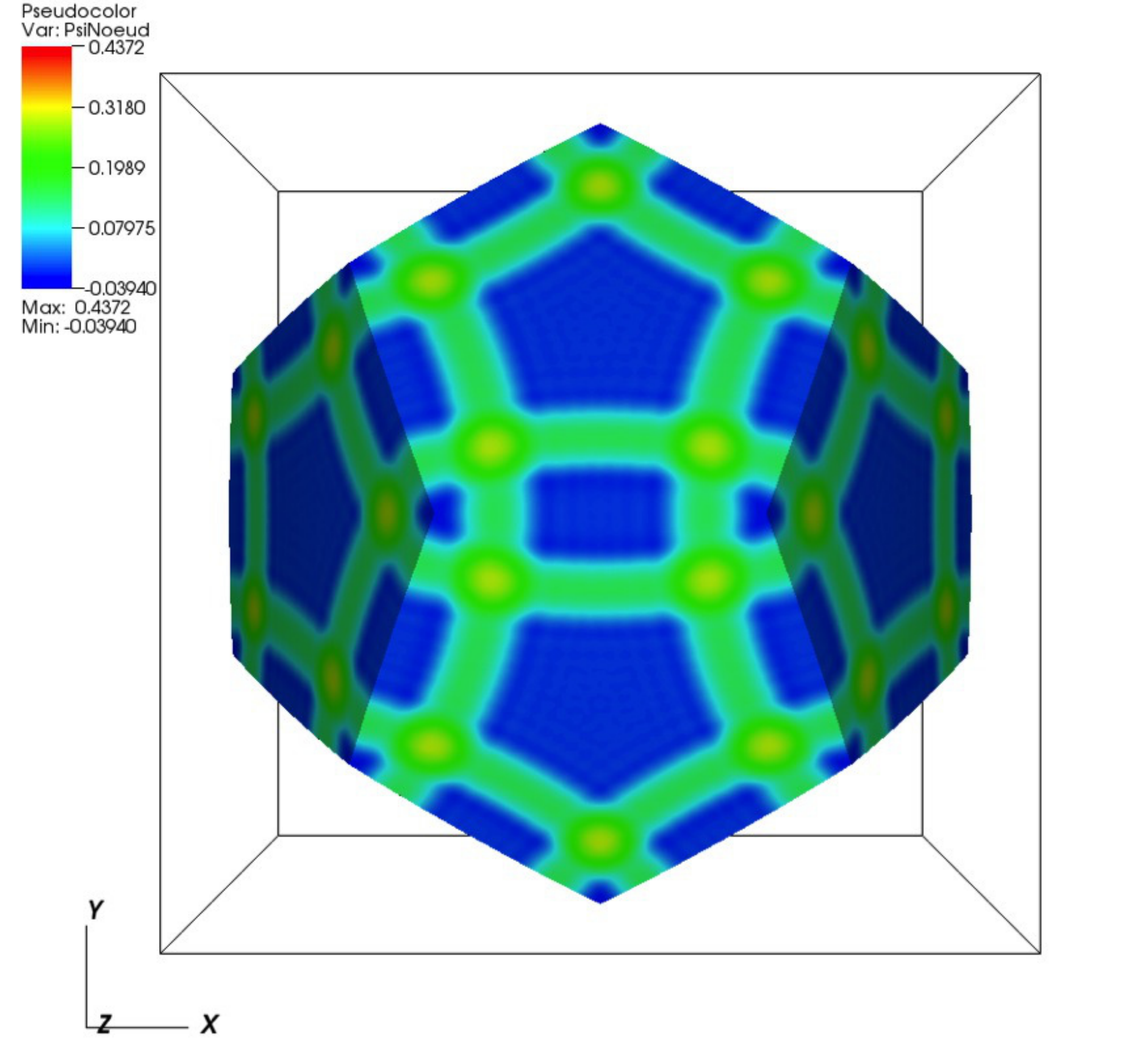}
\includegraphics[scale=0.25]{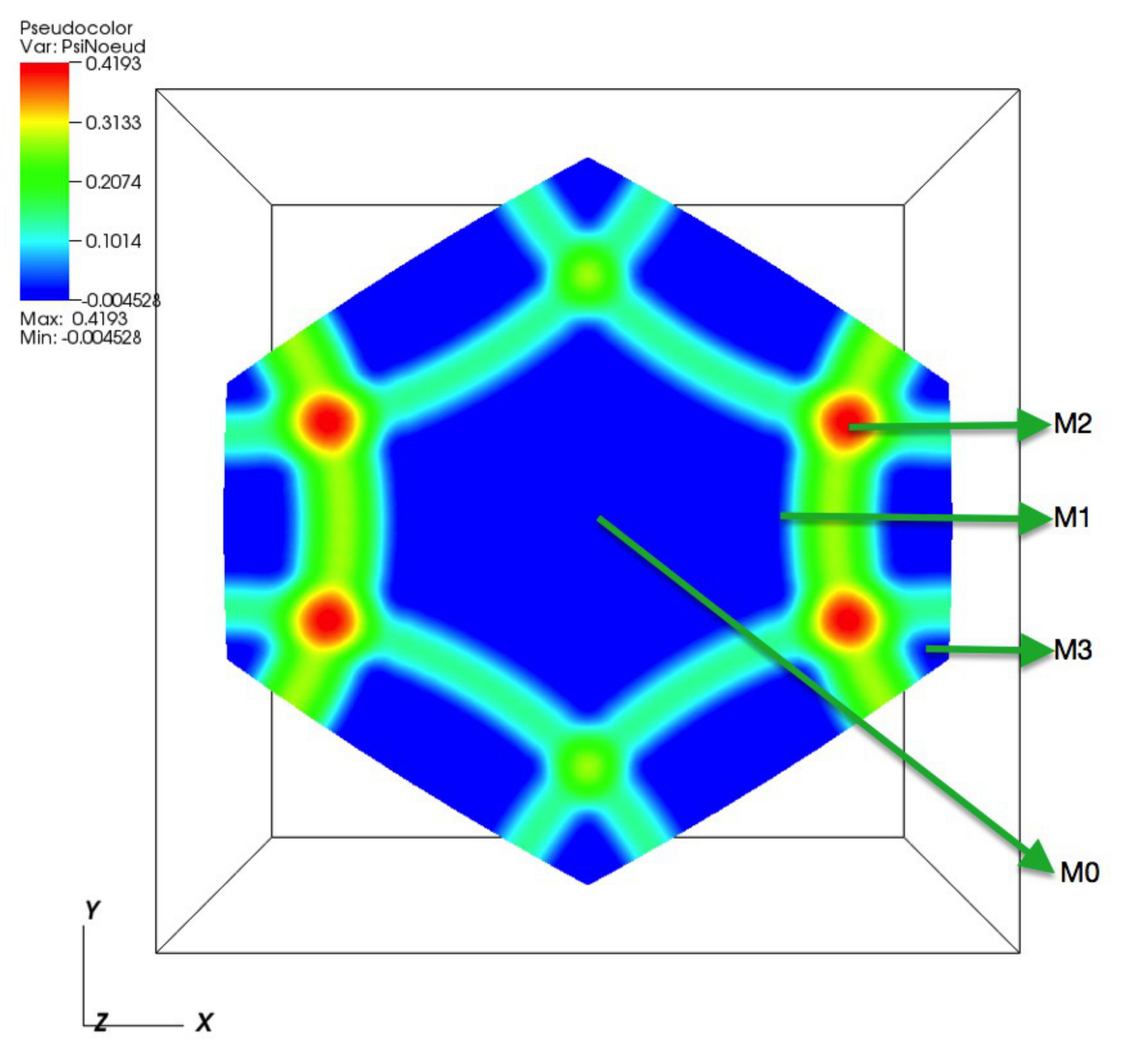}
\caption{\label{Init2ch}Solution observed at $t_{obs}=15.06$ for
  $Init_2$ and $t^{\star}=1.5$. Left: on the boundary of the
  dodecahedron. Right: on the slice at $z=0$.}
\end{figure}

\begin{figure}[!h]
\includegraphics[scale=0.3]{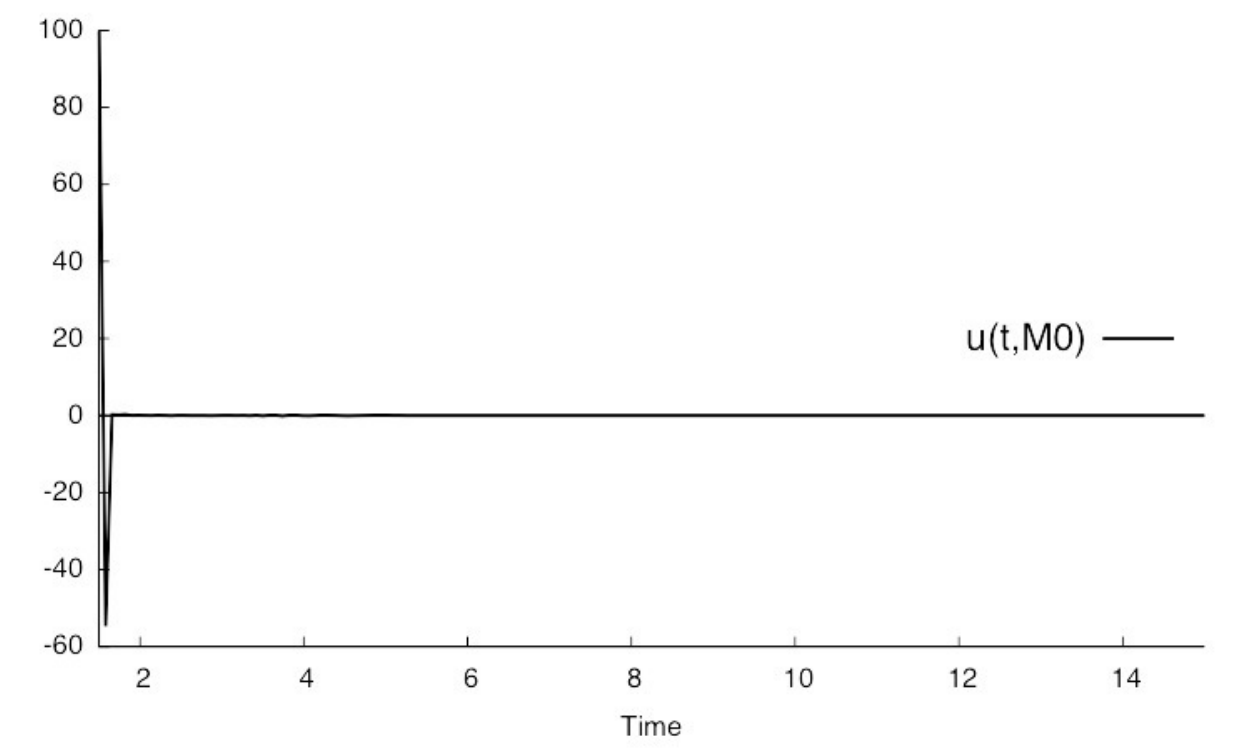}
\includegraphics[scale=0.3]{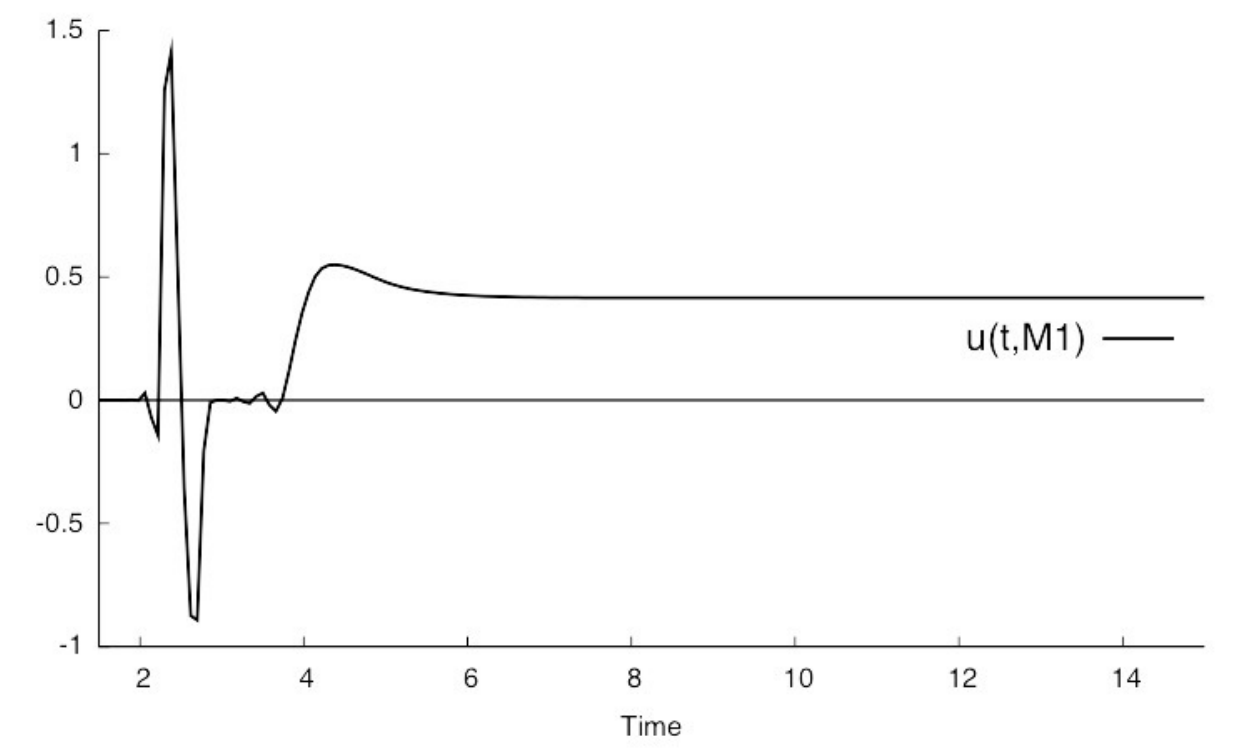}
\includegraphics[scale=0.3]{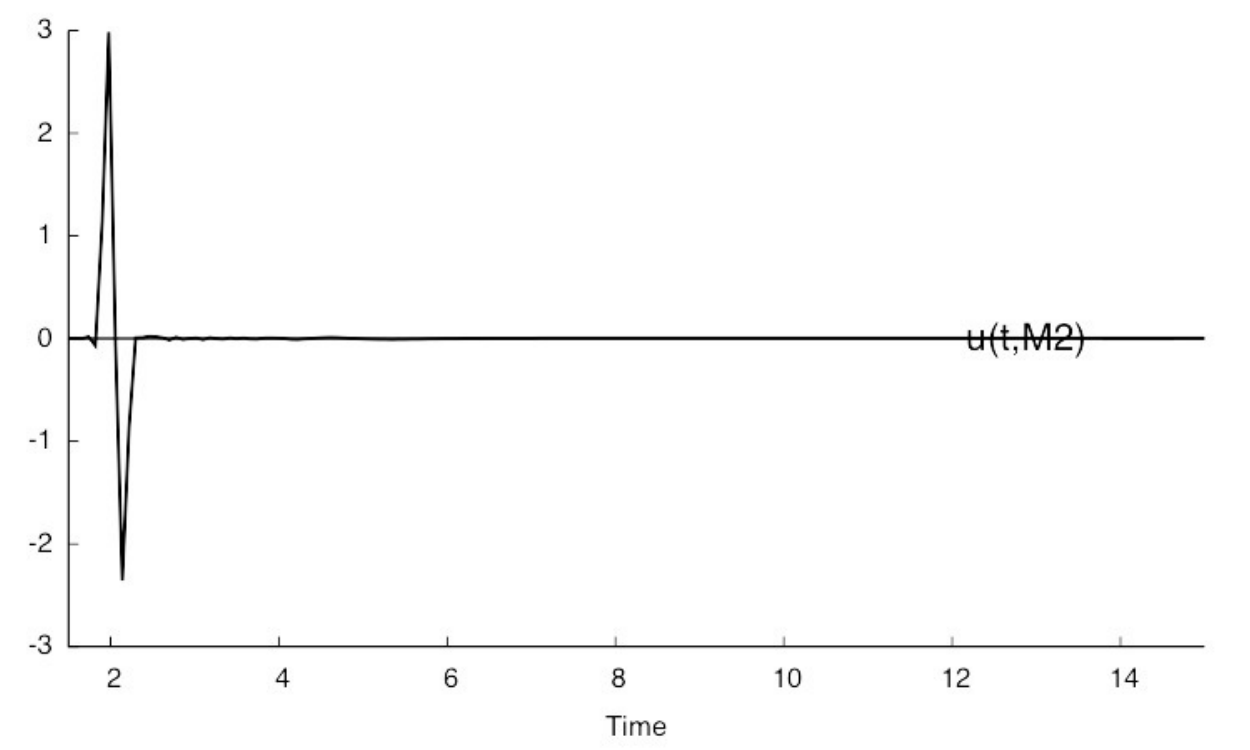}
\includegraphics[scale=0.3]{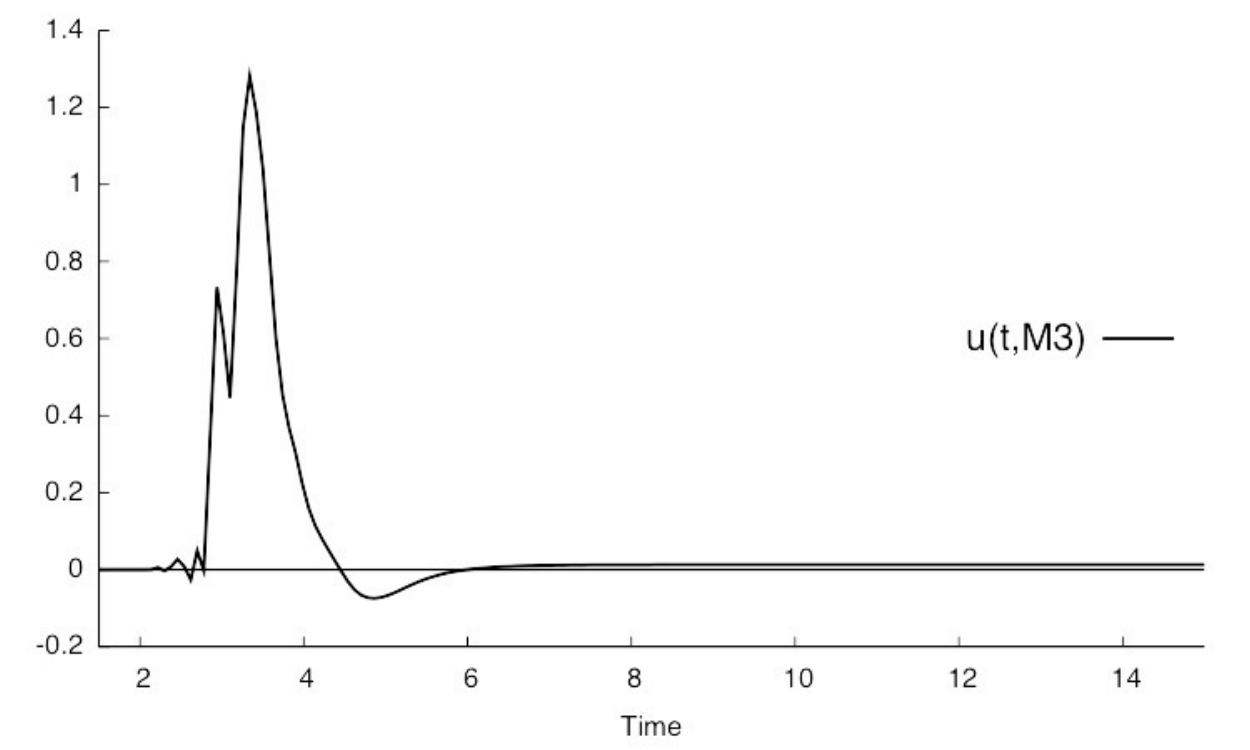}
\caption{\label{Init2ch-M}}{Solution $u(t,X)$ with initial data
  $Init_2$ at $t^{\star}=1.5$.}
\end{figure}

\begin{figure}[!h]
\begin{center}
\includegraphics[scale=0.25]{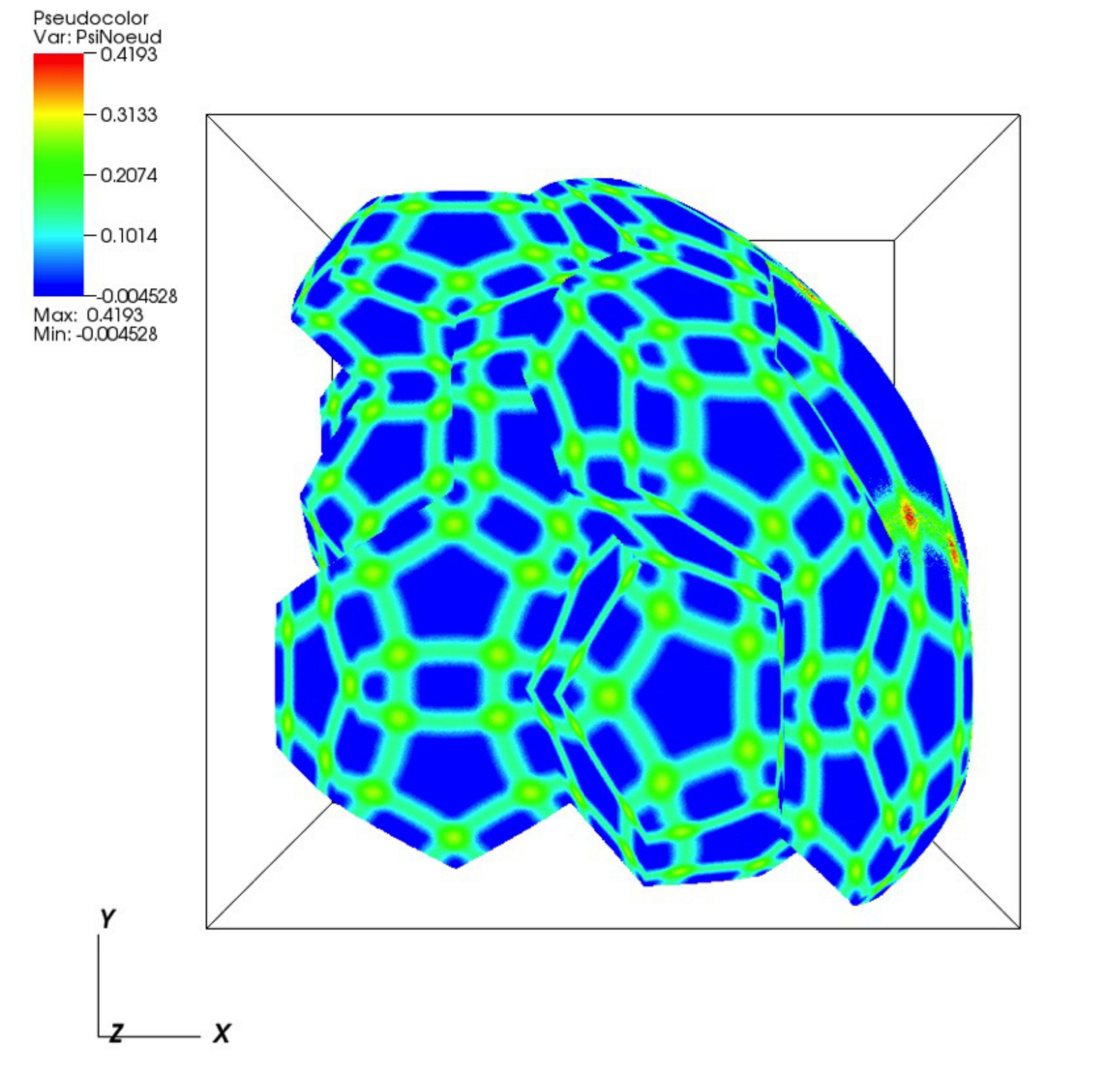}
\caption{\label{solutionpavage}}{Pull-back of the solution at $t_{obs}=15.06$ in a part of the universal
  covering $\mathcal{S}^3$ of $\mathbf{K}$.}
\end{center}
\end{figure}

Figure (\ref{Init2ch}) shows the  solution at time $t_{obs}=15.06$ on the boundary of the dodecahedron,
$\partial\mathcal{F}_v$, and in its interior, on the
slice $z=0$. In fact, at this time the asymptotic state is reached. We
can see this fast convergence on figure (\ref{Init2ch-M}) that shows the time evolution of the solution
at some points $M$ of $\mathcal{F}_v$ depicted in
figure (\ref{Init2ch}). Since the whole universe is included inside the
horizon sphere, it is interesting to present the pull-back of the
solution in the universal covering ${\mathcal S}^3$ of $\mathbf K$.
Figure (\ref{solutionpavage}) shows the solution on a partial tiling of
the unit ball $\mathcal{B}'$ (see figure (\ref{DODIM}) and the explanations after formula
equation (\ref{coordsphere}).
\begin{figure}[!h]
\begin{center}
\includegraphics[scale=0.32]{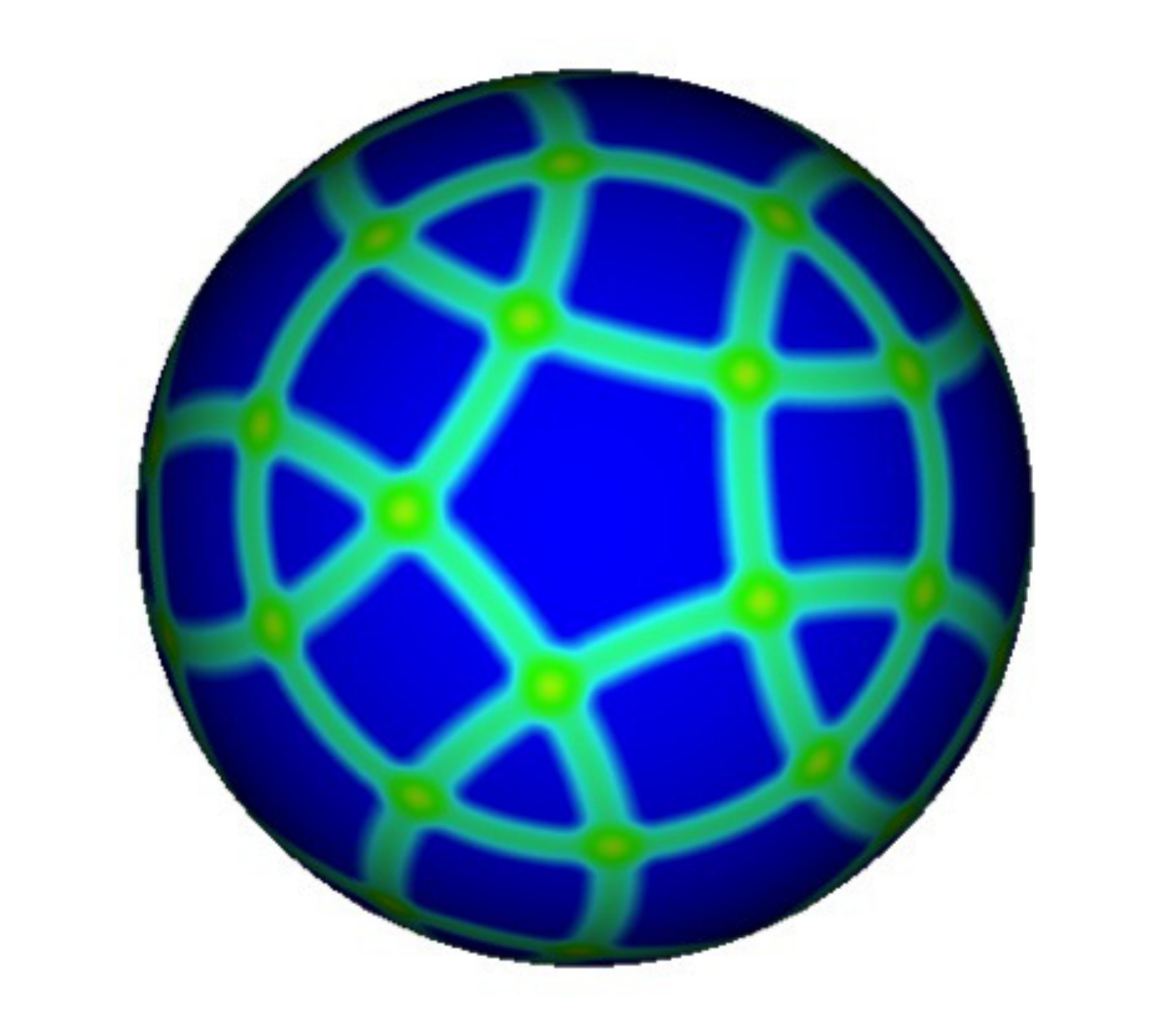}
\includegraphics[scale=0.27]{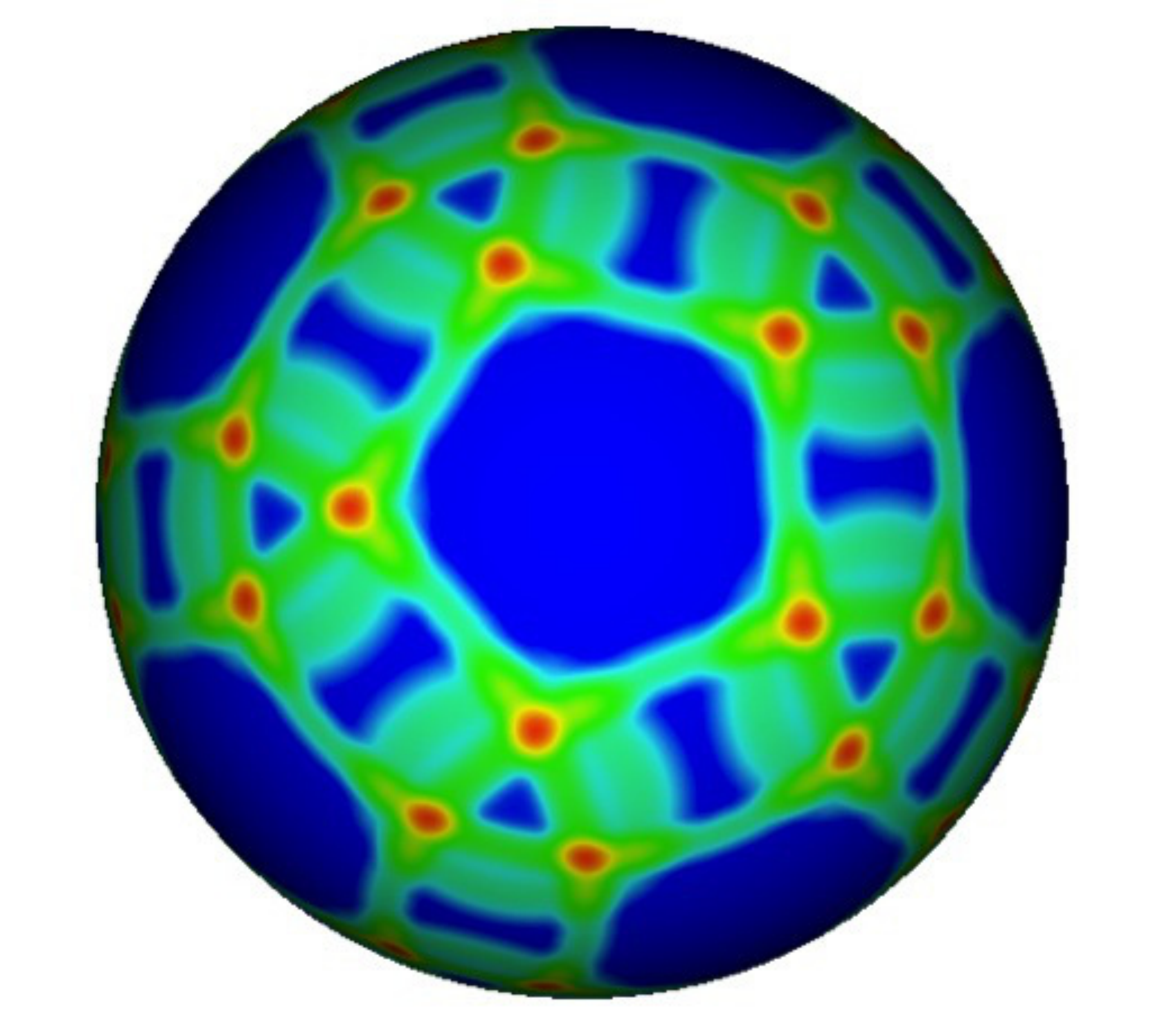}
\caption{\label{skyDOD}}{Deep skies: on
  the left $\chi=\arcsin(0.5)\simeq0.523$, on the right $\chi=\arcsin(0.6)\simeq0.643$.}
\end{center}
\end{figure}


We now show the application of our numerical method to the determination
of the temperature fluctuation in the deep sky. We consider an
observer located at time $t_0$ at the centre of the Poincar\'e dodecahedron,
$x=y=z=0$, that corresponds to the north pole of $\mathcal{S}^3$,
defined by $(x^0,x^1,x^2,x^3)=(1,0,0,0)$, or in spherical coordinates,
$\chi=0$. We assume the initial pertubation of the metric is given by
$Init_2$ at $t^*=1.5$ and $t_0$ is large enough for the asymptotic
state to be reached.  Then according to the Sachs-Wolfe formula equation (\ref{swapprox}),
the temperature fluctuation seen by this observer in the direction $(\theta,\varphi)\in\mathcal{S}^2$ is given by
$$
\frac{\delta T(\theta,\varphi)}{T}\simeq \frac{1}{3}\Psi_{\infty}(\chi,\theta,\varphi),
$$
where the angle $\chi$ is determined by $\chi=\eta_0-\eta_{ls}$ that is the conformal time between the time of
recombination $t_{ls}$ and $t_0$ ($t^*<t_{ls}<t_0$). Its value is
given by equation (\ref{horch}),
$$
\chi=2\arctan\left(e^{t_0}\right)-2\arctan\left(e^{t_{ls}}\right),
$$
and the sky seen by the observer corresponds to the horizon sphere centered at
the origin, included in $\mathcal{B}'$, with the radius
$R_{t_{ls},t_0}=\sin\chi$. Figure (\ref{skyDOD}) shows the deep sky for
$R_{t_{ls},t_0}=0.5$ and $R_{t_{ls},t_0}=0.6$ (we recall that by equation (\ref{rmax}) the
radius of the
smallest ball containing $\mathbf K$ is $R_{max}\simeq 0.378$). We
remark that the initial perturbation of the metric  leads to a
temperature fluctuation that shows spherical correlations along six
pairs of antipodal matched circles (the famous
{\it circles-in-the-sky}) that are a signature of a complex topology
with a positive curvature. In contrast, a complex topology and a
negative curvature lead to chaotic temperature fluctuations \cite{doug}.

\begin{figure}[!h]
\begin{center}
\includegraphics[scale=0.29]{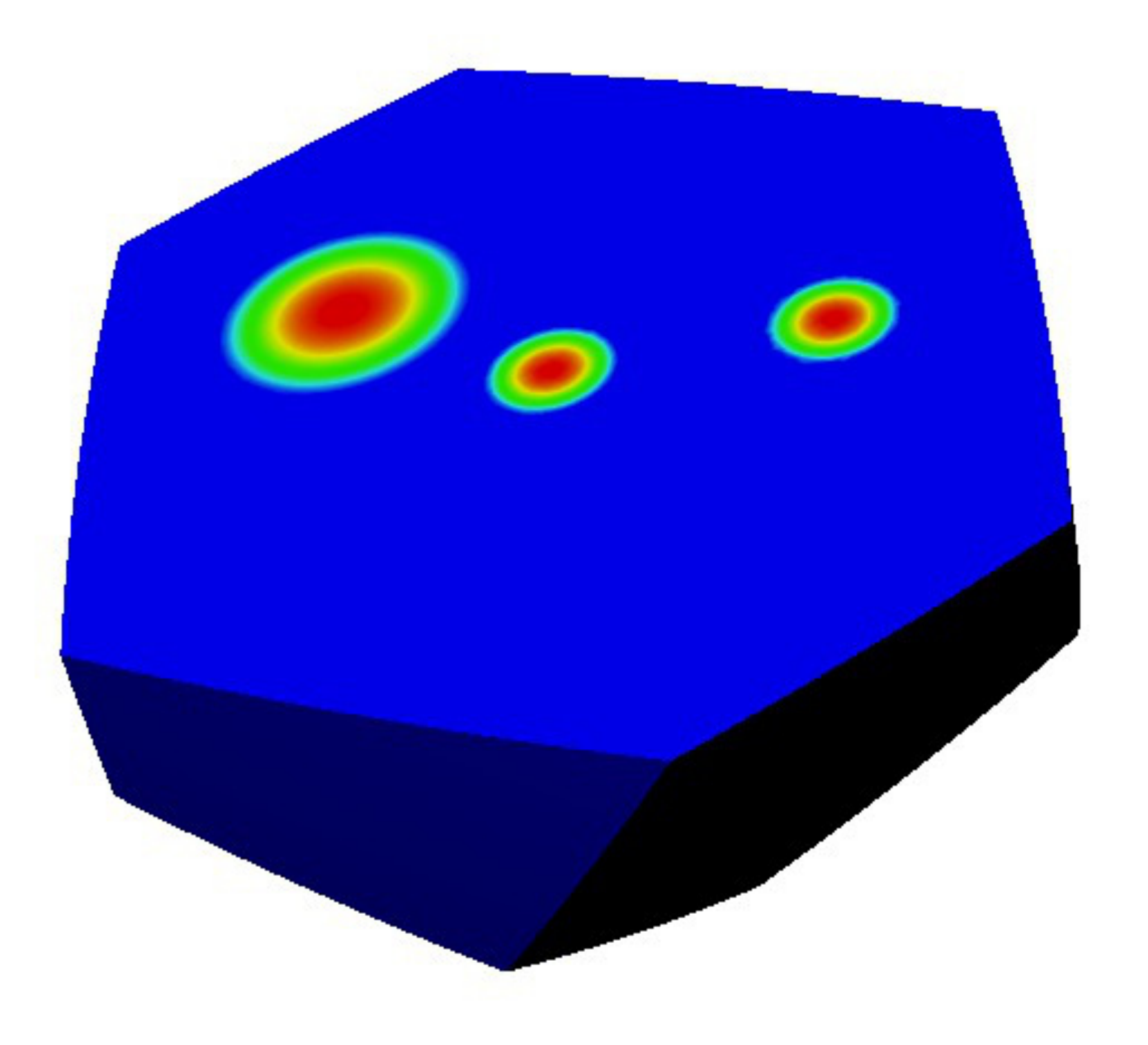}
\includegraphics[scale=0.29]{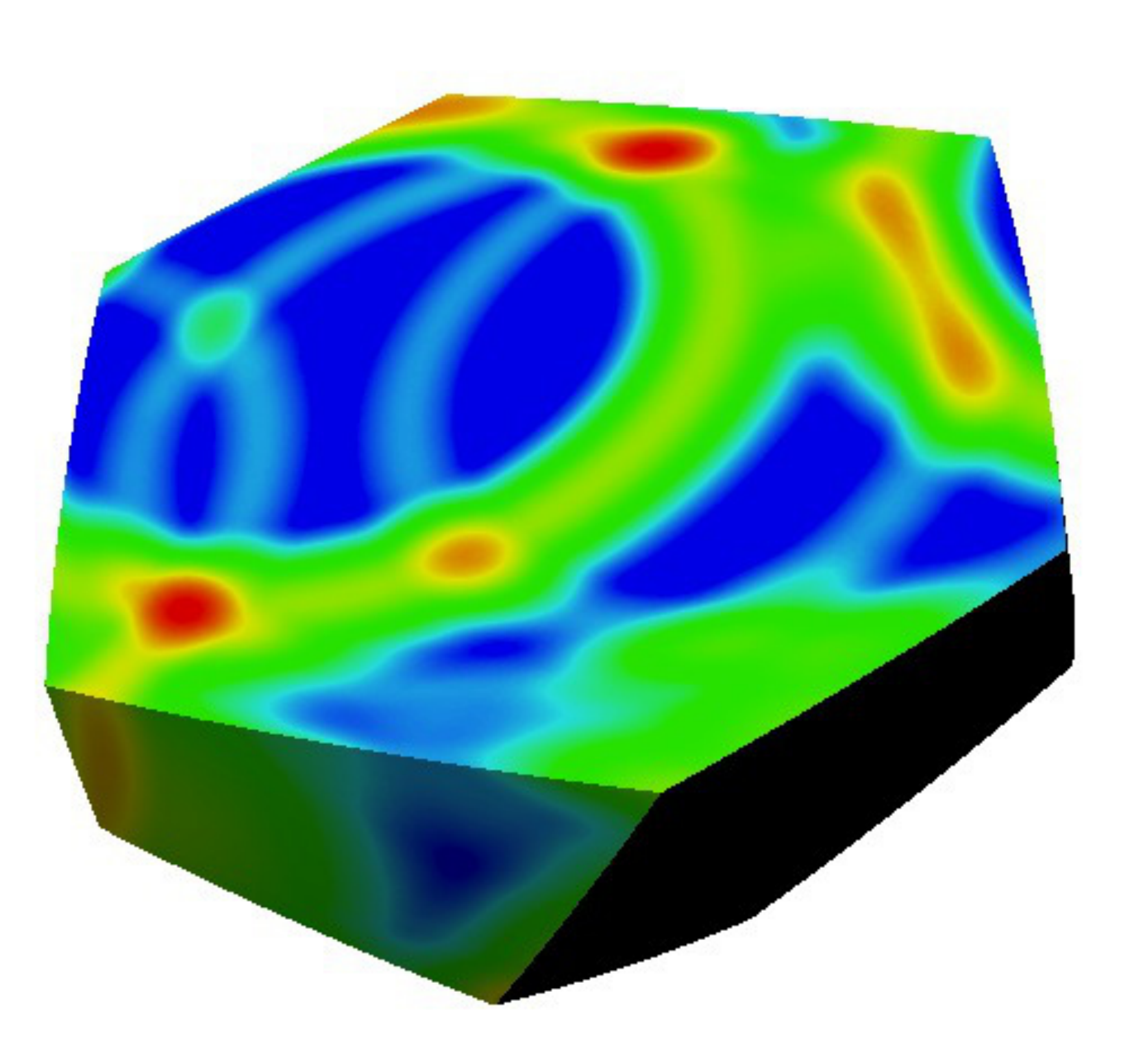}
\caption{\label{InitskyDOD2}}{Solution for $Init_3$ on the cut
  $10x+5y-15z=0$: on the left at $t^*=2$, on the right,
  the asymptotic profile.}
\end{center}
\end{figure}

\begin{figure}[!h]
\begin{center}
\includegraphics[scale=0.23]{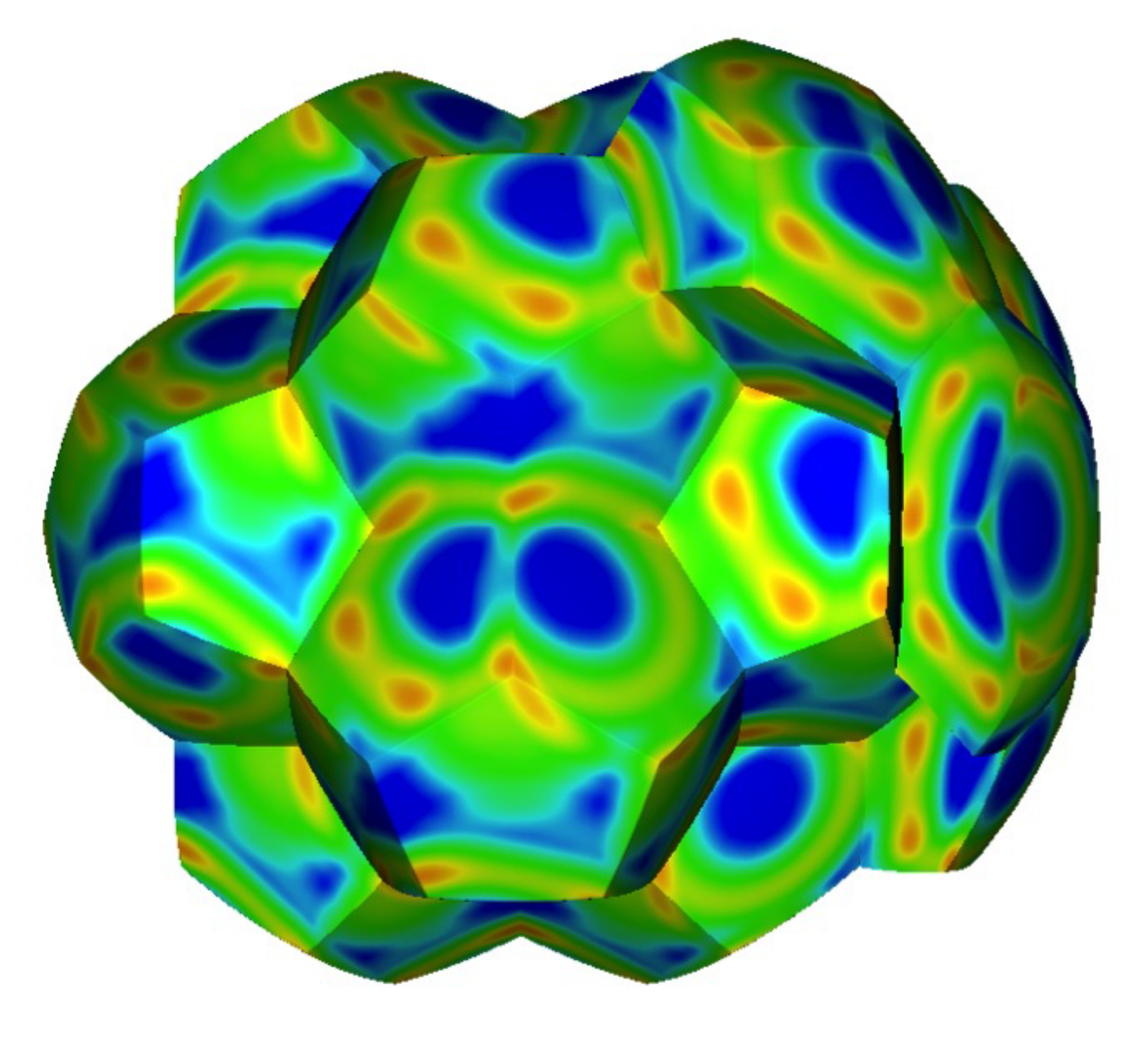}
\includegraphics[scale=0.28]{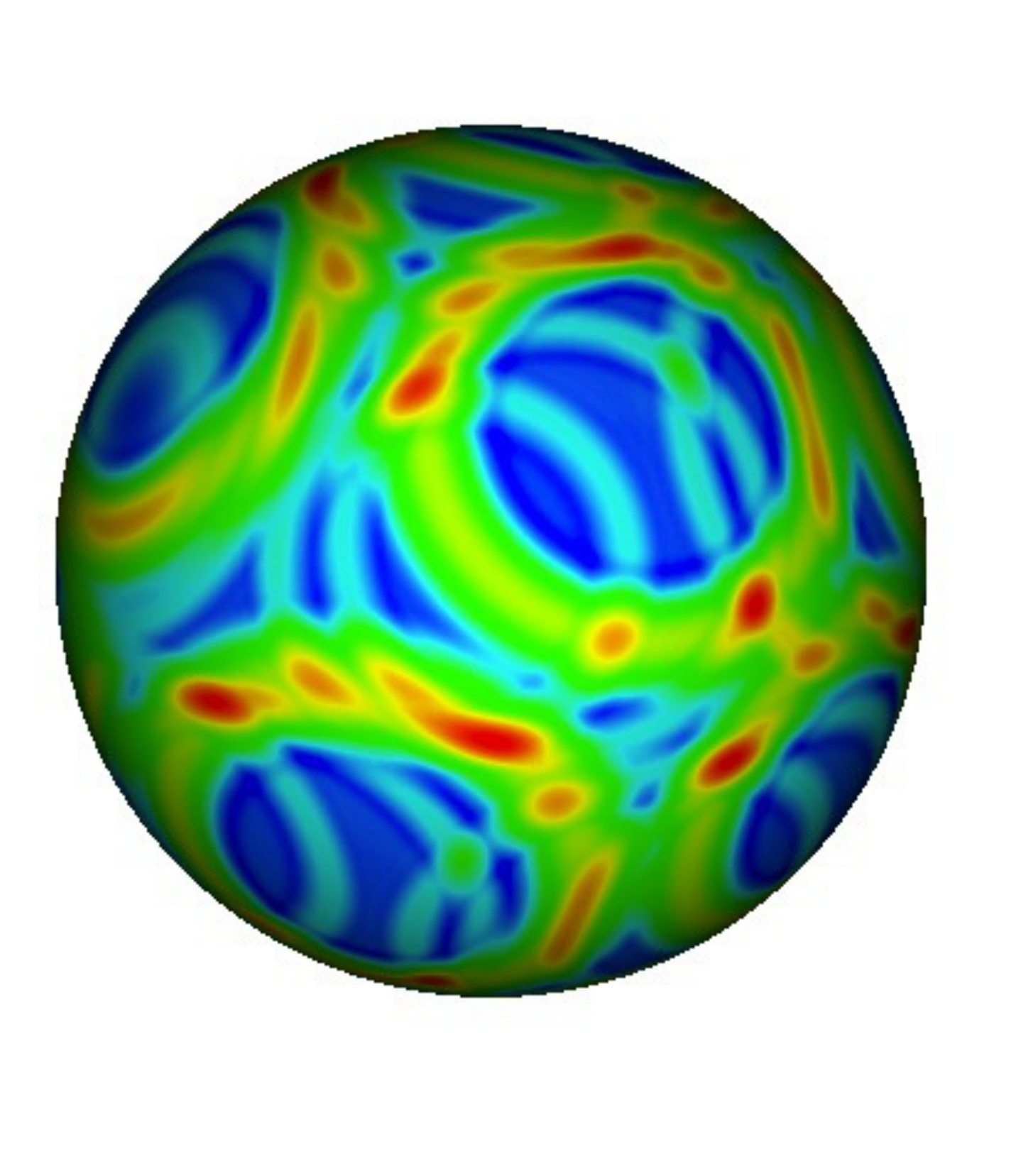}
\caption{\label{skyDOD2}}{On the left the asymptotic state in a part of the universal
  covering $\mathcal{S}^3$ of $\mathbf{K}$ around $(1,0,0,0)$. On
  the right the deep sky for $\chi=\arcsin(0.55)\simeq 0.582$.}
\end{center}
\end{figure}
\begin{figure}[!h]
\begin{center}
\includegraphics[scale=0.28]{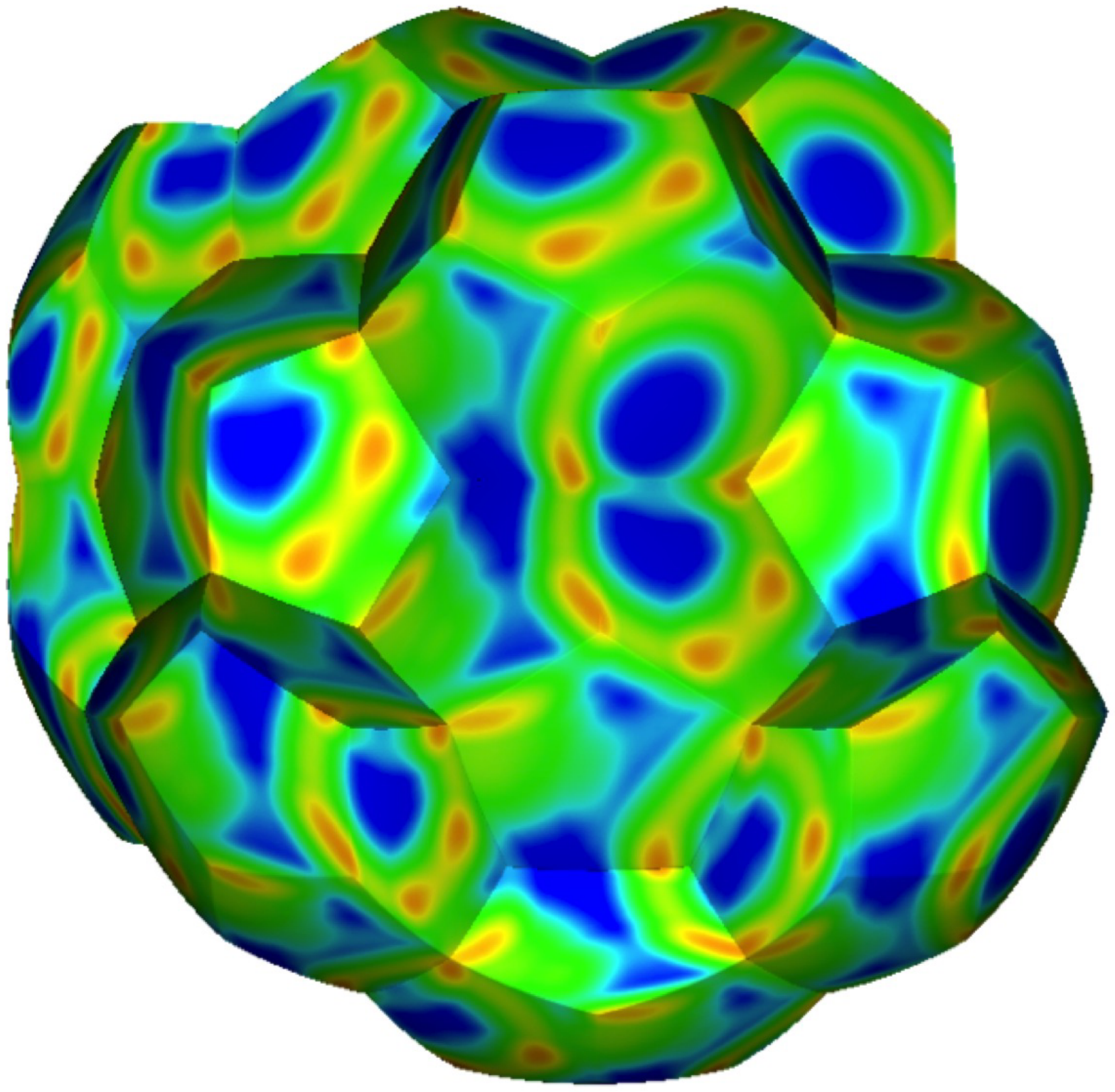}
\includegraphics[scale=0.28]{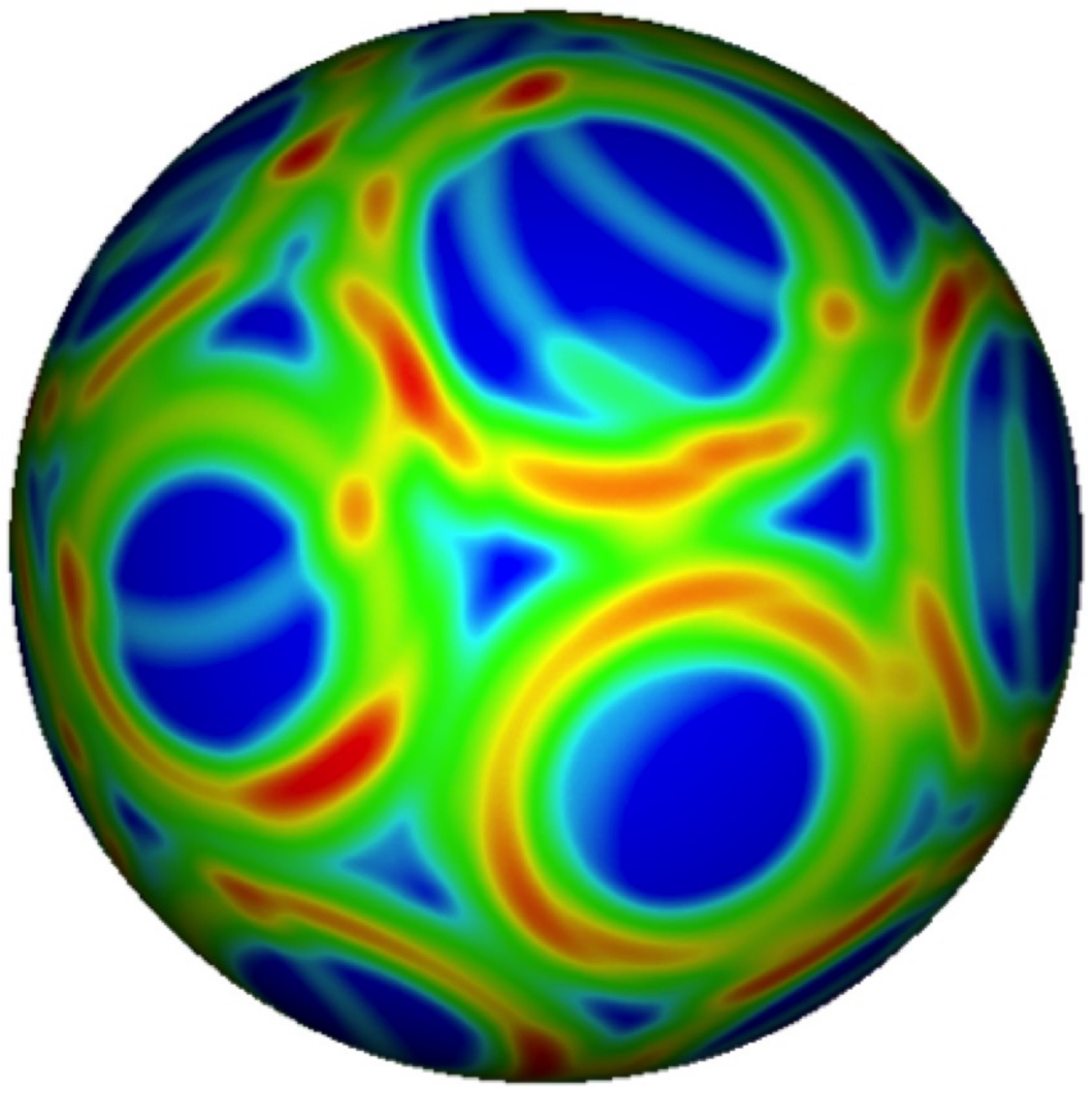}
\caption{\label{skyDOD3}}{On the left the asymptotic state in a part of the universal
  covering $\mathcal{S}^3$ of $\mathbf{K}$. On
  the right the deep sky  for $\chi=\pi-\arcsin(0.6)\simeq 2.498$.}
\end{center}
\end{figure}
\begin{figure}[!h]
\begin{center}
\includegraphics[scale=0.25]{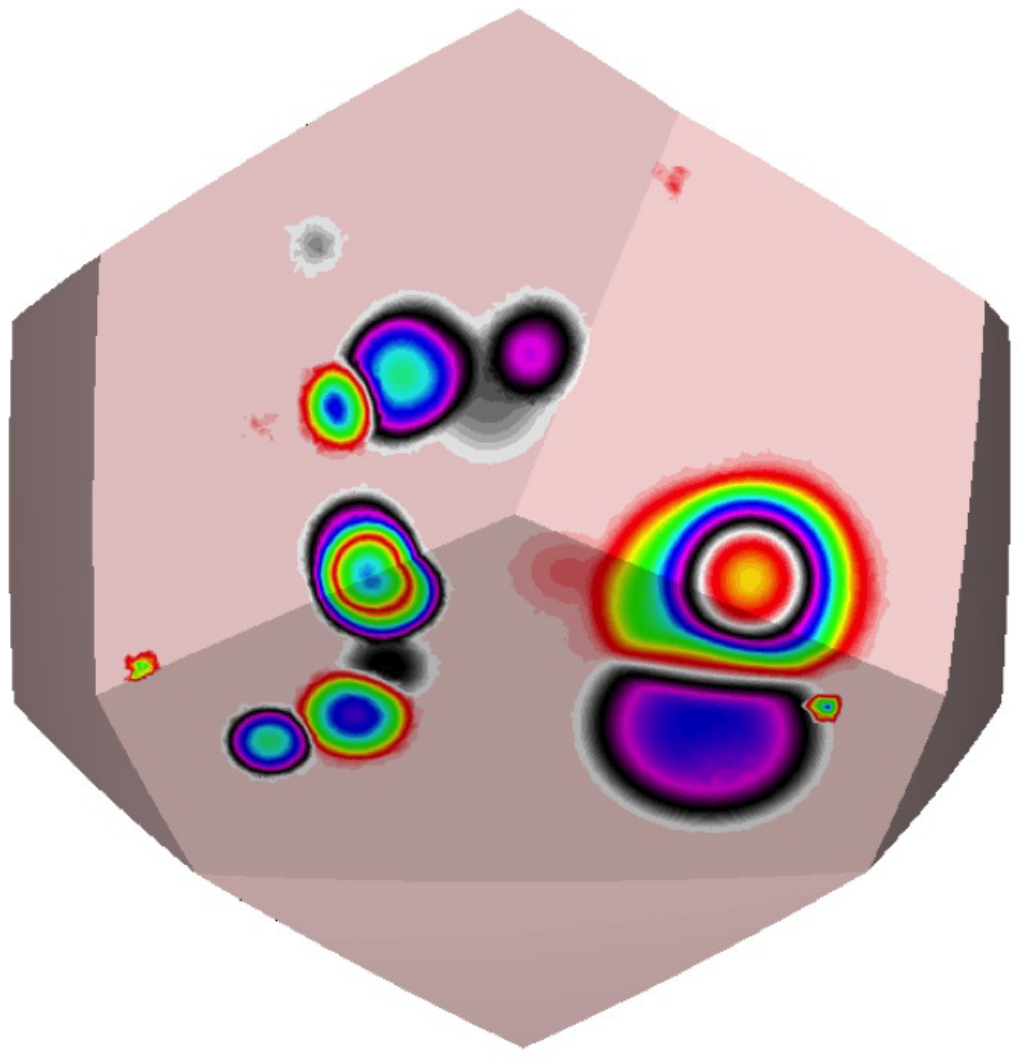}
\includegraphics[scale=0.24]{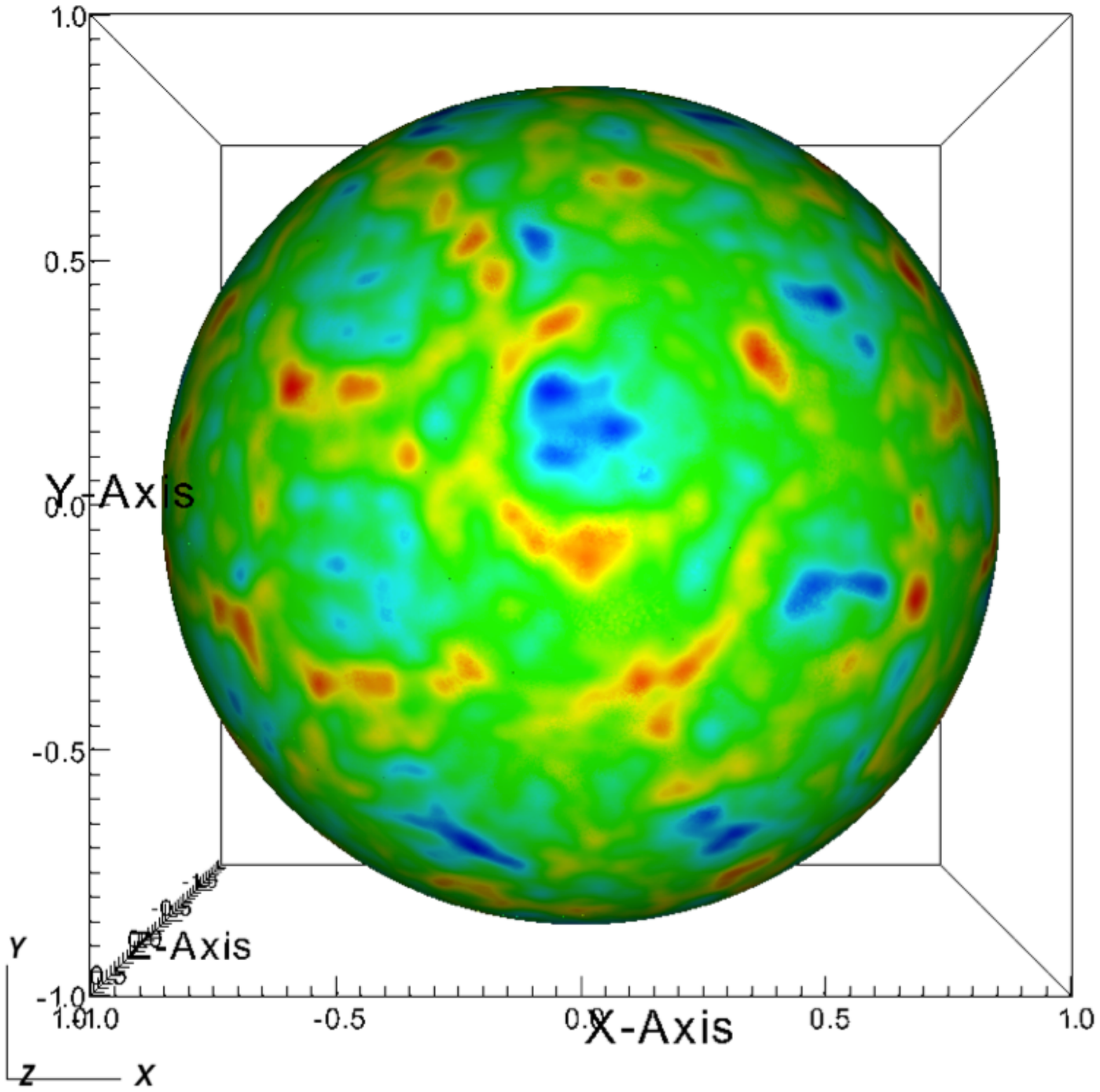}
\includegraphics[scale=0.255]{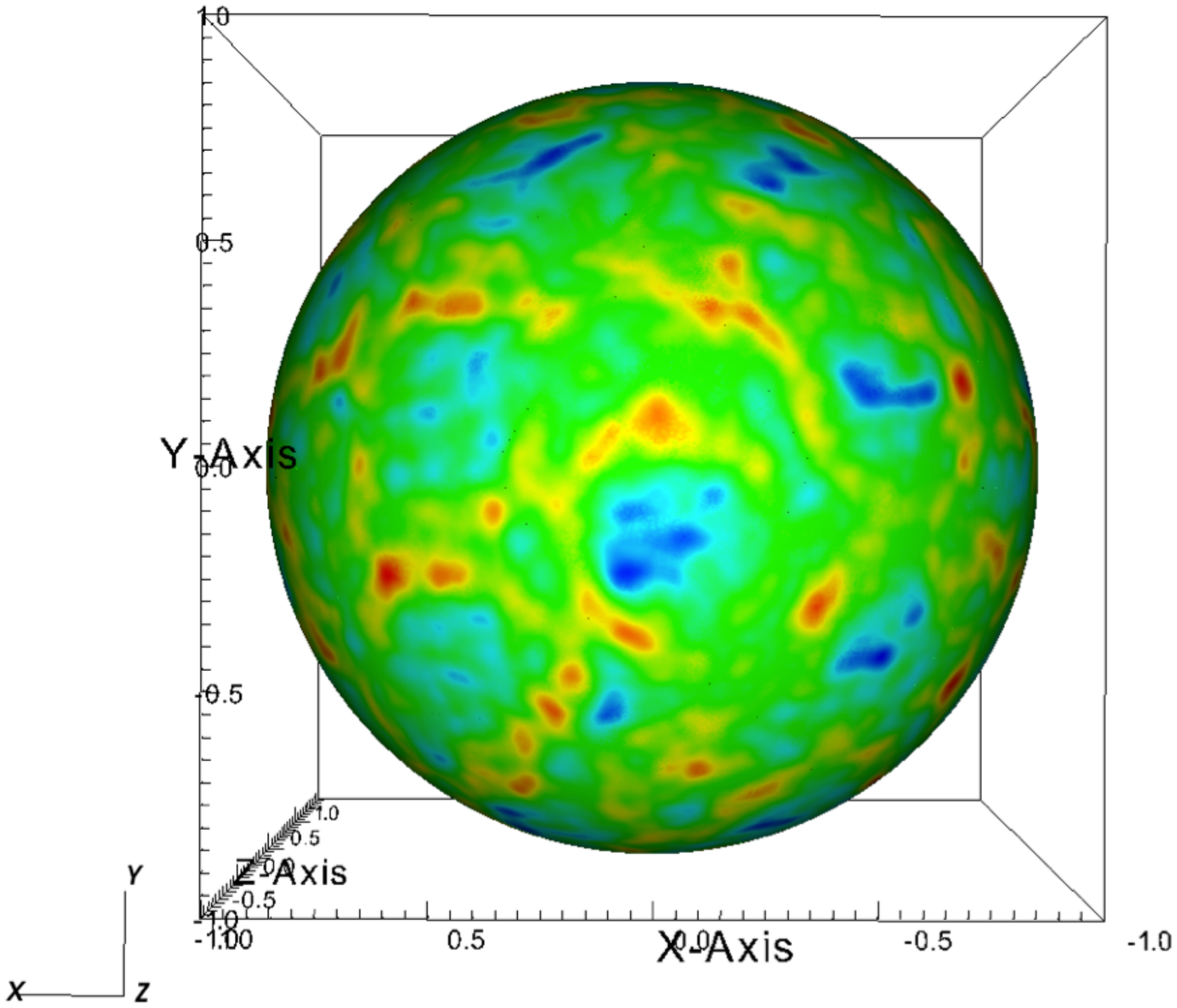}
\caption{\label{skyDOD4}}{An initial random fluctuation and two
  antipodal views of the deep sky for $\chi=\pi/2$.}
\end{center}
\end{figure}

We now present a more complex solution associated to an
initial data $u(t^*,.)=Init_3$, $\partial_tu(t^*,.)=0$ given at time $t^*=2$. $Init_3$ is composed of $Init_2$ to which we have added two another fields of the same type (see 
(\ref{Init})): one with $X_0=(0.1,0.1,0.1)$ and $R_0=0.1$, the other
with $X_0=(-0.15,0,-0.1)$ and $R_0=0.05$. We take $A_0=100$ again.
Figure \ref{InitskyDOD2} shows $Init_3$ and the asymptotic profile on
the two-plane $10x+5y-15z=0$. 
 The left figure \ref{skyDOD2} shows the pull-back of $\Psi_{\infty}$ on a part
 of the universal covering $\mathcal{S}^3$ of $\mathbf{K}$, included
 in $\mathcal{B}'$ (see (\ref{strois})), composed by
 $\mathcal{F}_v$, $p\left(g_i(\mathcal{F})\right)$ with $i=1,...,12$,
 $p\left(g_i\circ g_6(\mathcal{F})\right)$ with $i=1,...,5$,
 $p\left(g_i\circ g_3(\mathcal{F})\right)$ with $i=1, 2, 4, 5, 6$
 (see (\ref{projectionp}) and the Appendix). On the right the
 deep sky for $\chi\simeq 0.582$ is depicted. Similarly Figure
 \ref{skyDOD3} presents the case of a large conformal time. On the left, the part of the universal covering of
$\mathbf{K}$ is included in the second unit ball $\mathcal{B}''$
associated to $\chi\in[\frac{\pi}{2},\pi]$. This set is composed by $p(-\mathbf{1}\mathcal{F})$, $p\left(g_i(-\mathbf{1}\mathcal{F})\right)$ with $i=1,...,12$,
 $p\left(g_i\circ g_6(-\mathbf{1}\mathcal{F})\right)$ with $i=1,...,5$,
 $p\left(g_i\circ g_3(-\mathbf{1}\mathcal{F})\right)$ with $i=1, 2, 4, 5, 6$.
On the right the deep sky for $\chi\simeq 2,498$. We note the
correlations along six pairs of circles again.\\

We achieve this part with the more realistic case of a random initial
fluctuation ${\it Init_4}$ composed with 100
localized fluctuations of type (\ref{Init}). The radius $R_0$, the
centers $X_0$,  and the amplitudes $A_0$ are randomly choosen such
that $-100\leq A_0\leq 100$, $R_0\leq 0.1$. The left figure \ref{skyDOD4}
shows ${\it Init_4}$ on a cut of the dodecahedron along three planes. We compute
the solution for $u(t^*,.)={\it Init_4}$, $\partial_tu(t^*,.)=0$ given
at time $t^*=2$. The right figure \ref{skyDOD4} presents two antipodal
views of the deep sky for $\chi=\pi/2$. We can guess the correlations of the fluctuations along six pairs of matched antipodal circles.

\section{Conclusion}
In this paper we have considered models of dynamical spacetimes for
which the spatial section is the Poincar\'e dodecahedron. We have
solved the Cauchy problem for the D'Alembertian and expressed the
initial value problem in terms of variational formulation put on the
fundamental domain. We have established the existence of an asymptotic
state $\Psi_{\infty}$  at the time infinity for the finite energy
waves:
$$
\Psi_{\infty}:=\lim_{t\rightarrow\infty}\Psi(t).
$$
If $\Psi$ is the perturbation of the metric, this state describes the temperature fluctuations
in the deep sky
thanks to the
(ordinary) Sachs-Wolfe formula
$$
\frac{\delta T}{T}\simeq\frac{1}{3}\Psi_{\infty}.
$$
We have
obtained the analytic expression of this limit state for two important
cases of accelerating universes associated to the scale factors
$H^{-1}\cosh(Ht)$ and $e^t$. It turns out that $\Psi_{\infty}$ is given by a
pseudodifferential operator acting on the inital data of the
perturbation, $(\Psi(0),\partial_t\Psi(0))$ that is not convenient for
the computation.  Therefore we have performed a numerical method
for solving the time evolution problem. We also emphasize that this step is necessary if we
want to use the integrated Sachs-Wolfe formula involving $\Psi(t)$ for
any time $t$ since the last time of recombination. We employ the $\P_2$ type
finite elements for the sake of accuracy. The numerical experiments show
a good agreement with the theoretical results on the asymptotic
behaviours. That proves the robustness and the accuracy of our method
as well as its efficiency to compute the deep sky.
Its drawback is the lenght of the computation, usually we need one week to go from $t$ to $t +
1.2$ and this time of calculation would explode for a very refined
mesh. We can hope to be able to use
a parallel computing together with a domain decomposition on a
significantly refined mesh. Then we shall be able to treat the heavy
computations associated to random initial data and the {\it integrated}
Sachs-Wolfe formula.

\section*{acknowledgement}

This research was partly supported by the ANR funding
ANR-12-BS01-012-01.
\appendix 

\section*{Appendix} This appendix is devoted to a brief description of $\mathcal{F}$ and
$\mathcal{F}_v$. For more details on the construction of these domains
we refer to \cite{PDS}. 

First of all we recall that the binary
icosahedral group $\mathcal{I}^*$ is generated by two isometries $s$ and $\gamma$. Denoting by $\sigma=(1+\sqrt{5})/2$ the golden number we have: 
\[\begin{aligned}
\mathcal{I}^*&=\left< s,\gamma\; |\; (s\gamma)^2=s^3=\gamma^5\right>\\
&=\left\{\pm \mathbf{1},\pm \mathbf{ i},\pm \mathbf{j},\pm \mathbf{k},\frac12 (\pm \mathbf{1} \pm \mathbf{i} \pm \mathbf{j} \pm \mathbf{k}),\frac12(0 \mathbf{1} \pm \mathbf{i} \pm\frac{1}{\sigma} \mathbf{j} \pm\sigma \mathbf{k})\; (with\; even\; permutations) \right\}.
\end{aligned}
\]
where $\{\mathbf{1},\mathbf{i},\mathbf{j},\mathbf{k\}}$ are the four
basic quaternions of the set $\HH$ of all quaternions. Twelve Cliffort
translations $g_i$ belonging to $\mathcal{I}^*$ are involved in the construction
of $\mathcal{F}$:
\begin{eqnarray}
 \ g_1:=\frac12 \sigma \mathbf{1}+\frac12 \frac{1}{\sigma} \mathbf{i}+\frac12 \mathbf{j},\quad &\; g_2:=\frac12 \sigma \mathbf{1}+\frac12\mathbf{ i} -\frac12 \frac{1}{\sigma} \mathbf{k},\quad &\; g_3:=\frac12 \sigma \mathbf{1}+\frac12 \frac{1}{\sigma}\mathbf{ j}-\frac12 \mathbf{k},\nonumber\\
 \ g_4:=\frac12 \sigma \mathbf{1}-\frac12 \frac{1}{\sigma}\mathbf{ i}+\frac12 \mathbf{j},\quad &\; g_5:=\frac12 \sigma \mathbf{1}+\frac12 \frac{1}{\sigma}\mathbf{ j}+\frac12 \mathbf{k},\quad &\; g_6:=\frac12 \sigma \mathbf{1}+\frac12 \mathbf{i}+\frac12 \frac{1}{\sigma} \mathbf{k}. \nonumber
\end{eqnarray}
They are such that: $\quad \forall i \in \{1,...,6\},\;
g_i(F_i)=F_{i+6}$ ($F_i$ denote faces of $\mathcal{F}$). The inverses of these six first translations are the six other translations:
\begin{eqnarray*}
\ g_7:=(g_1)^{-1}=\frac12 \sigma \mathbf{1}-\frac12 \frac{1}{\sigma} \mathbf{i}-\frac12 \mathbf{j},&\; g_8:=(g_2)^{-1}=\frac12 \sigma \mathbf{1}-\frac12\mathbf{ i} +\frac12 \frac{1}{\sigma} \mathbf{k},\\
\ g_9:=(g_3)^{-1}=\frac12 \sigma \mathbf{1}-\frac12 \frac{1}{\sigma}\mathbf{ j}+\frac12 \mathbf{k},&\; g_{10}:=(g_4)^{-1}=\frac12 \sigma \mathbf{1}+\frac12 \frac{1}{\sigma}\mathbf{ i}-\frac12 \mathbf{j}, \\
\ g_{11}:=(g_5)^{-1}=\frac12 \sigma \mathbf{1}-\frac12 \frac{1}{\sigma}\mathbf{ j}-\frac12 \mathbf{k},&\; g_{12}:=(g_6)^{-1}=\frac12 \sigma \mathbf{1}-\frac12 \mathbf{i}-\frac12 \frac{1}{\sigma} \mathbf{k}. 
\end{eqnarray*}

\begin{figure}[!h]
\begin{center}
\includegraphics[scale=0.5]{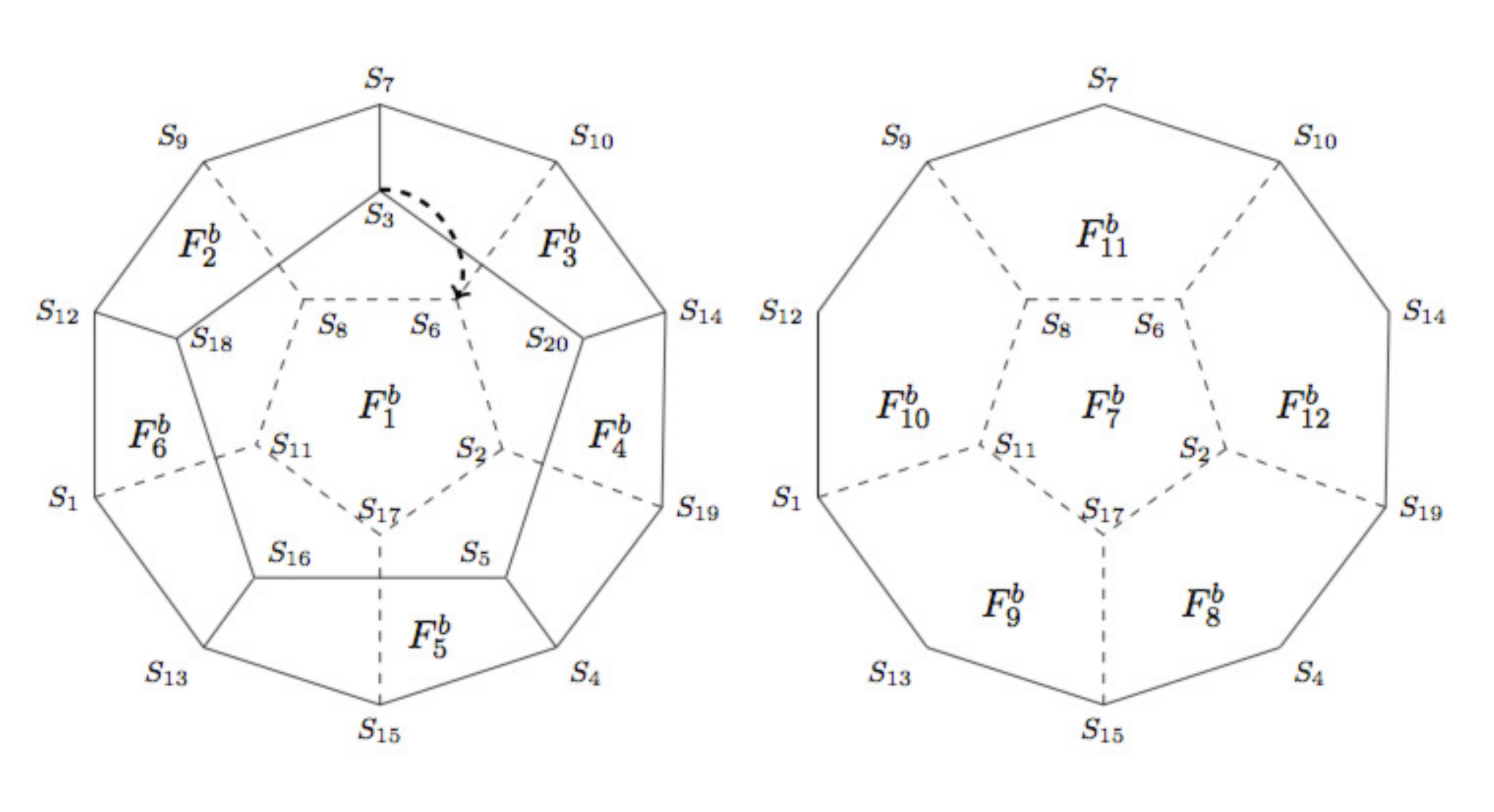}
\caption{Diagram of vertices $S_i$, edges and faces of $\mathcal{F}_v$.}
\label{DODSurf}
\end{center}
\end{figure}

We warn that Figure (\ref{DODSurf}) is not a faithful representation of
$\mathcal{F}_v$ as, for convenience, we have replaced the true
curved faces of $\mathcal{F}_v$
by the plane faces $F^b_i$ that are the projection on $\RR^3$ of the sets of all barycenters
in $\RR^4$ of the vertices $S_i^1,...,S_i^5$ of the faces $F_i$ of $\mathcal{F}$.

\begin{enumerate}
\item Coordinates of the vertices $S_i$ of $\mathcal{F}$:
\[
 \begin{array}{ll}
S_1=\frac{1}{2\sqrt{2}}\left(\sigma^2,-\frac{1}{\sigma},\frac{1}{\sigma},-\frac{1}{\sigma}\right),&S_2=\frac{1}{2\sqrt{2}}\left(\sigma^2,1,\frac{1}{\sigma^2},0\right),\\
S_3=\frac{1}{2\sqrt{2}}\left(\sigma^2,-\frac{1}{\sigma},-\frac{1}{\sigma},\frac{1}{\sigma}\right),&S_4=\frac{1}{2\sqrt{2}}\left(\sigma^2,\frac{1}{\sigma},-\frac{1}{\sigma},-\frac{1}{\sigma}\right),\\
S_5=\frac{1}{2\sqrt{2}}\left(\sigma^2,0,-1,-\frac{1}{\sigma^2}\right),&S_6=\frac{1}{2\sqrt{2}}\left(\sigma^2,\frac{1}{\sigma},\frac{1}{\sigma},\frac{1}{\sigma}\right),\\
S_7=\frac{1}{2\sqrt{2}}\left(\sigma^2,-\frac{1}{\sigma^2},0,1\right),&S_8=\frac{1}{2\sqrt{2}}\left(\sigma^2,0,1,\frac{1}{\sigma^2}\right),\\
S_9=\frac{1}{2\sqrt{2}}\left(\sigma^2,-\frac{1}{\sigma},\frac{1}{\sigma},\frac{1}{\sigma}\right),&S_{10}=\frac{1}{2\sqrt{2}}\left(\sigma^2,\frac{1}{\sigma^2},0,1\right),\\
S_{11}=\frac{1}{2\sqrt{2}}\left(\sigma^2,0,1,-\frac{1}{\sigma^2}\right),&S_{12}=\frac{1}{2\sqrt{2}}\left(\sigma^2,-1,\frac{1}{\sigma^2},0\right),\\
S_{13}=\frac{1}{2\sqrt{2}}\left(\sigma^2,-\frac{1}{\sigma^2},0,-1\right),&S_{14}=\frac{1}{2\sqrt{2}}\left(\sigma^2,\frac{1}{\sigma},-\frac{1}{\sigma},\frac{1}{\sigma}\right),\\
S_{15}=\frac{1}{2\sqrt{2}}\left(\sigma^2,\frac{1}{\sigma^2},0,-1\right),&S_{16}=\frac{1}{2\sqrt{2}}\left(\sigma^2,-\frac{1}{\sigma},-\frac{1}{\sigma},-\frac{1}{\sigma}\right),\\
S_{17}=\frac{1}{2\sqrt{2}}\left(\sigma^2,\frac{1}{\sigma},\frac{1}{\sigma},-\frac{1}{\sigma}\right),&S_{18}=\frac{1}{2\sqrt{2}}\left(\sigma^2,-1,-\frac{1}{\sigma^2},0\right),\\
S_{19}=\frac{1}{2\sqrt{2}}\left(\sigma^2,1,-\frac{1}{\sigma^2},0\right),&S_{20}=\frac{1}{2\sqrt{2}}\left(\sigma^2,0,-1,\frac{1}{\sigma^2}\right).
\end{array}
\]

\item Images by the Clifford translation $g_i$ of the face $F_i$ of $\mathcal{F}$ and of its edges. \\
$g_1$ maps $F_1$ to $F_7$, and we have for the vertices and the edges:\\
$g_1(S_{3}) =S_{6},\quad g_1( S_{18})=S_{8},\quad g_1(S_{16})=S_{11},\quad g_1(S_{5} )=S_{17},\quad g_1(S_{20})=S_{2}$, 

\noindent
$g_1(\wideparen{S_{3}S_{18}})=\wideparen{S_{6}S_{8}},\; g_1(\wideparen{S_{18}S_{16}})=\wideparen{S_{8}S_{11}},\; g_1(\wideparen{S_{16}S_{5}})=\wideparen{S_{11}S_{17}},\; g_1(\wideparen{S_{5}S_{20}})=\wideparen{S_{17}S_{2}},$

\noindent
 $g_1(\wideparen{S_{20}S_{3}})=\wideparen{S_{2}S_{6}}$.\\

$g_2$ maps $F_2$ to $F_8$, and we have for the vertices and the edges:\\
$g_2(S_{18}) =S_{15},\quad g_2( S_{12})=S_{17},\quad g_2(S_{9})=S_{2},\quad g_2(S_{7} )=S_{19},\quad g_2(S_{3})=S_{4}$, 

\noindent
$g_2(\wideparen{S_{18}S_{12}})=\wideparen{S_{15}S_{17}},\; g_2(\wideparen{S_{17}S_{2}})=\wideparen{S_{8}S_{11}},\; g_2(\wideparen{S_{9}S_{7}})=\wideparen{S_{2}S_{19}},\; g_2(\wideparen{S_{7}S_{3}})=\wideparen{S_{19}S_{4}},$

\noindent
$ g_2(\wideparen{S_{3}S_{18}})=\wideparen{S_{4}S_{15}}$.\\

$g_3$ maps $F_3$ to $F_9$, and we have for the vertices and the edges:\\
$ g_3(S_{3}) =S_{1},\quad g_3( S_{7})=S_{11},\quad g_3(S_{10})=S_{17},\quad g_3(S_{14} )=S_{15},\quad g_3(S_{20})=S_{13}$, 

\noindent
$g_3(\wideparen{S_{3}S_{7}})=\wideparen{S_{1}S_{11}},\; g_3(\wideparen{S_{7}S_{10}})=\wideparen{S_{11}S_{17}},\; g_3(\wideparen{S_{10}S_{14}})=\wideparen{S_{17}S_{15}},$\\ 
$g_3(\wideparen{S_{14}S_{20}})=\wideparen{S_{15}S_{13}},\;g_3(\wideparen{S_{20}S_{3}})=\wideparen{S_{13}S_{1}}$.\\

$g_4$ maps $F_4$ to $F_{10}$, and we have for the vertices and the edges:\\
$g_4(S_{20}) =S_{9},\quad g_4( S_{14})=S_{8},\quad g_4(S_{19})=S_{11},\quad g_4(S_{4} )=S_{1},\quad g_4(S_{5})=S_{12}$, 

\noindent
$g_4(\wideparen{S_{20}S_{14}})=\wideparen{S_{9}S_{8}},\; g_4(\wideparen{S_{14}S_{19}})=\wideparen{S_{8}S_{11}},\; g_4(\wideparen{S_{19}S_{4}})=\wideparen{S_{11}S_{1}},\; g_4(\wideparen{S_{4}S_{5}})=\wideparen{S_{1}S_{12}},$

\noindent
$g_4(\wideparen{S_{5}S_{20}})=\wideparen{S_{12}S_{9}}$.\\

$g_5$ maps $F_5$ to $F_{11}$, and we have for the vertices and the edges:\\
$g_5(S_{5}) =S_{10},\quad g_5( S_{4})=S_{6},\quad g_5(S_{15})=S_{8},\quad g_5(S_{13} )=S_{9},\quad g_5(S_{16})=S_{7}$, 

\noindent
$g_5(\wideparen{S_{5}S_{4}})=\wideparen{S_{10}S_{6}},\; g_5(\wideparen{S_{4}S_{15}})=\wideparen{S_{6}S_{8}},\; g_5(\wideparen{S_{15}S_{13}})=\wideparen{S_{8}S_{9}},\; g_5(\wideparen{S_{13}S_{16}})=\wideparen{S_{9}S_{7}},$

\noindent
$g_5(\wideparen{S_{16}S_{5}})=\wideparen{S_{7}S_{10}}.$\\

$g_6$ maps $F_6$ to $F_{12}$, and we have for the vertices and the edges:\\
$g_6(S_{16}) =S_{19},\quad g_6( S_{13})=S_{2},\quad g_6(S_{1})=S_{6},\quad g_6(S_{12} )=S_{10},\quad g_6(S_{18})=S_{14}$, 

\noindent
$g_6(\wideparen{S_{16}S_{13}})=\wideparen{S_{19}S_{2}},\; g_6(\wideparen{S_{13}S_{1}})=\wideparen{S_{2}S_{6}},\; g_6(\wideparen{S_{1}S_{12}})=\wideparen{S_{6}S_{10}},$\\
$g_6(\wideparen{S_{12}S_{18}})=\wideparen{S_{10}S_{14}},\;g_6(\wideparen{S_{18}S_{16}})=\wideparen{S_{14}S_{19}}$.
\end{enumerate}
Then we define the equivalence classes on $\mathcal{F}$ by:
\[
 \begin{array}{l}
 q \in \stackrel{\circ}{\mathcal{F}}\quad \Rightarrow \dot{q}=\{q\},\\
\exists (i,j,k), \; q\in F_i\cap F_j \cap F_k \quad  \Rightarrow \dot{q}=\{q,g_i(q),g_{j}(q),g_{k}(q)\},\\
(\exists (i,j),\; q\in F_i\cap F_j )\quad and \quad (\forall k \neq i,\; j,\;\; q\notin F_k) \quad  \Rightarrow \dot{q}=\{q,g_i(q),g_{j}(q)\},\\
(\exists i,\; q\in F_i)\quad and \quad (\forall j\neq i,\; q\notin F_j) \quad \Rightarrow \dot{q}=\{q,g_i(q)\}.
\end{array}
\]
Here the integers $i,j,k$ belong to $\{1,...,12\}$, and $F_i$ denotes a
face of $\mathcal{F}$.



\end{document}